\documentclass[aps,prx,reprint,superscriptaddress,nofootinbib]{revtex4-2}
\usepackage{blindtext}
\usepackage{lipsum}
\usepackage{graphics}
\usepackage{amsmath}
\usepackage{graphicx}
\usepackage{graphics}
\usepackage{amssymb}
\usepackage{verbatim}
\usepackage{framed}
\usepackage{float}
\usepackage{braket}
\usepackage{bm}
\usepackage{mathtools}
\usepackage{bbm}
\usepackage{multirow}
\usepackage[toc,page]{appendix}

\usepackage[colorlinks=true,citecolor=dodgerblue,linkcolor=dodgerblue,urlcolor=dodgerblue,pdftitle={2Hof}]{hyperref}

\makeatletter
\def\l@subsubsection#1#2{} 
\makeatother

\usepackage[dvipsnames]{xcolor}
\usepackage[normalem]{ulem}
\usepackage{wasysym} 
\usepackage[export]{adjustbox}

\makeatletter
\DeclareFontFamily{OMX}{MnSymbolE}{}
\DeclareSymbolFont{MnLargeSymbols}{OMX}{MnSymbolE}{m}{n}
\SetSymbolFont{MnLargeSymbols}{bold}{OMX}{MnSymbolE}{b}{n}
\DeclareFontShape{OMX}{MnSymbolE}{m}{n}{
    <-6>  MnSymbolE5
   <6-7>  MnSymbolE6
   <7-8>  MnSymbolE7
   <8-9>  MnSymbolE8
   <9-10> MnSymbolE9
  <10-12> MnSymbolE10
  <12->   MnSymbolE12
}{}
\DeclareFontShape{OMX}{MnSymbolE}{b}{n}{
    <-6>  MnSymbolE-Bold5
   <6-7>  MnSymbolE-Bold6
   <7-8>  MnSymbolE-Bold7
   <8-9>  MnSymbolE-Bold8
   <9-10> MnSymbolE-Bold9
  <10-12> MnSymbolE-Bold10
  <12->   MnSymbolE-Bold12
}{}

\let\llangle\@undefined
\let\rrangle\@undefined
\DeclareMathDelimiter{\llangle}{\mathopen}%
                     {MnLargeSymbols}{'164}{MnLargeSymbols}{'164}
\DeclareMathDelimiter{\rrangle}{\mathclose}%
                     {MnLargeSymbols}{'171}{MnLargeSymbols}{'171}
\makeatother

\usepackage{ulem}

\usepackage{amssymb}
\usepackage{pifont}
\newcommand{\cmark}{\ding{51}}%
\newcommand{\xmark}{\ding{55}}%

\newcommand{\mbbL}{\mathbb{L}}

\newcommand{\Top}{\mathsf{T}}
\newcommand{\Bot}{\mathsf{B}}

\newcommand{\br}{\mathbf{r}}
\newcommand{\bu}{\mathbf{u}}
\newcommand{\bk}{\mathbf{k}}
\newcommand{\bq}{\mathbf{q}}
\newcommand{\bxhat}{\hat{\mathbf{x}}}
\newcommand{\byhat}{\hat{\mathbf{y}}}
\newcommand{\bzhat}{\hat{\mathbf{z}}}
\newcommand{\phii}{\varphi}

\newcommand{\mcA}{\mathcal{A}}
\newcommand{\mcH}{\mathcal{H}}
\newcommand{\mcO}{\mathcal{O}}
\newcommand{\mcP}{\mathcal{P}}
\newcommand{\mcV}{\mathcal{V}}
\newcommand{\mcC}{\mathcal{C}}

\newcommand{\Isym}{\mathcal{I}}

\newcommand{\wilde}[1]{\widetilde{#1}}

\newcommand{\bG}{\mathbf{G}}
\newcommand{\Ad}{\mathrm{Ad}}

\newcommand{\EBZ}{\mathrm{EBZ}}

\newcommand{\n}[1]{\left| #1 \right|}
\newcommand{\st}[1]{\left\{ #1 \right\}}

\renewcommand{\v}[1]{\boldsymbol{#1}}

\definecolor{orange(ryb)}{HTML}{FFA500}
\definecolor{dodgerblue}{HTML}{1E90FF}
\definecolor{pinkerton}{HTML}{EC368D}
\definecolor{forest}{HTML}{6DD189}

\definecolor{edit}{HTML}{000000}

\usepackage{tikz}
\usepackage{tikz-cd}
\usetikzlibrary{arrows}
\usetikzlibrary{snakes}
\usetikzlibrary{intersections}
\usetikzlibrary{shapes.geometric}
\usetikzlibrary{decorations.pathmorphing, patterns,shapes}
\usetikzlibrary{decorations.markings}

\tikzset{
	partial ellipse/.style args={#1:#2:#3}{
		insert path={+ (#1:#3) arc (#1:#2:#3)}
	}
}

\tikzset{
	mid arrow/.style={postaction={decorate,decoration={
				markings,
				mark=at position .575 with {\arrow[#1]{stealth}}
	}}},
	near arrow/.style={postaction={decorate,decoration={
				markings,
				mark=at position .275 with {\arrow[#1]{stealth}}
	}}},
	far arrow/.style={postaction={decorate,decoration={
				markings,
				mark=at position .800 with {\arrow[#1]{stealth}}
	}}},
}

\pgfdeclarepatternformonly{south west lines}{\pgfqpoint{-0pt}{-0pt}}{\pgfqpoint{3pt}{3pt}}{\pgfqpoint{3pt}{3pt}}{
	\pgfsetlinewidth{0.4pt}
	\pgfpathmoveto{\pgfqpoint{0pt}{0pt}}
	\pgfpathlineto{\pgfqpoint{3pt}{3pt}}
	\pgfpathmoveto{\pgfqpoint{2.8pt}{-.2pt}}
	\pgfpathlineto{\pgfqpoint{3.2pt}{.2pt}}
	\pgfpathmoveto{\pgfqpoint{-.2pt}{2.8pt}}
	\pgfpathlineto{\pgfqpoint{.2pt}{3.2pt}}
	\pgfusepath{stroke}}

\makeatletter
\renewcommand\onecolumngrid{
\do@columngrid{one}{\@ne}%
\def\set@footnotewidth{\onecolumngrid}
\def\footnoterule{\kern-6pt\hrule width 1.5in\kern6pt}%
}

\renewcommand\twocolumngrid{
        \def\footnoterule{
        \dimen@\skip\footins\divide\dimen@\thr@@
        \kern-\dimen@\hrule width.5in\kern\dimen@}
        \do@columngrid{mlt}{\tw@}
}
\makeatother

\begin{document}

\title{Superconductivity in a Topological Lattice Model with Strong Repulsion}

\author{Rahul Sahay}
\thanks{These authors contributed equally to this work.}
\affiliation{Department of Physics, Harvard University, Cambridge, MA 02138, USA}
\author{Stefan Divic}
\thanks{These authors contributed equally to this work.}
\affiliation{Department of Physics, University of California, Berkeley, CA 94720, USA}
\author{Daniel E. Parker}
\affiliation{Department of Physics, University of California at San Diego, La Jolla, CA 92093, USA}
\affiliation{Department of Physics, Harvard University, Cambridge, MA 02138, USA}
\author{Tomohiro Soejima}
\affiliation{Department of Physics, Harvard University, Cambridge, MA 02138, USA}
\affiliation{Department of Physics, University of California, Berkeley, CA 94720, USA}
\author{Sajant Anand}
\affiliation{Department of Physics, University of California, Berkeley, CA 94720, USA}
\author{Johannes Hauschild}
\affiliation{Technical University of Munich, TUM School of Natural Sciences,
Physics Department, James-Franck-Str. 1, 85748 Garching, Germany}
\affiliation{Munich Center for Quantum Science and Technology (MCQST), Schellingstra{\ss}e 4, 80799 M\"unchen, Germany}
\author{Monika Aidelsburger}
\affiliation{Max-Planck-Institut f\"ur Quantenoptik, Hans-Kopfermann-Stra{\ss}e 1, 85748 Garching, Germany}
\affiliation{Fakult\"at f\"ur Physik, Ludwig-Maximilians-Universit\"at M\"unchen, Schellingstra{\ss}e 4, 80799 M\"unchen, Germany}
\affiliation{Munich Center for Quantum Science and Technology (MCQST), Schellingstra{\ss}e 4, 80799 M\"unchen, Germany}
\author{Ashvin Vishwanath}
\affiliation{Department of Physics, Harvard University, Cambridge, MA 02138, USA}
\author{Shubhayu Chatterjee}
\affiliation{Department of Physics, University of California, Berkeley, CA 94720, USA}
\affiliation{Department of Physics, Carnegie Mellon University, Pittsburgh, PA 15213, USA}
\author{Norman Y. Yao}
\affiliation{Department of Physics, Harvard University, Cambridge, MA 02138, USA}
\author{Michael P. Zaletel}
\affiliation{Department of Physics, University of California, Berkeley, CA 94720, USA}
\affiliation{Material Science Division, Lawrence Berkeley National Laboratory, Berkeley, CA 94720, USA}

\date{\today}

\begin{abstract}

The highly tunable nature of synthetic quantum materials--—both in the solid-state and cold atom contexts--—invites examining which microscopic ingredients aid in the realization of correlated phases of matter such as superconductors.
Recent experimental advances in moir\'e materials suggest that unifying the features of the Fermi-Hubbard model and quantum Hall systems creates a fertile ground for the emergence of such phases.
Here, we introduce the ``double Hofstadter'' model, a minimal 2D lattice model that incorporates exactly these features: time-reversal symmetry, band topology, and strong repulsive interactions.
By using infinite cylinder density matrix renormalization group methods (cylinder iDMRG), we investigate the ground state phase diagram of this model. We find that it hosts an interaction-induced quantum spin Hall insulator and demonstrate that weakly hole-doping this state gives rise to a  superconductor at a finite circumference, with indications that this behavior persists on larger cylinders.
At the aforementioned circumference, the superconducting phase is surprisingly robust to perturbations including additional repulsive interactions in the pairing channel.
By developing a technique to probe the superconducting gap function in iDMRG, we phenomenologically characterize the superconductor.
Namely, we demonstrate that it is formed from the weak pairing of holes atop the quantum spin Hall insulator. 
Furthermore, we determine the pairing symmetry of the superconductor, finding it to be $p$-wave---reminiscent of the unconventional superconductivity reported in experiments on twisted bilayer graphene (TBG).
Motivated by this, we elucidate structural similarities and differences between our model and those of TBG in its chiral limit.
Finally, to provide a more direct experimental realization, we detail an implementation of our Hamiltonian in a system of cold fermionic alkaline-earth atoms in an optical lattice.

\end{abstract}

\maketitle

\tableofcontents

\section{Introduction} \label{sec-intro}

Superconductors are remarkable phases of matter wherein charge-$e$ fermions---which microscopically repel one another---instead bind into charge-$2e$ Cooper pairs that then condense.
Traditionally, these phases are best understood in the context of weak interactions~\cite{mineev1999introduction, OG_BCS, KLsuperconductivity}.
For example, in the paradigm put forth by Bardeen, Cooper, and Schrieffer (BCS), phonons mediate pairing between electrons near the Fermi surface of a weakly-correlated Fermi liquid~\cite{OG_BCS}.
Despite its widespread applicability, the limitations of the phonon-based theory are suggested by the experimental discovery of ``unconventional'' superconductors, which display a plethora of exotic phenomena ranging from high transition temperatures, to non-standard pairing symmetries, and parent states with anomalously low carrier densities~\cite{mineev1999introduction, Sigrist_Unconventional_SC}.
Such superconductors arise in surprisingly distinct settings, e.g., the cuprate compounds~\cite{shen2008cuprate,anderson1997theory,LeeNagaosaWenRMP,Sachdev_cuprate_RMP,keimer2014high}, iron pnictides~\cite{wen2011materials,si2016high,fernandes2022iron}, and possibly even in twisted bilayer graphene (TBG)~\cite{cao2018unconventional,yankowitz2019tuning,lu2019superconductors,codecido2019correlated, arora2020superconductivity, nuckolls2023quantum} and other graphene multilayers~\cite{NadjPerge2022, park2021tunable, kim2021Alternating, kim2022evidence, pablo2022multilayer, Young2022Bernal, NadjPerge2023}.
This diversity motivates a fundamental question that is especially relevant in the present era of highly-tunable solid-state and cold atom synthetic quantum materials.
Namely, what microscopic ingredients aid the realization of unconventional superconducting ground states?

For the past several decades, investigations into unconventional superconductivity have brought to the forefront the ingredient of strong interactions.
This is encapsulated in the celebrated Fermi-Hubbard model~\cite{Hubbard1963, auerbach1998interacting,HMreview,HMreview2,HMreview3}, which consists of spinful fermions occupying tightly-localized orbitals coupled by strong repulsion.
These minimal ingredients underlie several proposed mechanisms for unconventional superconductivity~\cite{anderson1997theory,LeeNagaosaWenRMP,Sachdev_cuprate_RMP,keimer2014high} and give rise to numerous other exotic phenomena (see e.g.~\cite{HubbardSubir,LeeLee,WuPRX,scheurer2018topological,Sachdev_RPP,Szasz2020}).
As a consequence, the Fermi-Hubbard paradigm has been the subject of decades of sustained interest in condensed matter physics, pushing the forefront of theory and motivating new experimental paradigms such as cold atom quantum simulation~\cite{chiu2018quantum,mazurenko2017cold,mitra2018quantum,scherg2021observing,BOHRDTReview,spar2022realization,hartke2023direct,brown2019bad}.

Despite their success, the simple ingredients in the Fermi-Hubbard model are insufficient to describe strongly-interacting phenomena reliant on band topology, such as the celebrated quantum hall effects~\cite{girvin2002quantum,Hansson_QHReview}.
This setting of topological bands induced by strong magnetic fields, with accompanying time-reversal breaking, is naively unfavorable for superconductivity in the presence of repulsive interactions. However, a number of works have suggested regimes where superconductivity may emerge in topological Hofstadter-type models, such as~\cite{Shaffer2022selfsimilar, Tu2018Competing, Maska2001Upper, Iskin2017BCS, Iskin2017HofHub}. Recently, 
experiments on TBG and other graphene moir\'e multilayers---in which topology is widely believed to play a crucial role---present a related time-reversal-invariant context~\cite{Zou_TBG_Fragile, Po_Origin_of_Mott_PRX} in which superconductivity can and does appear~\cite{cao2018unconventional}.
As such, various studies have sought to demonstrate the presence of an all-electronic route to superconductivity in models of TBG and, more broadly, in electronic Hamiltonians with the ingredients of time-reversal symmetry, strong repulsive interactions, and topological bands.
However, owing to the complexity of the many-body problem, theoretical evidence for such a route has so far only derived from approximate or perturbative methods, e.g., mean-field theory~\cite{vlad_nick_SCgoldstone,KaneMeleSlaveBoson}, large-$N$ expansions~\cite{eslamskyrmeSC}, and effective continuum descriptions~\cite{shubsskyrme, eslamskyrmeSC, Christos_2020}.
Elucidating the existence and character of superconductivity in a fully-microscopic model with these ingredients, therefore, remains an important and outstanding challenge.

In this work, we introduce a minimal model---the ``double Hofstadter'' model--- that incorporates precisely these ingredients and employ unbiased matrix product state techniques to study its phase diagram on finite-circumference cylinders.
Our study reveals that their interplay leads to a plethora of exotic phenomena and culminates in the numerical demonstration of an unconventional superconducting ground state at a finite circumference, with further evidence that this behavior may persist on larger cylinders.



\begin{figure}[!t]
    \centering
    \includegraphics[width = 247 pt]{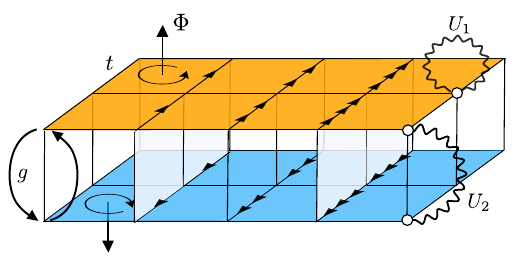}
    \caption{\textbf{The Double Hofstadter Model.} This work considers a simple lattice model with three essential ingredients: (1) narrow topological bands with (2) strong repulsive interactions and (3) time-reversal symmetry. These properties are realized in a bilayer lattice model of spinful fermions made by joining together two copies of the Harper-Hofstadter model with \textit{opposite magnetic fields in opposite layers}. Bonds represent electron hoppings with in-layer magnitude $t$ and inter-layer magnitude $g$. Peierls phases of $e^{i n\pi/2}$, graphically depicted via $n$ arrows along the direction of hopping, yield a magnetic flux of $\pm \Phi = \pm \pi/2$ per plaquette in the top and bottom layers, respectively. Finally, there are repulsive density-density interactions (wavy lines) with magnitude $U_1$ onsite and $U_2$ between the layers.
    }
    \label{fig:2HofSchematic}
\end{figure}

\section{Summary of Key Results} \label{sec-Key}

This work studies the \textit{double Hofstadter model}, a simple lattice model with the following properties:
\begin{enumerate}
    \item[1.] Short-range repulsive interactions (akin to the Hubbard model),
    \item[2.] Narrow or ``flat'' topological bands (akin to quantum Hall and some moir\'e systems),
    \item[3.] Time-reversal symmetry (facilitating superconducting phase coherence).
\end{enumerate}
To bake all of these ingredients into a single Hamiltonian, we start with the familiar Harper-Hofstadter model of spinful fermions on a square lattice with a magnetic flux $\Phi = \Phi_0/q$ per plaquette ~\cite{Hofstady,Harper_1955}.
This is a lattice model whose lowest band carries a Chern number and limits to the lowest Landau level as $q\to \infty$, but is not time-reversal symmetric.
We therefore consider a bilayer Hofstadter model (Fig.~\ref{fig:2HofSchematic}) with opposite magnetic fields in opposite layers, whereupon the model becomes time-reversal symmetric.
Finally, we allow tunneling between the layers and add strong local repulsive interactions, as in the Hubbard model. 
We analyze the resulting \textit{double Hofstadter model} and argue that it contains an unconventional $p$-wave superconductor of purely repulsive origin.
After defining the model precisely in Section~\ref{subsec-TheModel}, our key results are five-fold and culminate in connections to contemporary experiments.

\textbf{Flat Bands \& Fragile Topology:} First, in Section~\ref{sec:model}, we present the bandstructure of our model (See Fig.~\ref{fig:2HofBandStructure}).
Crucially, the lowest four bands are narrow and possess ``fragile topology"~\cite{PoFragileTopo,AhnPRX,SlagerFragile,CanoFragile,Else_Fragile}.
Even though their total Chern number vanishes, fragile topology implies that no symmetric, exponentially-localized Wannier basis for these bands exists (without the addition of trivial filled bands to the model).
This precludes any tight-binding description for the flat band subspace consisting of symmetric and exponentially-localized orbitals.

\textbf{Correlated Insulator at Half-Filling:} Next, in Section~\ref{sec-LAF}, we investigate the ground state phase diagram of the double Hofstadter model at half-filling of the flat band subspace. 
We first demonstrate analytically that the ground state at strong in-layer coupling is a generalized quantum Hall ferromagnet, which is a spontaneously-generated quantum spin Hall insulator.
We refer to this state as a layer antiferromagnet (``LAF'') as it is characterized by ferromagnetism within each layer but antiferromagnetism between layers [See Fig.~\ref{fig:LAF_and_competing}(a)].
We confirm this analytical understanding numerically using infinite cylinder density matrix renormalization group (cylinder iDMRG) methods.
Our simulation of the model is enabled by recent advances in the compression of matrix product operators~\cite{parker_2020}. 

\textbf{Unconventional Superconductivity:} Third, in Section~\ref{sec:SC}, we provide numerical evidence that hole-doping the LAF produces an unconventional superconducting ground state.
In this purely-repulsive setting, our evidence is based on state-of-the-art parallelized iDMRG numerics and is strongest at circumference $L_y = 5$.
There, we conclusively demonstrate both (algebraically) long-ranged superconducting correlations \textit{and} a finite charge gap.
Despite the exponential scaling of simulation complexity in $L_y$, we attempt a similar analysis on larger cylinders where we find preliminary---but ultimately inconclusive due to bond dimension limitations---evidence for a superconducting ground state.
We conclude by verifying that the superconducting phase at $L_y = 5$ is robust, with stability to several perturbations including additional repulsive interactions, and always coexists with LAF order.

In Section~\ref{sec-Gap_and_PairingSym}, we devise a novel and widely applicable method to probe the superconducting gap function in charge-conserving cylinder iDMRG, which we use to systematically characterize the putative superconductor.
We find that it is well-described by the weak pairing of quasi-holes above the strongly-correlated LAF insulator (i.e., it is ``BCS'' like) with a $p$-wave pairing symmetry.
We emphasize that this weak-pairing phenomenology emerges in a setting where electronic repulsion dominates the kinetic energy and is, moreover, unearthed using non-perturbative numerics that do not presuppose a mean-field description.
This observation then enables us to estimate the superconducting transition temperature. 

\textbf{Optical Lattice Realization:} We conclude by discussing connections between the double Hofstadter model and contemporary experiments in both cold atoms and the solid state.
First, drawing on recent developments in the quantum control of alkaline-earth atoms in optical lattices~\cite{zhang_spectroscopic_2014,mancini_observation_2015,miranda_site-resolved_2015,yamamoto_ytterbium_2016,hofrichter2016direct,kolkowitz_spinorbit-coupled_2017,heinz_2020,taie_observation_2020,oppong_2022, hofer2015observation, pagano_strongly_2015}, we develop a concrete experimental blueprint to realize the double Hofstadter model by encoding its degrees of freedom in the clock and hyperfine states of such atoms.

\textbf{Connection to Solid-State Experiments:} Next, we present a link between the double Hofstadter model and models of moir\'e materials. 
Explicitly, we define a mapping between the symmetries, Hamiltonian, and even some ground states of our model and those of chiral twisted bilayer graphene~\cite{TarnoChiral, kang2019strong, bultinck2020ground, vafek2020renormalization, BernevigTBGIV}.

\section{The Double Hofstadter Model}
\label{sec:model}

We now introduce the double Hofstadter model and its basic properties, showing how it incorporates repulsive interactions, narrow topological bands, and time-reversal symmetry. 
We define the model precisely in Sec.~\ref{subsec-TheModel} and summarize its key symmetries in Sec.~\ref{subsec-Symmetries}, focusing on time-reversal and magnetic translations.
Going to momentum space in Sec.~\ref{subsec-BandStructure}, we show that the lowest four bands of the model are energetically narrow, well-isolated, and topologically non-trivial.
Finally, Sec.~\ref{subsec-LayerPolarizedBasis} describes the projected model within the flat band subspace.
We identify a ``layer-polarized'' basis for this space, which will be the most natural basis to describe the many-body states, both numerically and analytically.

\subsection{The Model} \label{subsec-TheModel}

The double Hofstadter model consists of spinful fermions in a bilayer square lattice (See Fig.~\ref{fig:2HofSchematic}).
At each lattice site $\br = (x, y) \in \mathbb{Z}^2$, define a row vector of many-body creation operators,
\begin{equation} \label{eq-fermionicrow}
\hat{\psi}^\dag_\br =  \begin{pmatrix}
\hat{\psi}^{\dagger}_{\br\Top\uparrow} &
\hat{\psi}^{\dagger}_{\br\Top\downarrow} &
\hat{\psi}^{\dagger}_{\br\Bot\uparrow} &
\hat{\psi}^{\dagger}_{\br\Bot\downarrow}
\end{pmatrix}.
\end{equation}
We use notation $\ell~\in~\{ \Top, \Bot \}$ for layer and $s~\in~\{\uparrow, \downarrow\}$ for spin, with associated Pauli matrices $\ell^{0,x,y,z}$ and $s^{0,x,y,z}$.

The Hamiltonian has a few simple components. 
First, it contains nearest-neighbor hopping within each layer. Second, fermions in the top (bottom) layer experience an upwards (downwards)-pointing magnetic field that induces a $2\pi/q$ flux per plaquette, where $q$ is a fixed positive integer.
This is implemented in a tight-binding model via the Peierls substitution with a choice of electromagnetic gauge field whose value in each layer is $\textbf{A}_{\mathsf{T}/\mathsf{B}} = (0, \pm (2\pi/q) x,0)$.  %
Concretely, the in-layer hopping Hamiltonian consists of two time-reversal-related copies of the Harper-Hofstadter model with a $1/q$ flux fraction:
\begin{equation} \label{eq-2HofHopping}
    \hat{h}_t = -t \sum_{\br \in \mathbb{Z}^2} \hat{\psi}^{\dagger}_{\mathbf{r} + \hat{\mathbf{y}}} \, e^{\frac{2\pi i x}{q} \ell^z} \hat{\psi}_{\mathbf{r}} + \hat{\psi}^{\dagger}_{\mathbf{r} + \hat{\mathbf{x}}} \hat{\psi}_{\mathbf{r}} + \text{h.c.}, 
\end{equation}
where we use the convention that the top and bottom layers correspond to $\ell^z=+1$ and $-1,$ respectively.
Next, electrons may tunnel vertically between the two layers:
\begin{equation} \label{eq-2Hofg}
    \hat{h}_g = -g \sum_{\br \in \mathbb{Z}^2} \hat{\psi}^{\dagger}_\br\, \ell^x\, \hat{\psi}_\br.
\end{equation}
Finally, the Hamiltonian contains both an in-layer and inter-layer Hubbard interaction, given by
\begin{equation} \label{eq-2Hof_U1_and_U2}
\hat{V}_1 = \frac{U_1}{2} \sum_{\mathbf{r} \in \mathbb{Z}^2} :\hat{n}_{\mathbf{r}\Top}^2 + \hat{n}_{\mathbf{r}\Bot}^2: \quad \hat{V}_2 = U_2 \sum_{\mathbf{r} \in \mathbb{Z}^2} : \hat{n}_{\mathbf{r} \mathsf{T}} \hat{n}_{\mathbf{r}\mathsf{B}} :,
\end{equation}
where $\hat{n}_{\br \ell} = \sum_{s} \hat{n}_{\br \ell s}$ and where the colons indicate normal-ordering of the fermion operators.
With these three ingredients, the full Hamiltonian is (see Fig.~\ref{fig:2HofSchematic}):
\begin{equation}\label{eq-full_hamiltonian}
    \hat{H} = \hat{h}_t + \hat{h}_g + \hat{V}_1 + \hat{V}_2 = \hat{h} + \hat{V},
\end{equation}
where $\hat{h}$ is the non-interacting portion of the model and $\hat{V}$ consists of interaction terms.
In what follows, we set the magnitude of the in-layer hopping strength to $t=1$ and scale the other couplings to be in these units.
Furthermore, throughout the rest of this work, we will fix the flux fraction to $1/q = 1/4$ (the case depicted in Fig.~\ref{fig:2HofSchematic}).
We remark that spinless variants of this model have been explored in previous works for both bosons and fermions~\cite{goldman2010, aidelsburger_realization_2013, kennedy2013soc, Iskin2017HofHub, Iskin2017BCS, zhang2023synthetic}. However, we emphasize that most of the physics revealed in this work crucially depends on the presence of both the layer and spin degrees of freedom.

\subsection{Symmetries}
\label{subsec-Symmetries}

We now discuss the symmetries of the model beyond charge $U(1)_Q$ and spin $SU(2)_S$, which are manifest.
We will find that the coupling of two opposite Hofstadter layers makes both the magnetic point group and the translation symmetries of the model slightly non-trivial.

We first note that the model, while not directly invariant under time-reversal (which would flip the magnetic field penetrating either layer), is invariant under a modified symmetry $M_z \mathcal{T}$.
This operator consists of the usual time-reversal operator $\mathcal{T}$, that flips time and electronic spin, followed by an in-plane mirror flip $M_z$ that flips the layer.
Furthermore, the point group of $\hat{H}$ can be determined by considering the model's embedding into 3D space (see Fig.~\ref{fig:2HofSchematic}).
In particular, the model is invariant under the dihedral group of two-fold rotations around all three cartesian axes (though not four-fold, see below).
This group is generated by $\pi$-rotations around the $\bzhat$ and $\bxhat$ axes, which we denote by $C_{2z}$ and $C_{2x}$, respectively.
The latter naturally exchanges the two layers and is chosen to also flip spin.
The combination of the modified time-reversal symmetry with the point group defines the ``magnetic point group.'' As its generators, we take:
\begin{subequations}\label{eq:symmetry_actions}
\begin{align}
    \hat{C}_{2z} \hat{\psi}_{\br}^\dagger \hat{C}_{2z}^{-1} &= \hat{\psi}_{-\br}^\dagger \\
    \hat{C}_{2x} \hat{\psi}_{\br}^\dagger \hat{C}_{2x}^{-1} &= \hat{\psi}_{C_{2x} \br}^\dagger \ell^x s^x \\ 
    \hat{\mathcal{I}} \hat{\psi}_{\br}^\dagger \hat{\mathcal{I}}^{-1} &= \hat{\psi}_{-\br}^\dagger \ell^x s^x
\end{align}
\end{subequations}
where we have defined a \textit{spacetime inversion} symmetry $\hat{\mathcal{I}} = \hat{C}_{2z} \cdot \hat{M}_z \hat{\mathcal{T}}$, which has the convenient property that it preserves the electronic crystal momentum, and where $C_{2x}\br = (x,-y)$.
We remark that the spacetime inversion symmetry is anti-linear ($\hat{\mathcal{I}} i \hat{\mathcal{I}}^{-1} = -i$) while the ordinary point group symmetries are linear.

Next, the translation symmetries of $\hat{H}$ are also governed by the magnetic field.
Define the operators
\begin{align}\label{eq-decoupled_mag_trans}
    \hat{t}_{x}\hat{\psi}_{\br}^\dagger \hat{t}^{-1}_{x} = \hat{\psi}_{\br+\bxhat}^\dagger e^{\frac{2\pi i y}{q} \ell^z},\quad  \hat{t}_{y}\hat{\psi}_{\br}^\dagger \hat{t}^{-1}_{y} &= \hat{\psi}_{\br+\byhat}^\dagger.
\end{align}
These are \emph{not} symmetries (unless $g = 0$), but will be used to define the magnetic translation group. Their non-trivial commutator $\hat{t}_{y}^{-1} \hat{t}_{x}^{-1} \hat{t}_{y} \hat{t}_{x} = e^{\frac{2\pi i}{q} \ell^z}$ encodes the layer-dependent Aharonov–Bohm phase accrued by an electron hopping counter-clockwise around a plaquette.

The translation symmetry of the Hamiltonian is substantially reduced by the inter-layer tunneling, Eq.~\eqref{eq-2Hofg}.
While the magnetic flux through each layer remains spatially uniform, the tunneling induces a flux pattern in the region \textit{between} the layers that is non-uniform;\footnote{We remark that this non-uniform flux pattern is precisely why our model does not have a $C_{4z}$ symmetry in spite of the square lattice geometry.} see the shaded vertical plaquettes shown in Fig.~\ref{fig:2HofSchematic}.
For even $q$, this pattern repeats every $q/2$ sites,\footnote{When $q$ is instead odd, the pattern of fluxes through the vertical plaquettes repeats every $q$ sites, and the generators are simply $(\hat{t}_x)^q$ and $\hat{t}_y$, which commute.} and thus the magnetic translation symmetries of $\hat{H}$ are generated by $\hat{m}_x = (\hat{t}_{x})^{q/2}$ and $\hat{m}_y = \hat{t}_{y}$, which anti-commute:
\begin{align} \label{eq:nontrivial_trans}
[\hat{H}, \hat{m}_{x}] = [\hat{H}, \hat{m}_{y}] = 0, \quad \hat{m}_x \hat{m}_{y} = -\hat{m}_{y} \hat{m}_x.
\end{align}
To define the Bravais lattice and Bloch states, we consider the largest abelian subgroup of translations.
Its generators
\begin{align} \label{eq-commuting-translations}
    \hat{T}_x = \hat{m}_x^2,\quad \hat{T}_y=\hat{m}_y,\qquad [\hat{T}_x,\hat{T}_y]=0
\end{align}
specify a Bravais lattice $\mathbb{L}$ with primitive lattice vectors $\mathbf{a}_x = q\bxhat$ and $\mathbf{a}_y = \byhat.$
The corresponding reciprocal lattice is then spanned by $\mathbf{G}_x = 2\pi\bxhat/q$ and $\mathbf{G}_y = 2\pi\byhat.$
Each unit cell contains $q$ evenly-spaced sites.
We describe positions in the underlying square grid by specifying a unit cell and sublattice:
\begin{align}
    \br = \bu + \sigma \hat{\mathbf{x}} \quad\ (\bu \in\mbbL,\ \sigma\in\{0,\dots,q-1\}),
\end{align}
and relabel the fermion operators as $\hat{\psi}^\dag_{\sigma \ell s}(\bu)$.

\subsection{Band Structure and Topology}\label{subsec-BandStructure}

\begin{figure}[!t]
    \centering
    \includegraphics[width = 240 pt]{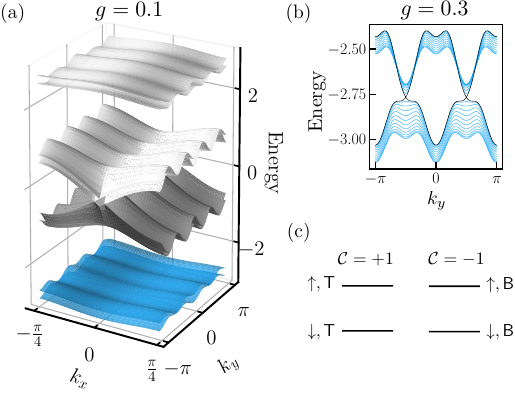}
    \caption{\textbf{Band Structure of the Double Hofstadter Model.} Panel (a) depicts the single-particle energy bands of the model at a flux fraction $1/q = 1/4$ and inter-layer tunneling amplitude $g = 0.1$.
    Including spin, there are 16 total bands.
    In this work, we focus on the lowest four (highlighted in blue), which have narrow bandwidth and are well-isolated from the remaining bands.
    (b) These narrow bands, shown here for $g=0.3$ in a smaller energy window, have $w_2=+1$ fragile topology and possess two Dirac cones that wind in the same direction.
    In (c), we show a schematic of the layer-polarized basis states, which span the flat band subspace and carry opposite Chern number $\mathcal{C} = \pm1$ in opposite layers. 
    When interactions are large, the energy difference between flat bands is irrelevant; in Section~\ref{subsec-LayerPolarizedBasis} we show that a maximally-layer polarized basis is more natural for studying ground state physics in the low-energy subspace.
    }
\label{fig:2HofBandStructure}
\end{figure}

Combining magnetic translations Eq.~\eqref{eq-commuting-translations} with Eqs.~\eqref{eq-2HofHopping} and \eqref{eq-2Hofg}, we obtain the non-interacting band structure of the double Hofstadter model.
The band structure is shown in Fig.~\ref{fig:2HofBandStructure}(a) and, accounting for two-fold spin degeneracy, features $2 \times 2 \times q = 16$ total bands.
The lowest four are narrow (``flat'') and energetically isolated by a large band gap.
Partially filling these narrow bands in the presence of interactions will be the primary interest of this study.
In what follows, it will be convenient to denote their occupation by a continuous value $0 \leq \nu \leq 4$.

In the absence of inter-layer tunneling, these narrow bands are those of the decoupled Hofstadter layers.
It is known that they are topological with Chern number $\mcC = +1 (-1)$ for the top (bottom) layers~\cite{TKNN}. 
In fact, in the limit of small flux $1/q \to 0$, these flat bands become time-reversal related copies of the lowest Landau level~\cite{harper2014perturbative, bauer2016quantum}.
Introducing non-zero tunneling $g$ hybridizes the lowest two bands in each spin sector.
Although their total Chern number is zero, the bands are nonetheless topologically non-trivial. They have a fragile topological invariant $w_2 = 1 \in \mathbb{Z}_2$, protected by spacetime inversion symmetry $\mathcal{I}$ and $U(1)_Q$~\cite{PoFragileTopo, AhnPRX, SlagerFragile, Else_Fragile,CanoFragile}.
A key implication of fragile topology is that the two $SU(2)_S$-symmetric bands possess two Dirac cones, as shown in Fig.~\ref{fig:2HofBandStructure}(b), that wind in the same direction.
At even $q$, the Dirac cones are pinned to $\mathbf{k}_{\pm} = \mathbf{G}_{x}/2 \pm \mathbf{G}_{y}/4$ by $C_{2z}, C_{2x}$ and $m_x$ (details in the Supplemental Material~\cite{SM}).

\subsection{Flat Band Hamiltonian}\label{subsec-LayerPolarizedBasis}

Let $\hat\mcP$ denote the many-body projection operator into the low-lying Hilbert space of the narrow bands of $\hat{h}$.
Provided that the interaction scale $U$ is smaller than the gap above the narrow Hofstadter bands,
we expect the physics at $\nu \leq 4$ to be accurately captured by the projected Hamiltonian $\hat\mcP \hat{H} \hat\mcP$.
This simplification is ubiquitous in studies of the quantum Hall effect at partial filling of the lowest Landau level~\cite{Bonesteel1995}. 
There, one finds that interactions enhance the gap to higher bands, thereby justifying the projection assumption beyond the naive regime of applicability.

\subsubsection{Layer-Polarized Basis for Flat Bands}

When the interaction strength $U$ greatly exceeds the bandwidth $W$ of the low-lying bands, the single-particle energy eigenbasis is no longer a physically useful way to organize the narrow-band subspace.
Drawing inspiration from the $g=0$ limit in which layer is a good quantum number, we instead work in a basis of states that are \textit{maximally-polarized} into either the top or bottom layer.
Crucially, these states will carry a layer-dependent Chern number, a fact that will underpin our analytical and numerical understanding of the many-body phenomena.
Concretely, we define a ``layer-polarized'' basis for the flat-band subspace as an eigenbasis of the projected layer operator:
\begin{align} \label{eq-renorm-layer-basis}
    \hat\mcP \hat\ell^z(\bk) \hat\mcP \ket{v_{ls}(\bk)} = \lambda_l(\bk) \ket{v_{ls}(\bk)},
\end{align}
and demand that the (real) eigenvalues carry a consistent sign across the Brillouin zone.
Here, $\hat\ell^z(\bk) = \hat\psi^\dagger(\bk) \ell^z \hat\psi(\bk)$ is defined in terms of the Fourier transformed microscopic fermion operators,
\begin{align}
    \hat{\psi}^\dag_{\sigma \ell s}(\bk) = \sum_{\bu\in\mbbL} \hat{\psi}^\dag_{\sigma \ell s}(\bu) e^{i\bk\cdot(\bu + \sigma \hat{\mathbf{x}})}.
\end{align}
In this basis, the creation operators for the layer-polarized orbitals take the form of Bloch-like states:
\begin{align}
    \hat{\phii}_{l s}^{\dagger}(\mathbf{k}) = \sum_{\sigma, \ell}\hat{\psi}_{\sigma \ell s}^{\dagger}(\mathbf{k}) v_{\sigma \ell, l}(\mathbf{k}),
\end{align}
For numerical convenience, we fix the residual gauge freedom by requiring that these layer-polarized orbitals be maximally localized in $x$ when Fourier transformed, resulting in hybrid Wannier states~\cite{qi2011generic, hybrid_Sgiarovello, hybridvandy1, hybridvandy2}.

We further remark that the two eigenvalues in Eq.~\eqref{eq-renorm-layer-basis} are opposite in sign and equal in magnitude.
For all parameters of interest, their magnitude is bounded below as $|\lambda_{\mathsf{T}/\mathsf{B}}(\bk)| > 0.95$, very near the maximum possible value of unity.
Like the Harper-Hofstadter bands of the layer-decoupled Hamiltonian $\hat{h}_t$, the layer-polarized basis states carry a Chern number $\mathcal{C} = \pm 1$ for $l = \mathsf{T}/\mathsf{B}$ respectively.
We emphasize that this holds in spite of the fact that layer is not a good quantum number in the presence of inter-layer tunneling.
We therefore refer to the label $l$ as the ``dressed layer'' index and remark that it is indistinguishable from the true layer index $\ell$, within the low-energy subspace, in the layer-decoupled limit.

\subsubsection{Flat Band Projected Hamiltonian}\label{subsec-FlatBandProjection}

We now write the projected Hamiltonian in terms of the layer-polarized basis for the flat-band Hilbert space:
\begin{align} \label{eq-proj-Hamiltonian}
    \hat\mcH = \hat\mcP \hat{H} \hat\mcP = \sum_\bk \hat\varphi^\dagger_\bk \delta{h}(\bk) \hat\varphi_\bk + \hat{\mathcal{V}}.
\end{align}
This decomposition of the Hamiltonian is chosen so that the interaction part is positive semi-definite:
\begin{align} \label{eq:strongcoupling_H}
    \hat{\mathcal{V}} &= \frac{1}{2} \sum_{\bq\in\text{EBZ}}V_{\ell\ell'}(\bq) \delta\hat\rho_{\bq\ell} \delta\hat\rho_{-\bq\ell'},
\end{align}
and $\delta{h}(\bk)$ consists of the remaining single-particle terms.
The latter can be further decomposed as $\delta h_0(\bk) + \delta h_g(\bk)$, where $\delta h_g(\bk) \sim \mathcal{O}(g)$ is purely off-diagonal in dressed layer and $\delta h_0(\mathbf{k}) \propto l^0$ is controlled by the bandwidth of the bare Hofstadter bands, up to a constant offset and $\mcO(g^2)$ corrections.

We refer to $\hat{\mathcal{V}}$ as the ``strong-coupling'' form of the interaction. 
Explicitly, $\delta\hat\rho_{\bq\ell} = \hat\rho_{\bq\ell} - \bar\rho_{\bq\ell} \hat{I}$ denotes the layer-resolved charge density relative to a translationally-invariant reference charge background that evenly fills the two layers, where (\textit{c.f.}~\cite{kang2019strong, bultinck2020ground})
\begin{align} \label{eq-projecteddensity}
    \hat\rho_{\bq\ell} &= \frac{1}{\sqrt{q N}} \sum_{\mathbf{k}} \hat{\phii}^{\dagger}_{\mathbf{k}} \Lambda_{\bq\ell}(\mathbf{k}) \hat{\phii}_{\mathbf{k} + \bq}, \\
    [\Lambda_{\bq \ell}(\mathbf{k})]_{l l'} &=  \sum_{\sigma} v_{\sigma \ell, l}^*(\mathbf{k}) v_{\sigma \ell, l'}(\mathbf{k} + \bq),
\end{align}
and $\bar{\rho}_{\bq\ell} = \frac{1}{\sqrt{qN}} \sum_{\mathbf{k}, \mathbf{G}} \delta_{\mathbf{q}, \mathbf{G}} \text{tr}\left[\Lambda_{\mathbf{G}\ell}(\mathbf{k})\right]$.
For purely on-site interactions, Eq.~\eqref{eq-2Hof_U1_and_U2}, the interaction kernel $V(\bq)$ is a matrix in the layer indices that can be expressed as:
\begin{equation}
    V(\mathbf{q}) = U_1 \ell^0 + U_2 \ell^x,
\end{equation}
which is crucially independent of $\mathbf{q}$.
Moreover, we note that the momentum $\bq$ lives in an extended Brillouin zone, or ``EBZ,'' which is defined relative to the microscopic square lattice as opposed to the Bravais lattice (see the Supplemental Material for more detail~\cite{SM}).

The symmetries of the double Hofstadter model strongly constrain its form factors. Namely, spacetime inversion symmetry gives $\Lambda_{\bq}(\bk) = \sum_\ell \Lambda_{\bq\ell}(\bk)$ the following functional form:\footnote{This form is obtained by passing to the layer-polarized basis in which spacetime inversion acts as $\hat\Isym \hat\varphi^\dag(\bk) \hat{\Isym}^{-1} = \hat\varphi^\dag(\bk)s^xl^x$.}
\begin{equation}
    \Lambda_{\bq}(\bk) = F_{\bq}^S(\bk) e^{i \Phi_{\bq}^S(\bk) l^z}
    + F_{\bq}^A(\bk) l^x e^{i \Phi_{\bq}^A(\bk) l^z},   
\label{eq:form_factor_decomposition}
\end{equation}
with real-valued
functions $F^{S/A}$ and $\Phi^{S/A}$. In the `decoupled' limit $g \to 0$, the $U(2)$ symmetry of the double Hofstadter model is enlarged to $U(2) \times U(2)$, corresponding to independent spin rotations and charge conservation in each layer. This implies $F^A$ vanishes when $g=0$, and in fact $F^A = \mcO(g)$ is always perturbatively small. The decoupled limit, therefore, serves as the starting point for the strong coupling theory of the double Hofstadter model that we develop in the next section.

\section{The Layer Anti-Ferromagnet} \label{sec-LAF}

We now investigate the phase diagram of the double Hofstadter model at electronic densities consistent with filling two of the four flat bands, i.e., $\nu=2$.
Our key finding in this section is a correlated insulating ground state, the \textit{layer anti-ferromagnetic} (LAF) state, that will later be shown to be the parent state for superconductivity.
The LAF is characterized by ferromagnetic order within each layer and anti-ferromagnetic order between layers [see Fig.~\ref{fig:LAF_and_competing}(a)], thereby spontaneously breaking the $SU(2)_S$ symmetry down to $U(1)_S$.
Topologically, the LAF is an interaction-induced quantum spin Hall insulator protected by a $U(1)_Q \times U(1)_S$ symmetry.\footnote{
Alternatively, it is protected by $U(1)_Q \rtimes \mathbb{Z}_2^T$, where $\mathbb{Z}_2^T$ is generated by an anti-unitary time-reversal symmetry of the double Hofstadter model $M_z \mathcal{T}'$, which acts on fermion operators as $(\hat{M}_z \hat{\mathcal{T}}') \hat\psi_{\mathbf{k}}^{\dagger} (\hat{M}_z \hat{\mathcal{T}}')^{-1} = \hat\psi^{\dagger}_{-\mathbf{k}} \ell^x s^y $. Here, $(M_z \mathcal{T}')^2 = (-1)^{P_F}$, with $P_F$ being the fermion parity~\cite{KaneMele}.}

\subsection{Analytic Arguments for LAF at Large $U_1$}
\label{subsec-strongcouplingtheory}

\begin{figure*}
    \centering
    \includegraphics[width = 494 pt]{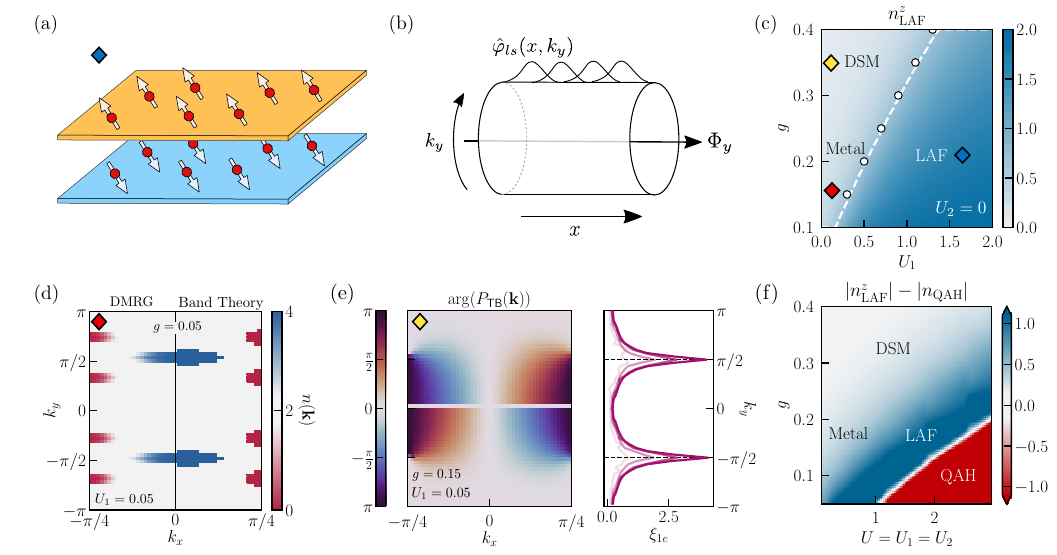}
    \caption{\textbf{The Layer Anti-Ferromagnet and Competing Phases.} 
    (a) Schematic depiction of the LAF at $\nu=2$, a quantum spin Hall insulator with opposite spin polarization in opposite layers.
    We will later argue that hole-doping this state gives rise to an unconventional superconductor.
    (b) We numerically evidence the LAF using cylinder iDMRG in a basis of hybrid Wannier orbitals indexed by unit cell $x$, momentum $k_y$, spin $s$, and (dressed) layer $l$.
    Threading flux $\Phi_y$ through the cylinder shifts the discrete $k_y$ momenta by $\Phi_y/L_y$, enabling continuous access of all momenta in the BZ.
    (c) In particular, working at $L_y = 5$, we evaluate the LAF polarization \eqref{eq-LAFOP} as a function of $g$ and $U_1$ (at fixed $\nu=2$ and $U_2 = 0$), finding the LAF phase in the majority of the strong coupling regime.
    The approximate phase boundary to a weak-coupling metal and Dirac semi-metal (DSM) is marked with a dotted line (See Sec.~\ref{subsubsec-LAFInlayer} for details). 
    %
    %
    (d) The metal is identified by examining the fermion occupations $n(\bk)$ across the BZ, where we find pockets of excess electrons or holes.
    We find near-quantitative agreement between the flux-threaded iDMRG data (left) and the occupations according to the non-interacting band structure (right).
    %
    %
    (e) In contrast, the DSM is characterized both by phase winding of $P_{\mathsf{T} \mathsf{B}}(\bk)$ [defined in Eq.~\eqref{eq-correlation-matrix}] and a peak in the $1e$ correlation length at $k_y = \pm \pi/2$ that grows with $\chi$. 
    From light to dark: $\chi = 128,\ 512,\ 1536,\ 3096$.
    %
    %
    (f) The presence of inter-layer repulsion modifies the phase diagram by introducing a quantum anomalous Hall insulator that competes with the LAF.
    }
    \label{fig:LAF_and_competing}
\end{figure*}

Let us begin by analytically arguing that the LAF state depicted in Fig.~\ref{fig:LAF_and_competing}(a) is the expected ground state when the inter-layer interaction $U_1$ is the dominant term in the Hamiltonian.
We do so by showing that interactions favor in-layer ferromagnetic order, while additional inter-layer hopping selects for inter-layer antiferromagnetic order.
We later confirm this picture numerically in Subsection~\ref{subsec-NumericalPD}, where we use extensive iDMRG computations to demonstrate the stability of the LAF and discuss a rich array of competing phases.

\subsubsection{In-Layer Ferromagnetism}

We start by working in the limit with vanishing inter-layer tunneling and inter-layer interactions, $g = U_2 = 0$, in which the two layers are completely decoupled.
We further take the limit where the remaining in-layer dispersion $\delta h_{0}(\mathbf{k})$ [Eq.~\eqref{eq-proj-Hamiltonian}] vanishes (up to an inconsequential momentum independent shift).
At the filling fraction $\nu = 2$, the in-layer interaction Hamiltonian favors states that occupy the two layers equally. All that remains, therefore, is to determine the behavior of spin in each layer.
For generic narrow, half-filled Chern bands, strong repulsive interactions will generally favor ferromagnetic spin alignment in a phenomenon known as \textit{quantum Hall ferromagnetism}~\cite{girvin2002quantum, repellin_2020}.
Intuitively, this behavior arises from the interplay between strong interactions and the spread of Wannier functions.
In a topological band, Wannier functions cannot all be atomically localized.
It is therefore advantageous for the electrons to spin polarize, which forces the many-body wavefunction to vanish when two electrons are brought together and results in a favorable direct exchange energy.
We remark that this is different from the paradigm of the Fermi-Hubbard model.
There, electrons can be perfectly localized to atomic lattice sites, resulting in a vanishing direct exchange energy and a dominant anti-ferromagnetic ``super-exchange'' interaction.

Consequently, the layer-decoupled double Hofstadter model should have \textit{bilayer quantum Hall ferromagnets} as favorable ground states, namely
\begin{align} \label{eq-BLFM-ansatz}
    \ket{\bm{n}_\Top, \bm{n}_\Bot} = \prod_\bk \hat\varphi^\dag_{\Top,\bm{n}_\Top}(\bk) \hat\varphi^\dag_{\Bot,\bm{n}_\Bot}(\bk) \ket{0}.
\end{align}
Here, each layer is associated with an order parameter characterizing its spontaneous magnetization:
\begin{align}
    \bm{n}_{\ell} \in S^2 \simeq SU(2)/U(1),
\end{align}
and electronic states whose spins are aligned to this axis:
\begin{equation}
    \hat\varphi^\dag_{\ell,\bm{n}_\ell} = \cos(\theta_\ell/2) \hat\varphi^\dag_{\ell, \uparrow} + e^{i \phi_\ell} \sin(\theta_\ell/2) \hat\varphi^\dag_{\ell, \downarrow},
\end{equation}
where $\phi_{\ell}$ and $\theta_{\ell}$ are the azimuthal and polar angles of $\bm{n}_{\ell}$, respectively. 
We can place this intuition on rigorous analytical footing by recognizing that the term $\delta \hat{\rho}_{\bq\ell} \delta \hat{\rho}_{-\bq\ell}$ appearing in the strong-coupling interaction $\hat{\mathcal{V}}$ [Eq.~\eqref{eq:strongcoupling_H}] is positive semi-definite for each $\mathbf{q}$.
This implies that any state annihilated by every $\delta\hat\rho_{\mathbf{q}\ell}$ is an exact ground state of this interaction. 
In particular, this is true of the ferromagnetic trial states in Eq.~\eqref{eq-BLFM-ansatz}:
\begin{align} \label{eq-deltarho-annihilation}
    \delta \hat\rho_{\bq\ell} \ket{\bm{n}_\Top, \bm{n}_\Bot} = 0,\quad \forall\,\bm{n}_{\Top/\Bot} \in S^2.
\end{align}
Analytically, this follows from the Pauli exclusion principle and the form factors $\Lambda_{\bq\ell}(\bk)$ being layer-diagonal in the $g=0$ limit, Eq.~\eqref{eq:form_factor_decomposition}, that is itself a consequence of layer $U(1)$ symmetry. 
This shows that the trial states \eqref{eq-BLFM-ansatz} are exact ground states of the strong-coupling Hamiltonian \eqref{eq:strongcoupling_H} and are therefore approximate ground states of the full Hamiltonian.

\subsubsection{Between Layer Anti-Ferromagnetism}

Upon including the inter-layer tunneling $g$, the above trial ground states, Eq.~\eqref{eq-BLFM-ansatz}, are expressed in the layer-polarized basis.
We then treat the residual single-particle terms perturbatively.
Similar to the original microscopic basis, the $\mathcal{O}(g)$ terms shuttle electrons between the dressed layer orbitals, or quantitatively:
\begin{align}
     \delta \hat{h}_g = \sum_\bk \hat\varphi^\dag_\bk \Big[ h_g^x(\bk) l^x + h_g^y(\bk) l^y \Big] \hat\varphi_\bk \approx \hat{\mathcal{P}} \hat{h}_g \hat{\mathcal{P}}. \label{eq:chern_off_diagonal_dispersion}
\end{align}
Since the bilayer quantum Hall ferromagnets are insulators, these virtual tunneling processes hybridize them with gapped excitations, generically lowering their energy due to level repulsion.
Since such tunneling processes are forbidden by Pauli exclusion when $\bm{n}_{\Top/\Bot}$ are aligned---as opposed to anti-aligned---the effect of $g$ is to introduce an anti-ferromagnetic ``super-exchange'' interaction between the top and bottom layers.
As a consequence of second-order perturbation theory in $g$, any trial state with $\bm{n}_\Top = -\bm{n}_\Bot$ is reduced in energy by
\begin{equation} \label{eq-JLAFthing}
    J = \bra{\text{LAF}} \delta \hat{h}_g \hat{\mathcal{V}}^{-1} \delta \hat{h}_g \ket{\text{LAF}} \sim \frac{g^2}{U_1}.
\end{equation}
This energetic advantage selects out LAF states, i.e., bilayer quantum Hall ferromagnets that spontaneously break the $SU(2)_S$ spin rotation symmetry down to $U(1)$ by condensing the vector order parameter
\begin{equation}
    \bm{n}_{\text{LAF}} = \bm{n}_{\mathsf{T}} - \bm{n}_{\mathsf{B}}.
\end{equation}
This establishes the LAF as a ground state of the full Hamiltonian in the limit of vanishingly-flat in-layer dispersion $\delta h_{0}(\mathbf{k})$ [Eq.~\eqref{eq-proj-Hamiltonian}] and perturbatively small $g$.
In the next section, we go beyond these limits via a large-scale numerical investigation.

\subsection{Numerical Phase Diagram at $\nu = 2$} \label{subsec-NumericalPD}

We now present the phase diagram of our model at the filling fraction $\nu = 2$, which we obtain using cylinder iDMRG methods that we describe below.
Consistent with our analytical expectations, we find a robust LAF phase at large values of the in-layer interaction $U_1$. 
Furthermore, we numerically characterize competing quantum orders, finding evidence for weakly-interacting metallic phases inherited from the band structure and a layer-polarized quantum anomalous Hall insulating state at large $U_2$.

\subsubsection{Cylinder iDMRG Numerical Method} \label{subsubsec-iDMRG}

Let us now describe the cylinder iDMRG numerical method that we will use to obtain the phase diagram at $\nu=2$ and, upon doping, superconductivity.
Since we are interested in the physics at partial filling of the isolated flat bands, we restrict to matrix product state (MPS) wave functions that live entirely within these bands, thereby reducing the number of degrees of freedom from all $4q$ bands to just 4, for layer and spin.
Moreover, since the fragile topology of the flat band subspace forbids an exponentially-localized basis that respects all the symmetries~\cite{PoFragileTopo,AhnPRX,SlagerFragile,CanoFragile,Else_Fragile}, we choose a computational basis of mixed position-momentum space states.
Such states, commonly referred to as ``hybrid'' Wannier states~\cite{WannierQi, marzari-vanderbilt, Motruk_mixedxk, hybridvandy1, hybridvandy2, hybrid_Sgiarovello}, are exponentially-localized \textit{along} the cylinder and are delocalized \textit{around} the cylinder.
A similar topological restriction arises in TBG~\cite{koshino_wannier_TBG, po_faithfultightbinding}, where a line of work by some of us has established the use of such states in cylinder iDMRG~\cite{soejima_2020, wang2022kekule, strain}.

This basis, which we encode into the MPS tensors, takes the following form:
\begin{equation}\label{eq:computational_basisp}
    \hat{\phii}_{ls}(j, k_y) = \frac{1}{\sqrt{N_x}} \sum_{k_x =0}^{2\pi/q} e^{i k_x qj} \hat{\phii}_{ls}(\mathbf{k}),
\end{equation}
and is depicted graphically in Fig.~\ref{fig:LAF_and_competing}(b).
Here, $j$ labels ``rings'' along the cylinder length, $k_y = (2\pi n + \Phi_y)/L_y$ labels discrete momentum cuts through the BZ that can be shifted by tuning the flux through the cylinder, $\Phi_y \in [0, 2\pi)$.
The basis is chosen such that its Fourier transform $\hat{\phii}_{ls}(\mathbf{k})$ is a layer-polarized basis (as described in Sec.~\ref{subsec-LayerPolarizedBasis}), permitting $\hat{\phii}_{l s}(j, k_y)$ to be labeled by dressed-layer and spin.
Moreover, the phase of each $\hat{\phii}_{ls}(\mathbf{k})$ is chosen such that $\hat{\phii}_{l s}(j, k_y)$ are maximally localized in the $\hat{\mathbf{x}}$ direction~\cite{marzari-vanderbilt, vanderbilt2018berry, SM}, giving hybrid Wannier states.  
Each ring of this computational cylinder, therefore, has a total of $2\times 2 \times L_y$ tensors whose physical legs encode the occupations of fermions at a given dressed-layer, spin, and momentum. Our numerics explicitly conserve electric charge ${U}(1)_Q$, spin ${U}(1)_S \subset {SU}(2)_S$, and $\mathbb{Z}/L_y \mathbb{Z}$ momentum around the cylinder.

Carrying out iDMRG on this model requires ``MPO compression''~\cite{parker_2020}.
When the Hamiltonian is resolved in the basis \eqref{eq:computational_basisp}, both the dispersion and interaction become long-ranged.
A naive matrix product operator (MPO) representation of this Hamiltonian has bond dimension $D \approx 10^6$, far beyond what is computationally feasible.
We therefore apply an MPO compression technique~\cite{soejima_2020, parker_2020} to create a faithful MPO representation of our Hamiltonian with bond dimension $D \approx 300$ and operator norm errors of $\delta < 10^{-3}t$ or better.
We consider cylinder circumferences $L_y = 5,6,7$. Since the the MPS bond dimension required to accurately describe ground states increases exponentially with $L_y$, we employ parallelized iDMRG simulations to reach MPS bond dimensions as large as $\chi = 12288$ using the open source \texttt{tenpy} library~\cite{hauschild2018efficient}.

\subsubsection{The LAF at Strong In-Layer Interactions} \label{subsubsec-LAFInlayer}

We now numerically verify the presence of LAF order at $\nu=2$ in the regime of large in-layer repulsive interactions $U_1 \gg t$.
First, however, we must address a subtlety associated with the cylinder geometry used for iDMRG.
Since the underlying infinite cylinder is quasi-1D and the candidate LAF state spontaneously breaks continuous spin rotational symmetry, Hohenberg-Mermin-Wagner (HMW) considerations imply that it must disorder due to quantum fluctuations~\cite{hohenberg1967existence, mermin1966absence, halperin2019hohenberg}.
The resulting quasi-1D state is symmetric with a spin gap that decays exponentially in the cylinder circumference~\cite{AffleckReview}.
Moreover, its LAF polarization, though finite at finite bond dimension, decays to zero as $\chi\to\infty$.
To estimate the extent of the LAF phase in the 2D limit, we take an approach analogous to a susceptibility experiment.
In particular, we add a weak Zeeman field to the Hamiltonian:
\begin{equation} \label{eq-Zeeman}
    \hat{H} \to \hat{H} + B_z \sum_{\mathbf{r} \in \mathbb{Z}^2} \hat{\psi}_{\mathbf{r}}^{\dagger} \ell^z s^z \hat{\psi}_{\mathbf{r}},
\end{equation}
and take $B_z = 10^{-2}$, which is much smaller than all other microscopic scales in the problem.
Such a Zeeman field explicitly breaks $SU(2)_S$ spin symmetry down to $U(1)_S$ but leaves all other symmetries unbroken.\footnote{In the presence of such a field, the LAF is no longer a symmetry-breaking phase but remains a topologically non-trivial quantum spin Hall insulator.}
We then approximately identify the phase extent of the LAF in the 2D limit by evaluating
\begin{align} \label{eq-LAFOP}
    n^z_{\text{LAF}}(\mathbf{r}) = \langle \hat\psi^{\dagger}_{\mathbf{r}}\ell^z s^z \hat\psi_{\mathbf{r}} \rangle
\end{align}
in iDMRG, which effectively measures the susceptibility to LAF ordering.
Fig.~\ref{fig:LAF_and_competing}(c) shows $n^z_{\text{LAF}}$ as a function of $g$ and $U_1$ at fixed $U_2=0$, with a heuristic phase boundary identified where $\partial n^z_{\text{LAF}}/\partial U_1$ has the largest slope.
In the Supplemental Material~\cite{SM}, we numerically verify that the LAF region of the phase diagram is insulating by computing the charge-$1e$ correlation length, $\xi_{1e}$, and demonstrating that it converges to a finite value as $\chi \to \infty$.
In summary, we have numerically identified the presence of a LAF insulating state at $\nu = 2$ that inhabits a broad portion of the strong-coupling regime, confirming our analytical expectations.

\subsubsection{Competing Phases at Weak Interactions} \label{subsubsec-Weak}

Next, we explore the weakly-interacting regime.
We first remark that the non-interacting band structure at $\nu=2$ exhibits two distinct phases as a function of $g$.
The first occurs at $g \ll 1$, where the nearly-degenerate, weakly-hybridized Hofstadter bands give rise to a metal with a one-dimensional Fermi surface.
In contrast, sufficiently large $g$ gaps out these bands except at the two Dirac cones protected by $\mathcal{I}$ and $SU(2)_S$ (see Sec.~\ref{subsec-BandStructure}), yielding a Dirac semimetal (DSM).

Using iDMRG, we confirm that these phases persist in the presence of weak interactions.
We diagnose the metallic phase by plotting the fermionic occupations
\begin{align}
    n(\mathbf{k}) = \sum_{l,s} \braket{\hat\phii_{ls}^{\dagger}(\mathbf{k}) \hat\phii_{ls}(\mathbf{k}) }. 
\end{align}
Since we work at finite cylinder circumference, we access only a finite number of cuts through the BZ.
Nonetheless, we can gain continuous momentum resolution by threading flux through our cylinder~\cite{He_dmrg_kagome}, as depicted in Fig.~\ref{fig:LAF_and_competing}(b).
The results of this are shown in Fig.~\ref{fig:LAF_and_competing}(d) for $g = 0.15$ and $U_1 = 0.05$.
We find that the occupancies obtained using iDMRG quantitatively match the band structure occupancies of the non-interacting limit.

Next, we confirm the DSM by two independent means.\footnote{Note that in the presence of $B_z$, the Dirac cones are generically gapped out, as explicitly breaking $SU(2)_S$ removes their topological protection. Nevertheless, since $B_z$ is weak compared to the single-particle energy scales, the DSM can still be identified using the present diagnostics.}
A defining characteristic of this state is that it possesses two Dirac cones that wind in the same direction within each spin sector.
This can be detected via the winding of $\arg P_{\mathsf{T} \mathsf{B}}(\mathbf{k}),$ where $P$ is the correlation matrix~\cite{soejima_2020}:
\begin{align} \label{eq-correlation-matrix}
    P_{ll'}(\mathbf{k}) = \sum_{s} \braket{\hat\phii_{l's}^{\dagger}(\mathbf{k}) \hat\phii_{ls}(\mathbf{k})}.
\end{align}
Fig.~\ref{fig:LAF_and_competing}(c) plots $\arg P_{\mathsf{T} \mathsf{B}}(\mathbf{k})$ over the flux-threaded BZ at $g=0.05$ with finite interactions $U_1=0.05$. We find that it indeed has the same winding around the points $\mathbf{k}_{\pm} = (\pi/4, \pm \pi/2)$.
That these points are Dirac cones---even at finite interaction strength---can be confirmed by computing the maximum charge-$1e$ correlation length $\xi_{1e}(k_y)$ as a function of the momentum $k_y$ around the cylinder, made continuous by threading flux.
In a weakly-interacting system, this quantity parameterizes the decay of correlations of the form $\braket{\hat\phii^{\dagger}_{x, k_y} \hat\phii_{0, k_y} } \sim e^{-x/\xi_{1e}(k_y)}$.
As such, when charge-$1e$ excitations are gapless, $\xi_{1e}$ should diverge.
Indeed, we find that it strongly peaks and grows rapidly with bond dimension precisely when $k_y$ passes through the isolated values $\pm\pi/2$, indicative of a semi-metallic gap closure [see Fig.~\ref{fig:LAF_and_competing}(e)].

\subsubsection{Robustness of LAF to Interlayer Interactions}\label{subsubsec-QAHInterlayer}

We further confirm the robustness of the LAF phase by re-introducing the inter-layer repulsion $U_2$.
Even at $U\equiv U_2=U_1$ (a regime more relevant to the experimental realizations in Sec.~\ref{sec-experimental_realizations} below), we find that the LAF continues to occupy a sizable phase space, but is accompanied by a layer-polarized correlated insulator at smaller values of $g$ [See Fig.~\ref{fig:LAF_and_competing}(f)].
In this phase, $\hat\psi^{\dagger}_{\mathbf{r}} {\ell}^z \hat\psi_{\mathbf{r}}$ acquires a non-zero expectation value, thereby breaking both $C_{2x}$ and $C_{2z} \Isym$ spontaneously.
Since the dressed-layer index is locked to Chern (Sec.~\ref{subsec-LayerPolarizedBasis}), then it also has a quantized Hall conductance of $\sigma_{xy} = 2 \times e^2/h $, making it a quantum anomalous Hall (QAH) phase.
We detect the presence of both the LAF and QAH phase simultaneously by plotting
\begin{equation}
    |n_{\text{LAF}}(\mathbf{r})| - |n_{\text{QAH}}(\mathbf{r})| = |\langle \hat\psi_{\mathbf{r}}^{\dagger} \ell^z s^z \hat\psi_{\mathbf{r}} \rangle| - |\langle \hat\psi_{\mathbf{r}}^{\dagger} \ell^z \hat\psi_{\mathbf{r}} \rangle|
\end{equation}
as a function of $g$ and $U$, which is positive in the LAF phase and negative in the QAH phase.
In Fig.~\ref{fig:LAF_and_competing}(f), we observe a first-order phase transition between the LAF and the QAH phase at large $U$.

\section{Numerical Evidence for Superconductivity} \label{sec:SC}

In this section, we use large-scale cylinder iDMRG to argue that hole-doping the LAF gives rise to a robust finite-circumference superconducting phase, which pairs holes with opposite spin and between opposite layers.
In particular, in Subsection~\ref{subsec-SCscaling}, we undertake a careful double scaling analysis of the superconducting correlations in both bond dimension and cylinder circumference at a particular point in parameter space.
We find strong evidence for a superconducting state at $L_y = 5$ as well as indications, though ultimately inconclusive, that this behavior persists at larger circumference.
Subsequently, in Sec.~\ref{subsec-robustness}, we work at fixed circumference $L_y = 5$ and demonstrate that this superconducting ground state belongs to a robust and stable superconducting phase. 
In particular, we find that superconductivity is present across various parameter regimes and always coexists with LAF order, suggesting that the broad LAF phase at $\nu=2$ is the ``parent state'' of the putative superconductor.
Finally, we find that this $L_y=5$ superconducting order is stable to sizable inter-layer repulsive interactions and a finite spatial interaction range, both of which add additional repulsive interactions to the channels in which we find superconductivity.

\subsection{Existence of Superconducting Order} \label{subsec-SCscaling}

\begin{figure*}
    \centering
    \includegraphics[width = 494 pt]{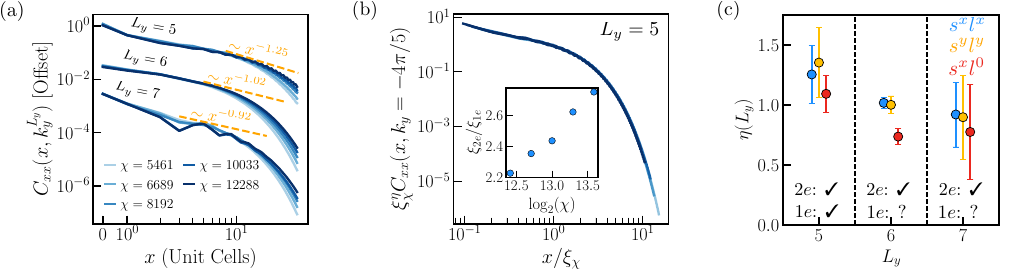}
    \caption{\textbf{Evidence for Superconductivity.} Working at fixed parameters $(g, U_1, U_2) = (0.15, 1.5, 0)$ and doping $\nu = 2 - 1/L_y$, we provide our numerical evidence for the existence of a superconducting state at $L_y = 5$ and algebraically-decaying pair correlations at all three circumferences.
    In (a), we examine the superconducting correlation function [Eq.~\eqref{eq-xxSC}] as a function of bond dimension $\chi$ for each circumference, finding that they each approach a power law as $\chi \to \infty$. 
    The power law obtained via a scaling collapse [see panel (b)] is shown with an orange dashed line.
    For ease of viewing, the correlations for $L_y=6,7$ are offset vertically by an arbitrary amount.
    (b) Power law behavior is quantitatively demonstrated by performing a scaling collapse [Eq.~\eqref{eq:xxSC_scaling_collapse}], shown here for $L_y=5$.
    To differentiate this power law behavior from that of a gapless metallic state, in the inset we show that the ratio $\xi_{2e}/\xi_{1e}$ increases as a function of bond dimension, a hallmark of a superconducting state [See Sec.~\ref{subsubsec-SCinCylinder}].
    (c) Power-law exponents of the superconducting correlations for all three channels as a function of cylinder circumference $L_y$.
    The powers $\eta(L_y)$ decrease as $L_y$ increases, consistent with the survival of superconductivity in the 2D limit.
    At the bottom of the panel, we summarize our confidence in the requisite gaplessness of $2e$ excitations and gap to $1e$ excitations at each circumference (See Sec.~\ref{subsubsec-CircScale} for more details).
    }
    \label{fig:SC_evidence}
\end{figure*}

In 2D systems, superconductivity typically manifests as a charge-$2e$ operator $\hat{\Delta}(\mathbf{r})$ having long-range order:
\begin{equation} \label{eq-LRO}
	\lim_{\mathbf{r} \to \infty} |\langle \hat{\Delta}^{\dagger}(\mathbf{r}) \hat{\Delta}(0) \rangle| \to \text{constant}, \quad\ \text{(2D)}
\end{equation}
thereby spontaneously breaking $U(1)_Q$ symmetry down to $\mathbb{Z}_2$.
However, to diagnose a superconducting state with iDMRG, two important subtleties arise due to the finite circumference and bond dimension of the simulations.
This subsection gives a brief technical discussion of some of these subtleties and presents numerical evidence for the existence of a superconductor in the double Hofstadter model.
Readers more interested in the phenomenological properties of the purported  superconducting phase may skip to the next section.

\subsubsection{Superconductivity in Cylinder iDMRG} \label{subsubsec-SCinCylinder}

Let us discuss how superconductivity manifests in in general in cylinder iDMRG (see also~\cite{gannot2023quantum, shubsskyrme}).
First, since iDMRG uses a quasi-1D cylinder geometry with finite circumference $L_y$, we must understand how to diagnose the quasi-1D analog of the superconductor---the \textit{Luther-Emery liquid}---and characterize its approach to a superconductor as $L_y\to\infty$.
The Luther-Emery liquid is characterized by the presence of fractionalized excitations---namely gapless ``chargeons'' and gapped ``spinons''--- that independently carry the electron's charge and spin (see e.g.~\cite{seidel2005luther}).
As a consequence, charge-$2e$ excitations (with vanishing total $\hat{S}^z$) have correlations that exhibit algebraic long-range order:
\begin{equation} \label{eq-algebraic-corr}
    |\langle \hat\Delta^{\dagger}(\mathbf{r}) \hat\Delta(0) \rangle| \sim |\mathbf{r}|^{-\eta(L_y)}
	\quad\ \text{($L_y$-cylinder).}
\end{equation}
As the cylinder circumference is increased, the realization of true long-range superconducting order in 2D is tied to the behavior $\eta(L_y) \to 0$ as $L_y \to \infty$.
In the Luther-Emery liquid, furthermore, correlation functions of electrons (gapped states comprised of both the chargeon and spinon) exhibit exponential decay:
\begin{equation}
    \langle \hat\phii^{\dagger}(\mathbf{r}) \hat\phii(0) \rangle_{\mathrm{LE}} \sim e^{-|\mathbf{r}|/\xi_{1e}}.
    \label{eq:LEL_1e_gapped}
\end{equation}
At finite circumference, this behavior crucially distinguishes the Luther-Emery liquid from the Luttinger liquid, where such correlations are algebraic.

A second important subtlety concerns the finite bond dimension of the MPS, which only permits the expression of exponentially-decaying (connected) correlations, and introduces a length scale $\xi_{\chi}$ associated to the largest non-trivial eigenvalue of the MPS transfer matrix.
As a consequence, power law correlations of any charge-$Q$ operator are cut off at a distance $\xi_{Q, \chi} \leq \xi_{\chi}$ that scales with bond dimension according to~\cite{Tagliacozzo, pollmann2009theory, vanhecke2019scaling}:
\begin{equation} \label{eq:Q-chi-cutoff}
    \log(\xi_{Q, \chi}) \propto \log(\chi).
\end{equation}
However, in practice, this scaling behavior can also be found in \textit{gapped} systems when the true correlation length $\xi$, which dictates the exponential decay and is inversely related to the energy gap, greatly exceeds $\xi_{\chi}$. %
This makes it difficult to distinguish the Luttinger and Luther-Emery liquids discussed above, especially in scenarios where the latter has a very small spinon gap, as both may appear to show algebraic scaling of $\xi_{1e, \chi}$.
Nevertheless, even in such cases, the two could be distinguished by making use of standard arguments in the finite entanglement scaling theory of matrix product states~\cite{pollmann2009theory, Tagliacozzo, vanhecke2019scaling}.
In particular, when the MPS is attempting to capture a fully gapless state like the Luttinger liquid, $\xi_{\chi}$ is the only length scale present in the correlations of the state.
As such, we expect correlation functions of charge $Q$ operators will decay with a correlation length $\xi_{Q, \chi} = a_Q \xi_{\chi}$, where $a_Q$ (at asymptotically large $\chi$) is a constant of proportionality that does not depend on bond dimension.
For the Luttinger liquid, this leads to the sharp prediction that $\xi_{2e}/\xi_{1e}$ approaches a constant as $\chi \to \infty$.
This contrasts with the Luther-Emery liquid, in which $\xi_{2e}/\xi_{1e}$ is expected to diverge.

In summary, to demonstrate the existence of superconductivity in cylinder iDMRG, one must in principle perform a double scaling analysis.
At each fixed circumference, the state must approach a Luther-Emery liquid, characterized by algebraic correlations of charge-$2e$ operators and the divergence of $\xi_{2e}/\xi_{1e}$ with bond dimension.
In addition, in principle,  one must show that the exponent of algebraic decay for charge-$2e$ operators goes as $\eta(L_y) \to 0$ as $L_y \to \infty$.
However, since increasing the cylinder circumference results in exponential scaling of the simulation complexity, cylinder iDMRG studies can typically only demonstrate convergence to a Luther-Emery liquid for a few, relatively small circumferences~\cite{JiangDevereaux2019,Jiang2020PRL,Jiang2020PRR,Chung2020,Zhu2022}.

\subsubsection{Bond Dimension Scaling} \label{subsubsec-ChiScaling}

We begin by providing strong numerical evidence for Luther-Emery liquid physics at $L_y = 5$ by performing a careful scaling in the bond dimension.
We work at fixed hole doping $\nu = 2 - 1/5$ and at parameters $(g, U_1) = (0.15, 1.5)$, choosing for the moment to set $U_2  = 0$, though we emphasize that our results are robust to the inclusion of further repulsive interactions (as we illustrate in Subsection~\ref{subsec-robustness}).
As discussed above, we must demonstrate that the ground state exhibits algebraically-decaying correlation functions of charge-$2e$ operators.
To this end, we work in the computational basis of fermions defined in Eq.~\eqref{eq:computational_basisp} and introduce the following set of  zero $y$-momentum charge-$2e$ operators:
\begin{equation} \label{eq-defDelta}
\hat\Delta^{\dagger}_{\alpha \beta}(x; \delta x, k_y) = \hat\phii^T \Big(x + \frac{\delta x}{2}, k_y \Big) s^{\alpha} l^{\beta} \hat\phii \Big(x -\frac{\delta x}{2}, -k_y \Big),
\end{equation}
where $x$ is the center of mass position of the holes, $\delta x$ is their relative position,\footnote{Here, $\delta x$ is implicitly assumed to be even. For odd $\delta x$, we instead take the holes to be at $x + (\delta x + 1)/2$ and $x - (\delta x - 1)/2$ (see Supplemental Material~\cite{SM}).} and $\alpha,\beta \in \st{0,x,y,z}$.
We then consider correlation functions of such operators:
\begin{equation} \label{eq-xxSC}
    C_{\alpha\beta}(x, k_y) = N_{2h}^{-1} \langle \hat\Delta_{\alpha\beta}^{\dagger}(x; 0, k_y) \hat\Delta_{\alpha\beta}(0; 0, k_y) \rangle,
\end{equation}
which we choose to normalize by the average number of available hole pairs $N_{2h} = \n{2 - \nu} / 2$.
This correlation function is shown in Fig.~\ref{fig:SC_evidence}(a) for $k_y^{L_y = 5} = 4\pi/5$ and $\alpha \beta = x x$.
As the bond dimension increases, the correlation length increases and $C_{xx}$ approaches a power law (the dashed straight line on the log-log scale), signaling the onset of algebraic order.
To gain further quantitative confirmation, we perform a scaling collapse of the form
\begin{equation}
\label{eq:xxSC_scaling_collapse}
     C_{xx}^{(\chi)}(x, k_y) = \xi_{2e, \chi}^{-\eta}\  g_{xx}(x/\xi_{2e, \chi}, k_y),
\end{equation}
where $C_{xx}^{(\chi)}$ is the correlation function Eq.~\eqref{eq-xxSC} at bond dimension $\chi$ and  $\xi_{2e, \chi}$ is its correlation length.
This scaling collapse, shown in Fig.~\ref{fig:SC_evidence}(b), is achieved with a power law exponent $\eta(L_y = 5) = 1.3 \pm 0.2$.
This data confirms the existence of algebraic long-range order in $2e$ pair correlations at fixed $L_y$.

In addition, we demonstrate that the behavior of the $1e$ correlations is consistent with a Luther-Emery liquid and not a Luttinger liquid (see Sec.~\ref{subsubsec-SCinCylinder} for the distinction).
To do so, we investigate the ratio $\xi_{2e, \chi}/\xi_{1e, \chi}$ as a function of bond dimension, which we plot in the inset of Fig.~\ref{fig:SC_evidence}(b).
We find that the ratio increases rapidly with bond dimension which, per the discussion of Sec.~\ref{subsubsec-SCinCylinder}, is inconsistent with the behavior of a Luttinger liquid and suggests a gap in the $1e$ sector.
In fact, employing the recently developed methods of Refs.~\cite{rams2018precise,vanhecke2019scaling}, we obtain an extrapolated, and clearly finite, value $\xi^{-1}_{1e, \infty} = 0.11$ in units of inverse $4a$, the length of the unit cell.
In contrast, $|\xi^{-1}_{2e, \infty}| < 0.003$ is nearly zero, consistent with our above evidence for algebraic order in this charge sector.
This demonstrates that the ground state at $L_y=5$ is a Luther-Emery liquid, the quasi-1D analog of the superconducting phase.

We conclude with some important phenomenological observations about the superconducting correlations.
In particular, algebraic long-range order is most dominantly seen in the following pairing channels:
\begin{align} \label{eq-channels}
    s^x l^x,\ s^y l^y,
\end{align}
appearing with nearly equal strength, as well as the $s^x l^0$ channel, which is notably of much weaker strength.
As such, the pairing is between opposite spins and predominantly between opposite layers. %
The approximately equal strength of the $s^x l^x$ and $s^y l^y$ correlations are a consequence of an approximate $U(1)$ symmetry generated by $s^z l^z$.
This symmetry also explains the dominance of inter-layer pairing over $s^x l^0$, which we henceforth ignore.
In summary, the dominant condensing pairing channels imply that the superconducting order is due to the condensation of inter-spin, inter-layer hole pairs.
We return to this point in Sec.~\ref{subsec-pairingsymmetry} in our analysis of the superconducting pairing symmetry.

\subsubsection{Circumference Scaling} \label{subsubsec-CircScale}

Having established the presence of a Luther-Emery liquid state at finite circumference $L_y = 5$, we now investigate whether such a state tends to a \textit{bona fide} superconductor in the 2D limit by investigating it at larger circumferences.
This investigation is challenging for three distinct reasons.
First, as with all DMRG studies, the computational resources required increase \textit{exponentially} with the $2 \times 2 \times L_y$ tensors that wrap around the cylinder.
To partially compensate for this scaling, we use parallelized iDMRG for circumferences $L_y=6,7$ with bond dimensions up to $\chi=12288$.
As a remark, compared to the Fermi-Hubbard model at the same $L_y$, our projected four-flavor model has twice as many degrees of freedom, making it vastly more expensive to simulate. 
Second, since the electronic filling must be commensurate with the periodicity of the MPS, we are limited to the circumference-dependent densities $\nu = 2 - 1/L_y$.\footnote{
Since $L_y=5,6,7$ are coprime, prohibitively large unit cells would be required to have equal fillings at each circumference.}
This makes it difficult to directly compare the physics at each circumference.
Finally, we will later show in Sec.~\ref{sec-Gap_and_PairingSym} that the superconductor pairs holes at an emergent Fermi surface (or more accurately Fermi points at finite circumference).
Crucially, the number and location of Fermi points vary significantly for small and accessible $L_y$ and, unlike in the Hubbard model, are non-monotonic in cylinder circumference (see the Supplemental Material~\cite{SM}).
This makes studying the approach to the 2D limit particularly challenging in our model.
Nevertheless, in what follows we analyze the superconducting correlations for each accessible circumference, ensuring that the MPS truncation error is $3\times 10^{-5}$ or better at each bond dimension. 
Ultimately, we find signatures of large pair correlations at $L_y = 6, 7$ but are unable to rule out the possibility of a Luttinger liquid at these circumferences.

We first determine whether the states we find are gapless in the charge-$2e$ sector by examining correlation functions of the form Eq.~\eqref{eq-xxSC}.
These are plotted in Fig.~\ref{fig:SC_evidence}(a) at all three circumferences in the $\alpha\beta=xx$ channel, with the larger two $L_y$ vertically offset for display, and taken at $k_y^{L_y = 6} = - 2\pi/3$ and $k_y^{L_y = 7} = - 2\pi/7$.
Evidently, at each circumference, these correlations become algebraically long-ranged with increasing bond dimension.
The exponents of algebraic decay are then extracted by collapsing under a scaling hypothesis analogous to Eq.~\eqref{eq:xxSC_scaling_collapse} and are shown in Fig.~\ref{fig:SC_evidence}(c) for each circumference and for each of the dominant superconducting channels.
We find a quantitative decrease in $\eta(L_y)$ as $L_y$ increases for the $xx$ and $yy$ channels, consistent with the superconducting expectation that $\eta(L_y) \to 0$ as $L_y \to \infty$.
For the $x0$ channels, the superconducting correlations at $L_y = 6$ are found to have a quantitatively different exponent and behavior as compare with the other; we comment in the Supplemental Material that this coincides with it having a different pairing symmetry in the $x0$ channel at $L_y = 6$~\cite{SM}.

Next, we investigate the behavior of the $1e$-correlations, which we find to be more subtle and ultimately inconclusive.
We provide the core results here and give further details in the Supplemental Material \cite{SM}.
Recall that our previous analysis of $L_y = 5$ found that these charge-$1e$ correlations remained exponentially decaying even as $\chi \to \infty$.
This was evidenced by the growth of $\xi_{2e}/\xi_{1e}$ and by performing a systematic extrapolation~\cite{rams2018precise,vanhecke2019scaling} to infinite bond dimension that revealed a non-zero inverse correlation length $\xi_{1e, \infty}^{-1}(L_y = 5) \approx 0.11$.
The charge-$2e$ and $1e$ correlations therefore both matched the expectations for a Luther-Emery liquid, Eqs.~(\ref{eq-algebraic-corr},~\ref{eq:LEL_1e_gapped}), at $L_y=5$. This conclusively identified the $L_y=5$ ground state as a quasi-1D superconductor.

At larger circumferences, however, the situation is less clear.
At $L_y = 6$, we find that the behavior of the $1e$-correlations does not conclusively point to either a Luttinger liquid or a Luther-Emery liquid.
On the one hand, we find that the magnitude of $\xi_{2e}$ is nearly 2.5 times larger than $\xi_{1e}$ at $\chi=12288$, indicating large superconducting pair correlations in the state.
This is suggestive of a Luther-Emery liquid.
On the other hand, we find that $\xi_{2e}/\xi_{1e}$ is roughly constant with bond dimension, which fails to exclude a Luttinger liquid.
Generally, at this circumference, we conclude that the finite-bond dimension state is too far from a `scaling limit' to properly differentiate the two possibilities.
This is evidenced by extrapolating $\xi_{1e/2e, \infty}^{-1}(L_y = 6)$ using the method detailed in Ref.~\cite{rams2018precise} and finding that we are too far from convergence to obtain sensible (e.g. non-negative) values of these quantities.

We conclude by discussing the case of $L_y = 7$.
Here, unlike the previous two cases, we find that $\xi_{2e}$ is only slightly larger than $\xi_{1e}$ with a ratio $\xi_{2e}/\xi_{1e}$ that remains relatively constant across the largest bond dimensions we can access.
As such, at this circumference, both correlation lengths appear to diverge as a function of bond dimension, though much larger bond dimension would be required for a more definitive determination.
Hence, there is little evidence in direct support of a Luther-Emery liquid over a Luttinger liquid.
Indeed, the two may exhibit similar behavior when the finite-$\chi$ correlation length is too small to resolve the charge gap (see Sec.~\ref{subsubsec-SCinCylinder}).
Altogether, this highly circumference-dependent behavior prevents us from drawing firm conclusions about the 2D phase.

\subsection{Robustness of the Superconducting Phase} \label{subsec-robustness}

\begin{figure*}
    \centering
    \includegraphics[width = 480 pt]{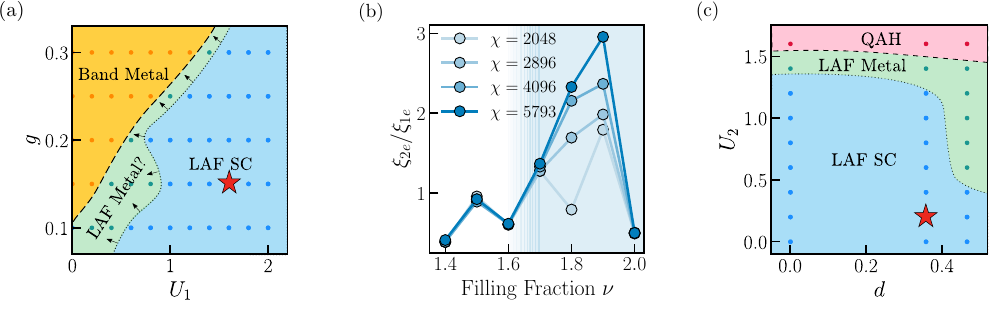}
    \caption{\textbf{Robustness of the $L_y=5$ Superconductor.} We investigate the extent of the $L_y = 5$ superconducting phase by examining the ratio $\xi_{2e}/\xi_{1e}$ (see Sec.~\ref{subsec-robustness} for more details).
    In panel (a), we depict the phase diagram at $\nu = 1.8$ in the absence of inter-layer repulsion. 
    A robust LAF-polarized superconducting region appears where the LAF order was found at integer filling, suggesting the LAF may be the parent state for the superconductor.
    We mark, with a star, the point at which the scaling analysis of Sec.~\ref{subsec-SCscaling} was performed and remark that the phase competes with a band metal and LAF-polarized metal, the latter of which retreats with increasing bond dimension (see Section~\ref{subsec:coexistence}).
    In panel (b), we fix $g = 0.1$ and $U_1 = 1.8$ and depict our heuristic diagnostic, the growth of the ratio $\xi_{2e}/\xi_{1e}$ vs. $\chi$, as a function of the electronic density $\nu \in [1.4, 2].$
    Superconductivity is found across the span $1.7 \lesssim \nu < 2$, with metallic LAF behavior beginning at $\nu=1.6$.
    In the Supplemental Material~\cite{SM}, we show that the system remains LAF-polarized throughout this range, further cementing the LAF as the superconducting parent state.
    In panel (c), we examine the robustness of our superconductor to additional repulsive interactions.
    In particular, fixing $(g, U_1) = (0.1, 1.4)$, we chart the phase diagram as a function of inter-layer repulsion $U_2$ and interaction range $d$ [See Eq.~\eqref{eq:longer_range_interactions}].
    We find the superconductor is remarkably robust, giving way initially to a LAF metallic phase when $U_2$ is sufficiently large, and then to a layer-polarized metallic phase when $U_2$ exceeds $U_1$.
    In the Supplemental Material, we performed a scaling analysis similar to that detailed in Sec~\ref{subsec-SCscaling} for the point marked with a red star~\cite{SM}.
    } \label{fig:SC_robustness}
\end{figure*}

Restricting to $L_y = 5$, where our detailed bond-dimension-scaling analysis conclusively demonstrated the existence of a superconductor, we now evaluate the extent and robustness of the finite-circumference superconducting phase.
%
To do so, we use a heuristic, but nonetheless principled, diagnostic of superconductivity that enables detailed investigation of the double Hofstadter phase diagram with the available computation resources.
We find that the superconductor appears exclusively upon doping the LAF insulating state at $\nu=2$ and that it is nearly maximally LAF polarized, leading us to posit that the LAF insulator is the parent state of the superconductor.
Moreover, the superconducting phase is found to exist across a broad range of hole densities $\nu < 2$ and is, rather surprisingly, substantially resilient to additional repulsive interactions, namely those between the layers and at a longer range.

Our diagnostic consists of examining the scaling of $\xi_{2e}/\xi_{1e}$ as a function of bond dimension.
In particular, following the discussion of Section~\ref{subsubsec-SCinCylinder}, the bond dimension scaling of this quantity can be used to distinguish a Luther-Emery liquid from a Luttinger liquid or any fully gapped quasi-1D phase.
In our numerics, we label portions of the phase diagrams ``superconducting'' if the ratio $\xi_{2e}/\xi_{1e}$ is both increasing as a function of bond dimension and, in particular, if it has exceeded unity at the largest bond dimension available.

\subsubsection{The Superconducting Phase: LAF Coexistence} \label{subsec:coexistence}

We now chart the portions of parameter space that host superconductivity at $L_y=5$.
We begin by studying the $(g, U_1)$ plane at a fixed electronic density $\nu=1.8$ and in the absence of inter-layer interactions.
Using the diagnostic described above, superconductivity is found to occupy a broad portion of the strong-coupling regime that hosts the LAF insulator at integer filling [see Fig.~\ref{fig:SC_robustness}(a)].
Intriguingly, the superconductor itself is always found to have sizable coexistent LAF polarization.
Moreover, its correlations are most long-ranged in the channels Eq.~\eqref{eq-channels}, suggesting a consistent pairing symmetry throughout the superconducting phase.
On the other hand, in the weak-coupling regime with small $U_1$, we identify a metallic state with nearly-vanishing LAF polarization, i.e. a band metal.
In between these two phases, we observe a metallic state with significant LAF polarization.

Similar to our identification of the LAF phase boundary at $\nu=2$, we demarcate the transition between the weak-coupling and LAF metallic phases by a dotted boundary corresponding to the largest slope $\partial n^z_{\text{LAF}}/\partial U_1$.
In contrast, the boundary between the LAF metallic and superconducting phases is more sensitive to bond dimension, even at the sizable value $\chi = 5793$.
This exemplifies a pattern that is pervasive in our numerical data: LAF metallic states commonly undergo a finite-$\chi$ transition into the LAF superconducting phase at a bond dimension that depends sensitively on parameters, whereas the reverse is exceedingly rare.
As a result, we expect that the set of points that meet the criteria for superconductivity will continue to grow with bond dimension, and that the LAF metallic phase may purely be an artifact of the finite accuracy of our simulations.

The intrinsic coexistence between superconductivity and LAF polarization is further elucidated by tuning the electronic density in the range $\nu \in [1.4, 2]$.
To access density increments of $1/10$ of a band, we quadruple the MPS unit cell.
In Fig.~\ref{fig:SC_robustness}(b), we plot our diagnostic for the superconducting order---the ratio $\xi_{2e}/\xi_{1e}$ as a function of bond dimension---and find it that indicates the persistence of  superconductivity in the range $1.7 \lesssim \nu < 2$.
Moreover, in the Supplemental Material, we demonstrate that the system remains substantially LAF-polarized across this parameter regime, with $n^z_\mathrm{LAF}$ decreasing only linearly in the hole doping.
At the remaining parameter points $1.4 \leq \nu \leq 1.6$, we find evidence of metallic correlations and possible translation symmetry-breaking along the $x$ direction, though we caution that a careful circumference scaling analysis would be required to identify the precise nature of this nearby phase.
Across the entire range of fillings, and in particular across the superconducting region, we find a persistent LAF polarization that nearly saturates the maximal value $\nu$, i.e., the total density of electrons.
This ubiquitous coexistence with LAF order leads us to hypothesize that the LAF insulator is the ``parent state'' of the superconducting state. 
We further corroborate this claim in Sec.~\ref{sec-Gap_and_PairingSym} where we identify the superconductor as a weak-pairing instability of holes at the LAF's mean-field Fermi surface.

\subsubsection{Stability to Further Repulsive Perturbations}

We now show that the superconductor is robust to both large inter-layer repulsion and a finite interaction range. 
To this end, we replace the microscopic on-site interactions with a longer-range Gaussian, parameterized as:
\begin{equation}
	\hat{V} = \sum_{\v{r}} e^{-\frac{\n{\v{r}-\v{r}'}^2}{2d^2}} :\left[
		\frac{U_1}{2} (\hat n_{\v{r}\Top}^2 + \hat n_{\v{r}\Bot}^2)
		+ U_2 \hat n_{\v{r}\Top} \hat n_{\v{r}\Bot}
	\right]:
 \label{eq:longer_range_interactions}
\end{equation}
where $d$ specifies the interaction range.
In Fig.~\ref{fig:SC_robustness}(c), we show the extent of the superconducting phase in the $(d,U_2)$ plane at the fixed parameters $(g,U_1)=(0.1,1.4)$.
Remarkably, in spite of the pairing being predominantly inter-layer (see Subsection~\ref{subsec-SCscaling}), we find the superconductor survives in the presence of substantial inter-layer repulsion $U_2 \sim \mathcal{O}(U_1)$ and is further resilient to simultaneously increasing the range of the interactions.
While the superconducting phase is globally identified by using the heuristic diagnostic introduced at the beginning of Sec.~\ref{subsec-robustness}, we cement its existence at $L_y=5$ in the presence of additional inter-layer repulsion by performing a detailed scaling analysis of the superconducting correlations, analogous to the one performed in Sec.~\ref{subsec-SCscaling}. We also see indications, via a slow but monotic increase in the ratio $\xi_{2e}/\xi_{1e}$, that this behavior may persist at $L_y = 6$.
These analyses are detailed fully in the Supplemental Material~\cite{SM} and are performed at the parameter point $(d, U_2) = (0.358, 0.2)$ [marked with a star on Fig.~\ref{fig:SC_robustness}(c)]. %

Despite the resilience of the superconducting phase, we find that when $d$ is increased sufficiently, the superconductor gives way to a translation-invariant, LAF-polarized metallic state.
Moreover, we remark that increasing $U_2$ in excess of $U_1$ induces a transition out of the LAF phase into a layer-polarized, translation-invariant metal that originates from the $\nu = 2$ QAH insulator.

\subsubsection{Stability to Changing the Zeeman Field}

We conclude by remarking briefly on the influence of the Zeeman field $B_z$ [Eq.~\eqref{eq-Zeeman}].
Recall that this weak pinning field is necessary in order to stabilize LAF polarization, which otherwise disorders due to the quasi-1D cylinder geometry (see Section~\ref{subsubsec-LAFInlayer}).
Though the bulk of our calculations are performed at the fixed value $B_z=10^{-2}$, a scan of the Zeeman field reveals superconductivity at values as small as $B_z=10^{-3}$.
For comparison, the bandwidth, the inter-layer tunneling and average interaction energy per electron are of order $10^{-1}$.
At some larger values of $B_z$, we encounter numerical errors in the limit of only in-layer Hubbard interactions, i.e., $U_2=d=0$, though we find that enlarging the MPS unit cell greatly improves convergence and favors a translation-invariant LAF metallic state. 
On the other hand, no convergence issues are encountered at the finite values $(U_2,d)=(0.2, 0.466)$, at which superconductivity appears to persist up to $B_z = 0.1$. %
Additional details and numerical data are provided in the Supplemental Material~\cite{SM}.

\section{Emergent ``Weak Pairing'' Picture of the Superconductor} \label{sec-Gap_and_PairingSym}
We now turn to characterizing the superconducting order reported in the previous section.
We do so by examining the so-called \textit{pair wavefunction}~\cite{mineev1999introduction}: 
\begin{equation} \label{eq-pairwavefunction}
    \Delta_{\alpha \beta}(\mathbf{k}) = \langle \hat\phii^T(\mathbf{k}) s^{\alpha} l^{\beta} \hat\phii(-\mathbf{k}) \rangle \equiv \langle \hat{\Delta}_{\alpha \beta}^{\dagger}(\mathbf{k}) \rangle,
\end{equation}
where we refer to the two-hole operator, colloquially, as the \textit{Cooper pair} operator.
In Sec.~\ref{subsec-numericalextractiongap}, we introduce a novel technique to probe this quantity in iDMRG numerics, which may be applied to a variety of 2D systems.
By examining $\Delta(\bk)$ across the Brillouin zone in Sec.~\ref{subsec-weakpairing}, we find that it peaks at specific locations that coincide with the Fermi surface of non-interacting holes above the LAF insulator predicted within self-consistent Hartree-Fock.
This phenomenology is characteristic of a (BCS-like) superconducting state formed from the pairing of holes near this LAF Fermi surface.
We emphasize that this weak-pairing phenomenology emerges in a context where the interaction scale is the largest energy scale present and is found via non-perturbative iDMRG numerics that do not presuppose a mean-field description of the state.

Moreover, the emergent mean-field description enables us to characterize other phenomenological features of the superconductor.
In particular, in Sec.~\ref{subsec-pairingsymmetry}, we use the pair wave function \eqref{eq-pairwavefunction} to uniquely characterize the pairing symmetry of the superconductor and find that, among other symmetry indices, it is $p$-wave.
Finally, in Sec.~\ref{subsec-transition-temperature} we leverage results from BCS theory to heuristically estimate the transition temperature of the superconducting state.

\subsection{Numerically Probing the Superconducting Pair Wavefunction} \label{subsec-numericalextractiongap}

The definition of the pair wavefunction ~\eqref{eq-pairwavefunction} implicitly assumes that the ground state wavefunction is a superconducting condensate in which the number of Cooper pairs in the ground state wavefunction is uncertain.
This presents an obstacle to extracting this quantity in our iDMRG calculations, as $U(1)$ particle number conservation causes $\Delta(\bk)$ to vanish uniformly.
Here, we provide a brief account of how to circumvent this difficulty, leaving a more complete and precise explanation to the Supplemental Material~\cite{SM}.
Broadly, we do so by introducing a \textit{proxy} pair wavefunction $\widetilde{\Delta}_{\alpha \beta}(\bk)$ that can be extracted within iDMRG and is proportional to $\Delta_{\alpha \beta}(\bk)$. 
Readers more interested in our physical results may skip to the next subsection where this quantity is used to probe the structure and pairing symmetry of the superconductor.

In the $L_y \to \infty$ limit, the definite $U(1)_Q$ charge of the iDMRG ground state implies that it will fail to satisfy cluster decomposition, defined as~\cite{weinberg1995quantum}:
\begin{equation}
    \lim_{|\mathbf{r}| \to \infty} \left( \langle \hat\Delta^{\dagger} (\mathbf{r}) \hat\Delta(0) \rangle  - \langle \hat\Delta^{\dagger}(\mathbf{r}) \rangle \langle \hat\Delta(0) \rangle\right) = 0
\end{equation}
As a result, long-range superconducting order [defined in Eq.~\eqref{eq-LRO}], does not imply a finite expectation value of the Cooper pair operator.
Nevertheless, let us assume that there exist cluster decomposition-satisfying ground states $\ket{\phii}$ of the 2D superconducting manifold for which
\begin{equation}
   \lim_{|\mathbf{r}| \to \infty} \lim_{L_y \to \infty} \left( \langle \hat\Delta^{\dagger} (\mathbf{r}) \hat\Delta(0) \rangle_{\psi(L_y)} -  \langle \hat\Delta^{\dagger} (\mathbf{r}) \rangle_{\phii} \langle \hat\Delta(0) \rangle_{\phii}\right) = 0,
\end{equation}
where $\langle \cdots \rangle_{a} = \bra{a} \cdots \ket{a}$ and $\ket{\psi(L_y)}$ is the iDMRG ground state at circumference $L_y$.

The pair wavefunction can then be extracted numerically as follows.
First, we define the \textit{proxy} pair wavefunction by the following correlation function:
\begin{equation} \label{eq-proxy2}
    \widetilde{\Delta}_{\alpha\beta,x,p_y}(\bk) = \langle \hat\Delta^\dagger_{\alpha\beta}(0; \bk) \hat\Delta_{\alpha\beta}(x; 0,p_y)\rangle_{\psi(L_y)}.
\end{equation}
Here $\hat{\Delta}_{\alpha \beta}(x; 0, p_y)$ is as defined in Eq.~\eqref{eq-defDelta}, while the $\hat\Delta^\dag$ term is defined by a Fourier transform in the relative position coordinate:
\begin{equation}
    \hat{\Delta}^{\dagger}_{\alpha \beta}(x; \mathbf{k}) = \sum_{\delta x} \hat{\Delta}^{\dagger}_{\alpha \beta}(x; \delta x, k_y) e^{i k_x \delta x},
\end{equation}
which, when averaged over the center of mass $x$, is equal to the Cooper pair operator defined in Eq.~\eqref{eq-pairwavefunction}.

The proxy pair wavefunction---which need not vanish in $U(1)$-conserving iDMRG since it involves the expectation value of a charge-neutral operator---should be thought of as a function of $\mathbf{k}$ (with fixed parameters $x$ and $p_y$).
Provided that we consider translationally-invariant states and take $x$ to be larger
than both the characteristic size of the Cooper pair (the ``coherence length'') $\xi_{\text{SC}}$ and the connected correlation length $\xi$, then for sufficiently large $L_y$ we expect that:
\begin{equation} \label{eq-proxy}
    \widetilde{\Delta}_{\alpha\beta, x, p_y}(\mathbf{k}) = C^{\alpha\beta}_{x,  p_y} \langle \hat\Delta_{\alpha \beta}^{\dagger}(\mathbf{k}) \rangle_{\phii} + \mathcal{O}\left(e^{- |x - \xi_{\text{SC}}|/\xi} \right),
\end{equation}
where the prefactor is a $\mathbf{k}$-independent constant and $\ket{\varphi}$ is a cluster decomposition-satisfying superconducting ground state of the 2D system.
As such, by examining the proxy pair wavefunction $\widetilde{\Delta}(\bk)$ as a function of momentum, we gain direct insight into the structure of the true pair wavefunction $\Delta(\bk)$.

\begin{figure*}
    \centering
    \includegraphics[width = 494 pt]{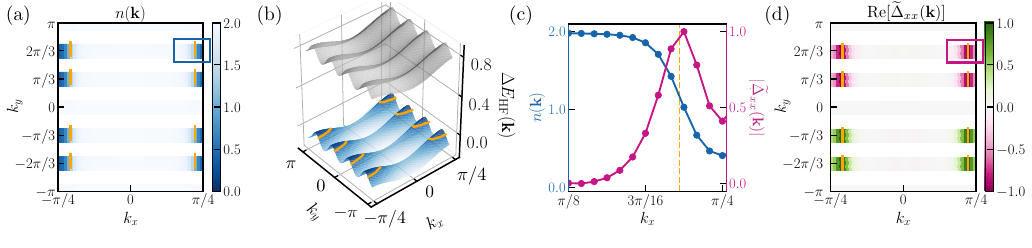}
    \caption{\textbf{Evidence for Weak Pairing at the LAF Fermi Surface.}
    (a) The electronic density of the putative superconducting iDMRG groundstate for the $L_y = 6$ cylinder, shown here at $\chi=12288$ and $(g, U_1, \nu) = (0.15, 1.5, 2-1/6)$, is uniform in the majority of the Brillouin zone but sharply depleted in isolated regions.
    The locations of these drops in the density match the mean-field Fermi surface of non-interacting holes above the LAF insulating state, marked in orange (in all subplots) and obtained using self-consistent Hartree-Fock (SCHF).
    In (b), we plot the SCHF band structure computed at $\nu = 2$ on $48 \times 48$ unit cells. 
     The energy $\Delta E_\text{HF}$ of the two-fold degenerate valence band is shaded in blue and defined relative to the valence band edge.
     For small hole-doping away from $\nu=2$, the Fermi surface (shown in orange) consists of four disconnected pockets.
     As SCHF enforces electronic $U(1)$ (disallowing pairing), this strongly-interacting effective bandstructure is a metallic ``parent state" for superconductivity.    
    %
    In (c), we zoom into one particular pocket---specified by blue and magenta boxes in (a) and (d), respectively---and plot both the electronic density (left axis) and the proxy pair wavefunction Eq.~\eqref{eq-proxy2} (right axis) obtained from iDMRG.
    The former drops off rapidly, while the latter sharply peaks, precisely at the location of the SCHF Fermi surface, strongly suggesting that the superconductor arises from a weak-pairing instability upon hole-doping the $\nu=2$ insulating LAF state.
    Panel (d) gives a global view of the proxy pair wavefunction, whose real part is orders of magnitude larger than its imaginary part. From its sign structure, we deduce the unconventional pairing symmetry of the superconductor, namely that it is $p$-wave and is phenomenologically consistent with a Bogoliubov gap function $\delta(\mathbf{k}) \sim \sin(k_y) ( s^+ l^+ + s^- l^-)$.
    }
    \label{fig:SC_kspace}
\end{figure*}

\subsection{Emergent Weak Pairing at the LAF Fermi Surface} \label{subsec-weakpairing}

In this section, we demonstrate that our superconductor arises from the momentum space ``BCS'' pairing of holes above the LAF insulator.
To be explicit, we will provide evidence that this emergent weak-pairing picture occurs as a two-step process.
First, strong interactions select LAF-polarized ground states, even in the presence of hole doping.
Such ``normal'' states are well-described by a LAF bandstructure that strongly renormalizes the four non-interacting bands into two valence bands and two conduction bands.
Second, holes---initially occupying a normal LAF metallic state---pair at the renormalized Fermi surface, triggering a superconducting instability.
Before continuing, we emphasize once again that this weak-pairing picture emerges in a setting where repulsive interactions are the dominant energy scale.

As preliminary evidence for this picture, we plot the density of electrons in the superconducting state as a function of momentum [see Fig.~\ref{fig:SC_kspace}(a)], finding that the density is depleted in localized ``pockets'' of momentum space but is otherwise uniform.
To understand the origin of these disconnected Fermi surfaces, we examine the mean-field band structure of the LAF insulator computed using standard (see e.g. \cite{bultinck2020ground}) self-consistent Hartree-Fock (SCHF). 
We show these bands in Fig.~\ref{fig:SC_kspace}(b), with the lower two degenerate bands corresponding to the occupied electronic states of the LAF insulator and the upper two degenerate bands corresponding to the spectrum of gapped single-electron excitations above the LAF.
At lowest order in $g$, these two sets of bands are merely the bare Hofstadter bands, but massively split by the interaction scale.
In Fig.~\ref{fig:SC_kspace}(a,b), we mark the Hartree-Fock Fermi energy at filling $\nu = 1.8$ with orange lines, finding that their locations coincide closely with drops in the iDMRG electronic density.

To further corroborate the BCS picture described above, we examine the structure of the pair wavefunction, Eq.~\eqref{eq-pairwavefunction}.
We first lay out our expectations. If BCS theory was applicable, we would expect the pair wavefunction to be related to the Bogoliubov gap $\delta_{\alpha \beta}(\mathbf{k})$ by
\begin{equation} \label{eq-pair-and-gap}
    \Delta_{\alpha \beta}(\mathbf{k}) = \frac{\delta_{\alpha \beta}(\mathbf{k})}{2\sqrt{ (E_{\text{HF}}(\mathbf{k}) -E_F)^2 +\delta_{\alpha \beta}^2(\mathbf{k}) }},
\end{equation}
where $E_{\text{HF}}(\mathbf{k})$ is the dispersion of holes above the normal LAF state---corresponding to the Koopman spectrum of hole excitations in self-consistent Hartree-Fock---and $E_F$ is the Fermi level.
Generally, we expect the superconducting gap $\delta_{\alpha \beta}$ to be smaller than the characteristic scale of the dispersion, namely the bandwidth. Within the validity of BCS mean field theory, we would therefore expect Eq.~\eqref{eq-pair-and-gap} to vanish sufficiently far from the Fermi surface.
On the other hand, the pair wavefunction would peak at the Fermi surface, at which it would be proportional to the phase of the gap function: $\Delta_{\alpha \beta}(\mathbf{k}) = \delta_{\alpha \beta}(\mathbf{k})/2|\delta_{\alpha \beta}(\mathbf{k})|$.

These expectations are borne out in our iDMRG numerics.
In particular, in Fig.~\ref{fig:SC_kspace}(c,d) we plot the proxy pair wavefunction \eqref{eq-proxy2} and find that it peaks precisely at the Fermi surface and decays quickly away from it.
In Fig.~\ref{fig:SC_kspace}(c), we zoom into one Fermi pocket and observe excellent agreement between the Fermi surface predicted from Hartree-Fock, the location at which the charge density drops off in iDMRG, and the peak in the proxy pair wavefunction.
In Fig.~\ref{fig:SC_kspace}(d), we plot the latter over the Brillouin zone and find that its magnitude is maximal at each of the four disconnected Hartree-Fock Fermi surfaces, providing further evidence that the superconductor arises as an instability of the LAF Fermi surface.
We leave the analysis and interpretation of the relative complex phase of $\Delta_{\alpha \beta}(\mathbf{k})$ to the next subsection.

\subsection{Pairing Symmetry} \label{subsec-pairingsymmetry}

In what follows, we characterize the pairing symmetry of the superconducting state by using the proxy pair wavefunction $\widetilde{\Delta}(\mathbf{k})$. 
To be precise, the pairing symmetry corresponds to the irreducible co-representation---the generalization of a representation to incorporate anti-unitary symmetries---of the magnetic space group under which the pair wavefunction transforms~\cite{mineev1999introduction, Sigrist_Unconventional_SC, ShafferHofSC, SM}.
In our case, since the Cooper pairs that condense to form the superconductor carry zero total momentum, we need only specify the transformation properties under the discrete rotations $C_{2z}$ and $C_{2x}$, the non-trivial magnetic translation $m_x$, and the anti-unitary spacetime inversion symmetry $\mathcal{I}$ (see Supplemental Material~\cite{SM}). 
Since these operators each square to the identity and mutually commute in the zero-momentum sector, these generators act on Cooper pairs as the group $(\mathbb{Z}_2)^3 \times \mathbb{Z}_2^{\mathsf{K}}$ with anti-unitary $\mathbb{Z}_2^{\mathsf{K}}$.
Given that the co-representations of $\mathbb{Z}_2^{\mathsf{K}}$ are trivial~\cite{SM}, specifying the irreducible representation amounts to identifying three $\mathbb{Z}_2$ characters for each condensed superconducting channel $s^\alpha l^\beta$.

Let $\hat{g}$ be the unitary operator corresponding to a symmetry $g \in \{C_{2x}, C_{2z}, m_x\}$.
We define the corresponding characters $\nu_g^{\alpha \beta}$ by~\cite{mineev1999introduction, SM, Sigrist_Unconventional_SC}:%
\begin{align} \label{eq-sym-index-def}
    \langle \hat{g}^{\dagger} \hat{\Delta}^{\dagger}_{\alpha \beta}(\mathbf{k}) \hat{g}  \rangle = \nu_g^{\alpha \beta} \Delta_{\alpha \beta}(\mathbf{k}),
\end{align}
where $\nu^{\alpha \beta}_g = \pm 1$ and $\alpha\beta$ (no implicit summation) correspond specifically to the dominant condensed channels specified in Eq.~\eqref{eq-channels}.
Intuitively, the characters capture how the pair wavefunction transforms under the application of symmetries on the ground state.
More generally, the role of these characters is played by matrices $\boldsymbol{\nu}_g$ that form a co-representation of the magnetic space group (see the Supplemental Material for a more formal and general treatment~\cite{SM}).

As an illustrative example, we compute the index for $g = C_{2z}$.
Its action on our Cooper pairs is given by:
\begin{equation}
\hat{C}_{2z} \hat{\Delta}^{\dagger}_{\alpha \beta}(\mathbf{k}) \hat{C}_{2z}^{-1} = \hat{\Delta}^\dagger_{\alpha \beta}(-\mathbf{k}).
\end{equation}
In the case $\alpha\beta = xx$, whose proxy pair wavefunction we have plotted in Fig.~\ref{fig:SC_kspace}(d), we readily observe that $\widetilde{\Delta}_{xx}(-\mathbf{k}) = - \widetilde{\Delta}_{xx}(\mathbf{k})$. From this and Eq.~\eqref{eq-sym-index-def}, we immediately find that:
\begin{equation}
    \nu_{C_{2z}}^{xx} = -1 \hspace{0.5cm} \text{ ($p$-wave symmetry)}.
\end{equation}
We also find that that $\nu_g^{yy} = \nu_g^{x0} = -1$, indicating that all condensed channels are $p$-wave in nature.

A similar analysis can be performed for the other symmetries.
In particular, inspection of the proxy pair wavefunction likewise leads to the conclusion
\begin{align}
    \nu_{C_{2x}}^{\alpha \beta} = \nu_{m_x}^{\alpha \beta} = -1, 
\end{align}
for both dominant pairing channels $\alpha \beta = xx, yy$.
Finally, for the anti-unitary symmetry $\mathcal{I}$, our choice of layer-polarized basis guarantees that $\hat{\mathcal{I}}\hat{\Delta}^{\dagger}_{\alpha \beta}(\mathbf{k}) \hat{\mathcal{I}}^{-1} =  \hat{\Delta}^{\dagger}_{\alpha \beta}(\mathbf{k})$. This implies that the pair wavefunction has a constant phase, assuming that $\mathcal{I}$ remains unbroken in the ground state. Indeed, our numerically-obtained proxy pair wavefunction is found to be purely real.

We conclude by remarking that our numerical observations are consistent with a superconductor whose Bogoliubov gap function $\delta_{\alpha \beta}(\bk)$ has the same symmetry properties as $\sin(k_y)$ for $\alpha \beta = xx, yy$.
Moreover, in the Supplemental Material~\cite{SM}, we provide numerical evidence that the pair wavefunction in the $\alpha \beta = xx$ channel appears with a relative minus sign compared to the $\alpha \beta = yy$ channel.
As such, the symmetry properties of a putative gap function can be summarized as:
\begin{equation}
\delta(\mathbf{k}) \sim \sin(k_y) ( s^+ l^+ + s^- l^-).
\end{equation}
Consistent with a parent LAF state, the gap function corresponds to non-unitary pairing with $\delta^{\dagger}(\mathbf{k}) \delta(\mathbf{k}) \propto (1 + s^z l^z)/2 = P_{\text{LAF}}$, the LAF projector.
Furthermore, while this form of the gap function typically implies the presence of symmetry-enforced nodes along the lines $k_y = 0, \pm \pi$, our superconductor instead remains gapped.
This is because the Fermi surface at which pairing occurs is disconnected and does not intersect either of these lines in momentum space (see Fig.~\ref{fig:SC_kspace}).

\subsection{Estimate of Transition Temperature} \label{subsec-transition-temperature}

The weak-pairing picture for the superconducting state, advocated for above, provides a heuristic framework for estimating the superconducting transition temperature.
To this end, we recall that the finite temperature transition out of a $(2+1)d$ superconductor occurs either due to: (\textit{i}) the loss of phase coherence and algebraically-long-ranged phase correlations, or (\textit{ii}) the unbinding of Cooper pairs.
The former is described by a Berezinskii–Kosterlitz–Thouless (BKT) transition corresponding to the proliferation of vortices in the superconductor and has a transition temperature that is typically upper bounded by the ground state superfluid stiffness $\rho_s$ as~\cite{mineev1999introduction, Sigrist_Unconventional_SC}:
\begin{equation}
T_c^{\text{BKT}} \lesssim \pi \rho_s/2.
\end{equation}
In contrast, the Cooper pair unbinding transition occurs at a temperature scale set by the maximum Bogoliubov gap $\delta_\mathrm{SC}$ at the Fermi surface~\cite{SM, mineev1999introduction, Sigrist_Unconventional_SC}:
\begin{equation} \label{eq-BCSTc}
 T_c^p \sim \delta_\text{SC}. 
\end{equation}
With input from self-consistent Hartree-Fock, we can estimate both scales.

In a BCS-like superconductor, the superfluid stiffness at $T = 0$ is set by the density of paired dopants divided by their effective mass, which in two dimensions scales as the Fermi energy~\cite{coleman2015introduction}.
As such, 
\begin{equation}
    T_c^\mathrm{BKT} \lesssim \pi \rho_s/2 \sim  \pi E_F/2 \approx 0.03\, t.
\end{equation}
On the other hand, the approximate magnitude of the Bogoliubov gap $\delta_\mathrm{SC}$ can be extracted from the pair wavefunction $\Delta_{\alpha \beta}(\mathbf{k})$.
In particular, for momenta around the Fermi surface $\mathbf{k} \approx \mathbf{k}_F$, Eq.~\eqref{eq-pair-and-gap} implies that
\begin{equation}
    4\Delta_{\alpha \beta}^2(\mathbf{k}) \approx \frac{\delta_{\alpha \beta}^2(\mathbf{k}_F)}{\delta_{\alpha \beta}^2(\mathbf{k}_F) + v_F^2(\mathbf{k} - \mathbf{k}_F)^2} .
\end{equation}
Here, we made the linear approximation $E_{\text{HF}}(\mathbf{k}) - E_F \approx v_F (\mathbf{k} - \mathbf{k}_F)$, with $v_F$ being the Fermi velocity, and took $\delta_{\alpha \beta}(\mathbf{k}) \approx \delta_{\alpha \beta}(\mathbf{k}_F)$.
As such, around the Fermi surface, $4 \Delta_{\alpha \beta}^2(\mathbf{k})$ is predicted to have a Lorentzian profile whose half-width-half-maximum is equal to $\delta_{\alpha \beta}(\mathbf{k}_F)/v_F$.
Maximizing over the dominant channels $\alpha\beta=xx,yy$ and taking $v_F = 0.17$ obtained from self-consistent Hartree-Fock at $L_y = 5$, we obtain the following estimate for the Bogoliubov gap:
\begin{equation}
    T_c^p \sim \delta_\mathrm{SC} \approx 0.01\, t.
\end{equation}
In particular, we find that the pair-breaking temperature $T_c^p$ is comparable to or less than the BKT temperature.
Though we expect the superconducting transition temperature to therefore be set by the energy scale $\delta_\text{SC}$, we carefully note the possibility of a BKT-type thermal transition out of the superconducting phase.
This is because of a reduction in the condensate fraction (equivalently, the Cooper pair density) at finite temperature due to thermal fluctuations, which can significantly decrease the superfluid stiffness $\rho_s$ as we approach the pair-breaking scale.
Consequently, as $T \to T^p_c$, the small value of $\rho_s$ can decrease the cost of vortices and cause them to proliferate, triggering a BKT transition that preempts $T_c^p$.

\section{Experimental Relevance: Cold Atoms and Solid-State} \label{sec-experimental_realizations}

Thus far, this work has focused on theoretically and numerically analyzing the double Hofstadter model: explicating its symmetries and topology, detailing its undoped phase diagram, and arguing for the emergence of a $p$-wave superconducting state at non-integer filling.
We now connect the double Hofstadter model to contemporary experimental platforms in both cold atoms and moir\'e materials.
First, Section~\ref{subsec-OLimplementation} provides an explicit experimental blueprint for realizing our model with alkaline-earth atoms in an optical lattice.
Second, Section~\ref{subsec-Relationship-to-Moire} elucidates a relationship to moir\'e materials via a ``dictionary" between the degrees of freedom and symmetries of the double Hofstadter model and models of magic-angle TBG.

\subsection{Optical Lattice Implementation} \label{subsec-OLimplementation}

To construct a concrete experimental blueprint for the double Hofstadter model in an optical lattice, we take advantage of two recent advances: the quantum control of alkaline-earth atoms, in particular state-dependent potentials~\cite{riegger_2018,heinz_2020,oppong_2022,hohn_state-dependent_2023}, and the engineering of synthetic gauge fields~\cite{aidelsburger_realization_2013,miyake_realizing_2013,kolkowitz_spinorbit-coupled_2017,cooper_topological_2019,aidelsburger_artificial_2018}. 
Explicitly, we consider a general species of fermionic alkaline-earth atom
living in a two-dimensional optical lattice and show how each term of our model can be engineered using established techniques.
Though our discussion below does not specialize to any particular atom, in the Supplemental Material we provide a survey of common species and comment on how their properties translate quantitatively into the couplings of our model~\cite{SM}.


The internal state of a general fermionic alkaline-earth atom is specified by a nuclear spin $I$ and two long-lived orbital ``clock'' states typically denoted by $\ket{\mathsf{g}}=$ $^{1}S_0$ and $\ket{\mathsf{e}}=$ $^{3}P_0$.
We encode the internal layer and spin degrees of freedom of the double Hofstadter model in these nuclear and clock states (see Fig.~\ref{fig:optical_double_Hofstadter}) as:
\begin{equation}
\begin{aligned}
    \ket{\mathsf{T} \uparrow} &= \ket{\mathsf{g}, \mathsf{a}}, \ &&\ket{\mathsf{T} \downarrow} = \ket{\mathsf{g}, \mathsf{b}}\\
    \ket{\mathsf{B} \uparrow} &= \ket{\mathsf{e}, \mathsf{a}}, \ &&\ket{\mathsf{B} \downarrow} = \ket{\mathsf{e}, \mathsf{b}}
\end{aligned}
\label{eq-nucspinencoding}
\end{equation}
where $\mathsf{a} \neq \mathsf{b} \in \{-I, \dots, I\}$ are two chosen nuclear hyperfine states.
%
The three types of terms in the double Hofstadter model---in-layer ``double'' Hofstadter hopping, inter-layer hopping, and on-site interactions---must be separately engineered. We describe each here, with full details provided in the Supplemental Material~\cite{SM}.

\subsubsection{Optical Engineering of Double Hofstadter Hopping}

\begin{figure}
    \centering
    \includegraphics[width = 247 pt]{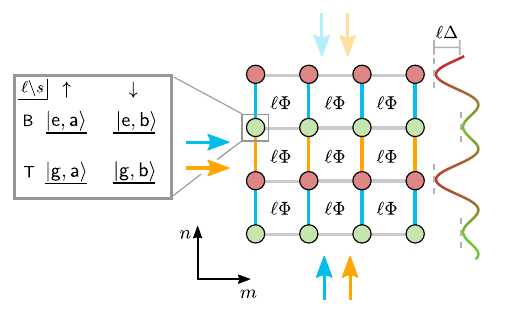}
    \caption{\textbf{Optical Lattice Implementation.} The double Hofstadter model is engineered by placing fermionic alkaline-earth atoms in a two-dimensional optical lattice.
    Layer and spin are encoded in the clock and nuclear spin states of the atom, respectively (see boxed levels).
    Furthermore, complex hoppings are generated by first creating a period-two staggered on-site potential (red and green sites; potential sketched on the right) to suppress bare hopping in the $y$-direction, labeled by $n$ in Eqs.~(\ref{eq:optical_lattice_hopping}-\ref{eq:optical_lattice_inter-layer_hoppings}).
    This period-two potential crucially has opposite signs for the opposite layers.
    Second, resonant hopping is restored separately on the even and odd bonds using two pairs of lasers (blue and orange arrows) that form a running wave in the $x$-direction and are retro-reflected (shown in lighter colors) to form a standing wave in the $y$-direction.
    These restored hoppings are then imbued with the spatially-dependent phases of the laser drives.
    In this scheme, atoms see an artificial magnetic flux $\ell \Phi = \ell \pi/2$  that is opposite in opposite layers, as desired.
    }
    \label{fig:optical_double_Hofstadter}
\end{figure}

To engineer the in-layer double Hofstadter hopping, consider placing atoms in a two-dimensional ``magic wavelength'' optical lattice with lattice spacing $a$~\cite{ye_2008,campbell_fermi-degenerate_2017}, created by superimposing two orthogonal standing waves.
To realize the requisite complex hopping, our strategy is to energetically suppress bare nearest-neighbor tunneling in the $y$-direction, and then restore it using Floquet driving~\cite{bukov_universal_2015,eckardt_colloquium_2017}, thereby imprinting the spatially-dependent phase of the drive onto the restored hoppings.
%
This approach combines two experimental protocols implemented by one of us~\cite{aidelsburger_realization_2013,aidelsberger_chern}, where the key ingredients are: (1) a homogeneous artificial magnetic field and (2) a layer-dependent direction of the field, which is realized using a state-dependent potential at the ``anti-magic wavelength'' of the two clock states.

To suppress hopping, we superimpose on the optical lattice an additional standing wave in the $y$-direction with lattice spacing $2a$ at the anti-magic wavelength, inducing opposite on-site potentials for the two clock states (see Fig.~\ref{fig:optical_double_Hofstadter}).
Neighboring lattice sites will therefore have a site-alternating energy mismatch of $\pm \Delta$, which is always opposite between atoms in the $\ket{\mathsf{e}}$ and $\ket{\mathsf{g}}$ states (i.e., $\Delta_\mathsf{e} = \pm \Delta = -\Delta_\mathsf{g}$).
This energetically suppresses nearest-neighbor hopping in the $y$-direction.
To restore it, we use the two-laser optical driving scheme introduced in Ref.~\cite{aidelsberger_chern}, which resonantly modulates the on-site potential on even and odd bonds in the $y$-direction independently.
Intuitively, since the different clock states see opposite staggered potentials, one will absorb energy from the drive while the other will emit energy into it.
As a result, their hoppings acquire conjugate complex phases.  
This intuition is borne out at zeroth-order in a Floquet-Magnus expansion~\cite{SM}, wherein this driving protocol gives an effective average hopping Hamiltonian:
\begin{equation}
    \hat{h}_t^{\text{OL}} = -t \sum_{m, n \in \mathbb{Z}_2} \hat\psi^{\dagger}_{m, n + 1} e^{i \phi_{mn}} \hat\psi_{m, n} + \hat\psi_{m + 1, n}^{\dagger} \hat\psi_{m, n} + \text{h.c.}
    \label{eq:optical_lattice_hopping}
\end{equation}
Here $\hat\psi^{\dagger}_{m, n}$ is the fermionic creation operator, which is a row vector in the spin and layer [e.g., see Eq.~\eqref{eq-fermionicrow}], at lattice site $m, n \in \mathbb{Z}^2$. The layer-dependent phases are:
\begin{equation}
    \phi_{m, n}^{\ell} = \pi \delta_{\ell \mathsf{B}} +  \frac{\pi \ell }{2}\begin{cases} (m + n) & n \text{ even} \\ (m + n + 1) & n \text{ odd,} \end{cases}
    \label{eq:optical_lattice_phases}
\end{equation}
where $\ell = +1\ (-1)$ for the top (bottom) layer.\footnote{To achieve these phases, we remark that the phases of the driving lasers must be locked both to one another and to the lattice potential (see the Supplemental Material for more details~\cite{SM}), which is demanding to realize experimentally.}
These phases produce the same flux pattern as in Eq.~\eqref{eq-2HofHopping}, namely a uniform $\pi/2$ ($-\pi/2$) flux through each plaquette in the top (bottom) layer (See Fig.~\ref{fig:2HofSchematic}).
%

\subsubsection{Optically Implementing the Interlayer Tunneling}

The inter-layer hopping term is relatively straightforward.
With a fixed encoding~\eqref{eq-nucspinencoding} for the internal degrees of freedom, such a term can be implemented using a laser addressing the optical frequency of the clock transition, which produces the requisite term,
\begin{equation}
    \hat{h}_g^{\text{OL}} = -g \sum_{m, n} \hat\psi^{\dagger}_{m, n} \ell^x \hat\psi_{m, n}.
    \label{eq:optical_lattice_inter-layer_hoppings}
\end{equation}
Together, Eqs.~\eqref{eq:optical_lattice_hopping}--\eqref{eq:optical_lattice_inter-layer_hoppings} specify the hoppings of the optical lattice. 
Although the pattern of complex phases in Eq.~\eqref{eq:optical_lattice_phases} differs from Eq.~\eqref{eq-2HofHopping}, one can explicitly check that the fluxes within each layer and the alternating flux pattern produced between the layers exactly matches the one shown in Fig.~\ref{fig:2HofSchematic} and detailed in Section~\ref{subsec-Symmetries}.
As such, the model of Eq.~\eqref{eq-2HofHopping} and the ``optical-lattice hopping model'' are related by an electromagnetic gauge transformation and are thus physically equivalent.

\subsubsection{Hubbard Interactions}

While the single-particle terms did not depend on the particular alkaline-earth atom used, the specific values of the Hubbard-$U$ couplings depend sensitively on this choice.
The coupling between two alkaline-earth atoms at the same lattice site is proportional to one of four independent scattering lengths $a_\mathsf{gg}, a_\mathsf{ee}, a_\mathsf{eg}^{+},$ and $a_\mathsf{eg}^{-}$, depending on the internal states of the atoms~\cite{GorkovSUN}. 
The first two of these specify the interaction when the two atoms are both in the $\ket{\mathsf{g}}$ and $\ket{\mathsf{e}}$ clock states, respectively, while the latter two correspond to symmetric or anti-symmetric occupation of the $\ket{\mathsf{g}}$ and $\ket{\mathsf{e}}$ states.
Crucially, these scattering lengths do not depend on the nuclear spin due to an approximate $SU(2I+1)$ symmetry~\cite{GorkovSUN,scazza2014observation,zhang_spectroscopic_2014}, making them independent of the nuclear spin states chosen for the encoding in Eq.~\eqref{eq-nucspinencoding}.
Translating the atomic labels to the language of the double Hofstadter model, the realized interacting Hamiltonian takes the form
\begin{equation}
\begin{aligned} \label{eq-opticallatticeinteractions}
    \hat{V}^{\text{OL}} &= \sum_{\ell; \mathbf{r} \in \mathbb{Z}_2} \frac{U_1^{\ell}}{2} \hat{n}_{\mathbf{r}}^{\ell} (\hat{n}_{\mathbf{r}}^{\ell} - 1) + U_2 \sum_{\mathbf{r} \in \mathbb{Z}_2} \hat{n}_{\mathbf{r}}^{\mathsf{T}} \hat{n}_{\mathbf{r}}^{\mathsf{B}} \\
    & + J_{\text{ex}} \sum_{s, s' \in \{ \uparrow, \downarrow\}}\sum_{\mathbf{r} \in \mathbb{Z}_2}  \hat\psi_{\mathbf{r} \mathsf{B} s}^{\dagger} \hat\psi_{\mathbf{r} \mathsf{T} s'}^{\dagger} \hat\psi_{\mathbf{r} \mathsf{B} s'} \hat\psi_{\mathbf{r} \mathsf{T} s},
\end{aligned}
\end{equation}
where $U_1^{\mathsf{T}/\mathsf{B}} \propto a_{\mathsf{gg}/\mathsf{ee}}$ controls the strength of two independent in-layer interaction strengths, $U_2 \propto (a_\mathsf{eg}^+ + a_\mathsf{eg}^-)$ controls the inter-layer interaction strength, and $J_\mathrm{ex} \propto (a_\mathsf{eg}^+ - a_\mathsf{eg}^{-})$ is an additional $SU(2)$-symmetric ``Heisenberg'' spin-exchange interaction between the layers which conventionally is ferromagnetic (anti-ferromagnetic) for $J_{\text{ex}}>0$ ($J_{\text{ex}} < 0$).

To obtain the LAF phase at integer filling and the superconductor upon doping, the numerical results of Sections~\ref{sec-LAF}~and~\ref{sec:SC} show that it is favorable to be in an interaction regime where in-layer interactions $U_1^{\mathsf{T}/\mathsf{B}}$ are large while inter-layer interactions $U_2$ are small.
Furthermore, if the additional $J_{\text{ex}}$ term is ferromagnetic, it will disfavor LAF states in favor of spin-polarized states and hence will need to be smaller than the perturbatively-generated ``super-exchange'' term that selects out the LAF (c.f. Eq.~\eqref{eq-JLAFthing}).
These requirements on the interactions place constraints on the possible atomic species that can be used to realize our proposal, as their scattering lengths must realize a favorable ratio between $U_1^{\mathsf{T}/\mathsf{B}}, U_2, \text{ and } J_{\text{ex}}$.
However, we now show that simply using an additional clock-state dependent optical potential can systematically suppress $J_{\text{ex}}, U_{2}$ (see Supplemental Material for more details~\cite{SM}), thereby ameliorating these constraints. 

To see this, recall that the lattice interaction parameters in Eq.~\eqref{eq-opticallatticeinteractions} arise from the overlap between the Wannier functions of the optical lattice potential.
In particular, $J_{\text{ex}}, U_2 \propto \int_{\mathbf{r} \in \mathbb{R}^3} |w_{\mathsf{e}}(\mathbf{r})|^2 |w_{\mathsf{g}}(\mathbf{r})|^2$ where $w_{\mathsf{e}/\mathsf{g}}(\mathbf{r})$ are the Wannier functions of the atoms in either clock state.
By using a clock-state-dependent potential to displace atoms in different clock states by a distance $z_{\delta}$, such overlaps are suppressed exponentially in $z_{\delta}$, enabling one to effectively eliminate these two couplings.\footnote{We remark that in general, separating the two species will similarly suppress the inter-layer tunneling $g$. However, the strength of this term can, in principle, be increased arbitrarily by increasing the strength of the laser drive.}
Indeed, under an approximation that the Wannier functions are the wavefunctions of the lowest harmonic oscillator (which we justify in the Supplemental Material), these overlaps fall off as $\sim e^{-z_{\delta}^2/\xi^2}/\xi$ where $\xi \sim 1/V_z^{1/4}$ is a localization length.~\cite{SM}.
Crucially, this leaves the strength of the in-layer couplings unchanged as they are set by on-site Wannier overlaps for each clock state independently.
%
%

%

\subsubsection{Summary}

Eqs.~\eqref{eq:optical_lattice_hopping}--\eqref{eq-opticallatticeinteractions} constitutes a concrete optical lattice realization of the double Hofstadter model:
\begin{equation}
    \hat{H}^{\text{OL}} = \hat{h}_t^{\text{OL}} + \hat{h}_g^{\text{OL}} + \hat{V}^{\text{OL}}.
    \label{eq:2hof_full_Hamiltonian_optical_lattice}
\end{equation}
The two key differences between the original Hamiltonian in Eq.~\eqref{eq-full_hamiltonian} and its optical lattice realization lie in the interaction terms.
First, the Hamiltonian naively contains an additional spin-exchange interaction parameterized by $J_{\text{ex}}$.
While such a term can potentially be deleterious to realizing the LAF-correlated insulator and superconductor predicted earlier in this work, we demonstrate that a simple state-dependent potential can be used to eliminate it completely.
Conveniently, this also suppresses the inter-layer interaction strength $U_2$, which is also unfavorable for these two states.
Second, the above Hamiltonian generically has an imbalance in the interaction strengths within the two layers (i.e., $U_1^{\mathsf{T}} \neq U_1^{\mathsf{B}}$) set by the atomic species.
While a small imbalance does not affect the realization of LAF order in the absence of inter-layer interactions, the model is most symmetric and faithfully realized when these terms are equal.
%
We find that this condition is best borne out for $^{87}\text{Sr}$ and $^{173}\text{Yb}$, for which $U_1^{\mathsf{T}}/U_1^{\mathsf{B}} \approx 0.55$ and $0.65$, respectively~\cite{hofer2015observation, scazza2014observation, Sr87, goban2018emergence}.


\subsection{Relationship to Moir\'e Materials}\label{subsec-Relationship-to-Moire}

\begin{figure}
    \centering
    \includegraphics[width = 247 pt]{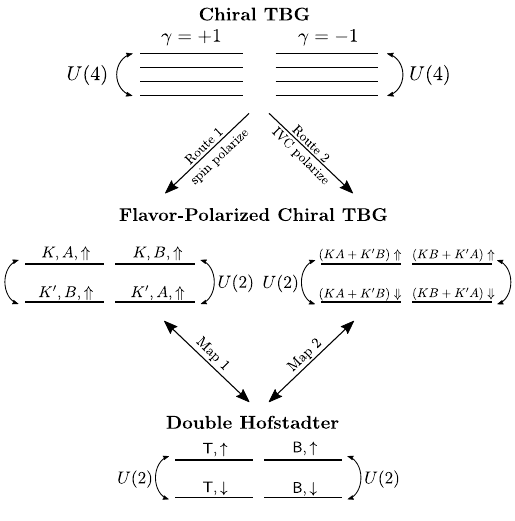}
    \caption{\textbf{Connection to Chiral Models of TBG.} (Top) Chiral TBG at its magic angle is typically described by a $U(4) \times U(4)$-symmetric model with eight internal states per momentum.
    These states can be partitioned into two $U(4)$-symmetric four-dimensional subspaces whose wavefunctions carry Chern number $\gamma = \pm 1$.
    (Middle) Upon flavor-polarizing, this model reduces to a four-band $U(2) \times U(2)$ symmetric model; the $U(2)$'s correspond to independent pseudo-spin rotations in opposite Chern sectors.
    (Middle, Left) When the flavor that is polarized is the electronic spin, this pseudo-spin roughly corresponds to the valley [See Eq.~\eqref{eq-psuedo1}].
    (Middle, Right) In contrast, when the electrons polarize into particular intervalley-coherent orbitals, this pseudospin corresponds to the electronic spin [See Eq.~\eqref{eq-pseudo2}].
    (Bottom) Both $U(2) \times U(2)$ symmetric models can be mapped to the strong-coupling double Hofstadter model [see Eq.~\eqref{eq:strongcoupling_H}].}
    \label{fig:chiral-TBG-comparison}
\end{figure}

Recall that we formulated the double Hofstadter model---and then developed its direct experimental implementation with cold atoms---in an effort to explore, in a minimal setting, the interplay between a set of ingredients commonly found in moir\'e materials (See Sec.~\ref{sec-intro}). 
Nevertheless, the presence of both correlated insulating states and unconventional superconductivity in this model, both appearing independently in experimental and theoretical studies of moir\'e graphene platforms, invites us to make precise the connection to such systems.
In this section, we reveal a close structural correspondence with the chiral model of magic-angle twisted bilayer graphene\footnote{For a pedagogical review of chiral TBG, we refer the reader to Ref.~\cite{pedagogical_patrick}.}~\cite{TarnoChiral, bultinck2020ground, BernevigTBGIV, kwan2021kekule}.


At the level of their band structures, the similarities between the two models are already striking.
In chiral TBG, the low-energy single-particle subspace also consists of energetically narrow bands with $w_2 = 1$ fragile topology in each flavor sector (i.e., valley and spin) and, consequently, hosts two Dirac cones of the same chirality per flavor~\cite{Po,Po2,ZouPo,song2019all,Hejazi,LiuDai,ahn2019Failure}.
Furthermore, just as the low-energy bands of the double Hofstadter model are naturally expressed in a Chern basis~\eqref{eq-renorm-layer-basis} of layer-polarized orbitals, the flat bands of chiral TBG are conveniently described by a basis of graphene sublattice-polarized Chern bands~\cite{bultinck2020ground}.

Going beyond single-particle physics, we will construct an explicit dictionary between the interacting models.
To do so, let us first recall some details of the strong-coupling theory of TBG.
The narrow bands of TBG are modeled by electrons carrying electronic spin $s$ and valley pseudospin $\tau$, with each electron occupying one of two narrow sublattice-polarized bands labeled by $\sigma$, for a total of eight flavors.
At the magic angle, these sublattice-polarized bands are exactly flat and the chiral model has a large $U(4) \times U(4)$ symmetry generated by independent spin and valley rotations in each Chern sector~\cite{bultinck2020ground,BernevigTBGIV}.
In Fig.~\ref{fig:chiral-TBG-comparison} (Top), we depict the division of the eight flat bands into two $U(4)$-symmetric subspaces with opposite Chern numbers.
Since experiments on TBG suggest some degree of flavor polarization in the vicinity of the superconductor (see e.g.~\cite{zondiner2020cascade, wong2020cascade,yu2022correlated}), many theoretical works reduce the eight degrees of freedom of the $U(4)\times U(4)$ model to four active flavors.
The result is a model with a $U(2) \times U(2)$ continuous symmetry (corresponding to charge conservation and independent rotations of a pseudospin) within each Chern sector $\gamma^z = \sigma^z \tau^z$~\cite{bultinck2020ground, Yves-phonons}.
These flavor-polarized models can be mapped explicitly to the double Hofstadter model.

Here, we consider two particular routes to flavor polarization studied in the literature [See Fig.~\ref{fig:chiral-TBG-comparison} (Middle)].
In the first route, the electronic spin is assumed to be fully polarized, and the resulting pseudospin $\bm{\eta}$ is related to the valley.
Each $U(2)$ symmetry is then generated by the electric charge, along with a pseudospin~\cite{pedagogical_patrick}
\begin{equation}\label{eq-psuedo1}
    \boldsymbol{\eta} = (\sigma^x \tau^x, \sigma^x \tau^y,  \tau^z).
\end{equation}
In the second route, recently posited in Ref.~\cite{Yves-phonons}, electron-phonon coupling causes the electrons to polarize into particular intervalley-coherent orbitals [See Fig.~\ref{fig:chiral-TBG-comparison} (Middle, Right)].
The resulting pseudospin is then just the electronic spin
\begin{equation}\label{eq-pseudo2}
    \widetilde{\boldsymbol{\eta}} = (s^x, s^y , s^z),
\end{equation}
whose components, along with electric charge, similarly serve as generators for each $U(2)$ symmetry.

To map these two flavor-polarized models to the double Hofstadter model, we recall from Sec.~\ref{subsec-FlatBandProjection} that the latter has an enhanced $U(2) \times U(2)$ continuous symmetry in the decoupled ($g = 0$) limit.
We may then identify the Chern index and pseudospin of the chiral models with the layer and electronic spin of the decoupled double Hofstadter model, respectively:
\begin{equation} \label{eq:TBG_2hof_mapping}
    l^z \longleftrightarrow \gamma^z, \qquad \bm{s}\longleftrightarrow \begin{cases} \boldsymbol{\eta} & \text{(Map 1)} \\ \widetilde{\boldsymbol{\eta}} & \text{(Map 2)} \end{cases}.
\end{equation}
Map 1 and Map 2, summarized in Fig.~\ref{fig:chiral-TBG-comparison} (Bottom), are bijections of both the degrees of freedom and continuous symmetries between the double Hofstadter model and flavor-polarized models of chiral TBG.

This identification extends to the Hamiltonian level.
Since the dispersion vanishes at the magic angle in the chiral limit, the projected Hamiltonian $\hat{H}^{\text{TBG}} = \sum_{\bq} V_{\bq} :\hat{\rho}_{\bq} \hat{\rho}_{-\bq}:$ with $\hat{\rho}_{\bq} = \sum_{\bk} \hat{c}_\bk^\dagger \Lambda^{\text{TBG}}_{\bq}(\bk) \hat{c}_{\bk+\bq}$ is entirely specified by the form factors $\Lambda^{\text{TBG}}$ [\textit{c.f.} Eq.~\eqref{eq-projecteddensity}].
Symmetries constrain their matrix structure to be $\Lambda^{\text{TBG}}_{\bq}(\bk) = F_{\bq}(\bk) e^{i \Phi_{\bq}(\bk) \gamma^z}$~\cite{pedagogical_patrick}.
For the double Hofstader model, the form factors~\eqref{eq:form_factor_decomposition} in the decoupled limit are
$\Lambda_{\bq}(\bk) = F^S_{\bq}(\bk) e^{i \Phi^S_{\bq}(\bk) l^z}$, which therefore have exactly the same structure.
Moreover, departure from the magic angle in chiral TBG introduces a non-zero dispersion $h \gamma^{x/y}$ that hybridizes the two Chern sectors (see Fig.~\ref{fig:chiral-TBG-comparison})~\cite{bultinck2020ground,pedagogical_patrick, eslamskyrmeSC}.
Such a term maps directly onto the inter-layer tunneling $g l^{x/y}$ of the double Hofstadter model.\footnote{In the double Hofstadter model, finite inter-layer tunneling $g$ also introduces an $\mcO(g)$ interaction term from $F^A$ in Eq.~\eqref{eq:form_factor_decomposition}.}
Maps 1 \& 2 therefore give an almost term-by-term correspondence in matrix structure between the Hamiltonians.

The similarities between the Hamiltonians suggest that their ground states may also be analogous.
To check this, we apply the above mappings to the LAF ground state at $\nu=2$.
Recall that the LAF state occupies the layer-polarized bands in the pattern:
\begin{equation}
    \ket{\text{LAF}} = \begin{tikzpicture}[scale = 0.6, baseline={([yshift=-.5ex]current bounding box.center)}]
    \draw[black, thick] (0, 1/2) -- (2, 1/2);
    \draw[black, thick] (3, 1/2) -- (5, 1/2);
    \draw[black, thick] (0, -1/2) -- (2, -1/2);
    \draw[black, thick] (3, -1/2) -- (5, -1/2);
    \node at (1, 1.2) {\small $\mathcal{C} = +1$};
    \node at (4, 1.2) {\small $\mathcal{C} = -1$};
    \node at (1, -1.2) {\small $\ $};
    \node at (4, -1.2) {\small $\ $};
    \node at (-2/3, 1/2) {\small $\ket{\uparrow}$};
    \node at (-2/3, -1/2) {\small $\ket{\downarrow}$};
    \foreach \i in {0, ..., 4}{
        \draw[color = orange(ryb), fill=orange(ryb)] (0 + \i * 0.5 , 1/2) circle (0.15);
        \draw[color = dodgerblue, fill=dodgerblue] (3 + \i * 0.5 , -1/2) circle (0.15);
    }
    \end{tikzpicture}
\end{equation}
Under Map 1, the LAF becomes either the Valley Hall (VH) state or the Kramer's Intervalley Coherent State (KIVC) (which are symmetry-equivalent in chiral TBG).
Strikingly, these are the ground states of the strong coupling model of chiral TBG at $|\nu|=2$~\cite{bultinck2020ground,BernevigTBGIV}.
Under Map 2, the LAF instead corresponds to the ``TIVC-QSH'' state, which has intervalley coherence and forms a spontaneous quantum spin Hall insulator in the microscopic electronic spin.
Studies of TBG where electron-phonon terms favor the relevant flavor polarization of Route 2 indeed find a TIVC-QSH ground state~\cite{Yves-phonons}.
The LAF therefore maps to ground states of flavor polarized models of TBG. This is remarkable.
Indeed, this suggests that the double Hofstadter model---which minimally incorporates the ingredients introduced in Sec.~\ref{sec-Key}---may yield valuable conceptual insight into the interacting phenomena present in chiral TBG.

We conclude this section by highlighting notable differences between chiral TBG and the double Hofstadter model.
Manifestly, the unit cells are quite different: hexagonal with $C_{6z}$ symmetry for unstrained TBG, versus rectangular in the double Hofstadter model.%
Chiral TBG also has an exact particle-hole symmetry~\cite{pedagogical_patrick}, whereas particle-hole is only approximately satisfied in the lowest-band-projected double Hofstadter model, becoming exact only in the $q\to \infty$ limit.
Another obvious difference lies in the interactions, which are local in the double Hofstadter model but longer ranged (screened Coulomb) in models of TBG.
While the range of interactions is expected to affect the energetics of excitations above the aforementioned insulating ground state~\cite{skyrmionswithoutsigma}, the relevance of the different discrete symmetries is less obvious and requires further investigation.

However, the key difference between the two models lies in the gauge-invariant properties of the Bloch wavefunctions~\cite{blount1962formalisms}.
These include the Berry curvature, or more generally the ``quantum geometry'' (reviewed in~\cite{pedagogical_patrick, liu2022recent}), and determine numerous physical quantities like the interaction-generated (Hartree-Fock) dispersion~\cite{bultinck2020ground}.
A key feature of TBG is that all these quantities are sharply concentrated near the $\Gamma$ point, which is known to strongly affect the ground state phenomenology (see e.g.~\cite{ShangNSM, parker2021field, wang2022narrow, song2022magic}).
In the double Hofstader model, by contrast, these quantities are distributed almost uniformly across the Brillouin zone, and the quantum geometry is much closer to an ideal ``vortexable'' limit~\cite{ledwith2022vortexability} (comparable to TBG at a chiral ratio $w_0/w_1 \sim 0.3$; details in Supplemental Material~\cite{SM}). Incorporating such an inhomogeneity into the double Hofstadter model could help gain a ``bottom-up'' view of the phenomenology of TBG and is an interesting direction for future research.
Nevertheless, given that experiments in magic-angle TBG~\cite{oh2021evidence} and its multi-layer generalizations~\cite{kim2022evidence} point towards nodal pairing~\cite{LakePheno}, the emergence of $p$-wave unconventional superconductivity in the related double Hofstadter model makes it a valuable departure point to explore pairing in more realistic moir\'e models.

\section{Discussion}

In this paper, we introduced the double Hofstadter model, a simple lattice model featuring strong repulsive interactions, narrow topological bands, and time-reversal symmetry.
By using state-of-the-art cylinder iDMRG, we investigated the ground state phase diagram of this model.
We found strong numerical evidence for the existence of a robust $p$-wave superconductor at circumference $L_y=5$ upon hole-doping a quantum spin Hall (QSH) insulator appearing at half-filling of the flat band subspace.
This correlated insulator, which we termed the Layer Anti-Ferromagnet (LAF), has opposite spin polarization in opposite, time-reversal-related Hofstadter layers.
Using a novel technique for probing the gap function of the superconducting condensate, we deduced both its pairing symmetry and characterized it as a ``BCS''-like superconductor comprised of Cooper pairs near the mean-field LAF Fermi surface.
We also presented an experimental blueprint for implementing the double Hofstadter model in near-term optical lattice setups and elucidated a close correspondence to chiral models of twisted bilayer graphene (TBG).

Before discussing our work in a broader context, we comment on the extent of our evidence for superconductivity and routes toward a more definitive verification.
Recall that our results were enabled by exploiting recent developments in the compression of matrix product operators~\cite{parker_2020}, extensive parallelization, and working at the largest accessible bond dimensions.
Although this enabled us to find definitive superconducting signatures at $L_y = 5$, the story at the larger circumferences, though promising, remains unclear.
Indeed, this is a consequence of a fundamental limitation of the cylinder iDMRG method, namely the exponential complexity of scaling the circumference.
An important and interesting direction for future work is to more conclusively demonstrate superconductivity in the 2D limit, either by developing new numerical methods to enable larger scale exploration \textit{within} cylinder iDMRG, or by developing and utilizing \textit{intrinsically} 2D numerical methods~\cite{dai2022fermionic}.

Caveats aside, one of the most intriguing features of this superconducting state is the persistence of superconductivity down to a very small LAF Zeeman pinning field, Eq.~\eqref{eq-Zeeman}.
While we introduced this as a numerical convenience---namely for stabilizing the LAF phase---its numerical value in this study is smaller than all other microscopic energy scales.
It is therefore likely that superconductivity is stabilized in the absence of any Zeeman field.
In this limit, the LAF is a \textit{spontaneous} symmetry-breaking state whose various low-energy fluctuations may play an essential role in the pairing mechanism.

This phenomenological feature of the model suggests intriguing connections to previous literature on superconductivity in doped, spontaneous QSH systems~\cite{ GroverSenthilSkyrmionSC,AbanovWiegmann} and important qualitative difference from prior numerical studies of superconductivity in models where the spin rotational symmetry is broken explicitly by strong spin-orbit coupling, e.g., interacting Kane-Mele systems~\cite{KaneMele,Yuan2012,ma2015triplet,Lee2019}. 
In the former case, superconductivity arises from the condensation of low-lying charge-$2e$ skyrmions of the QSH order parameter. A recent microscopic treatment of this mechanism demonstrated pairing within a purely repulsive model, which was proposed as a theory for superconductivity in TBG~\cite{shubsskyrme, eslamskyrmeSC,CBZ2020}.
The skyrmion mechanism was studied numerically both in coupled opposite-field Landau levels using iDMRG~\cite{shubsskyrme} and in models with effective local interactions amenable to QMC simulation~\cite{AssaadQMCskyrmeSC,Assaad2021}, as well as within effective field theory approaches~\cite{Christos_2020}.

An interesting future direction would be to study the relationship between a skyrmion superconductor and that which we have presented here.
In particular, while we have argued that our superconductor is ``BCS''-like and is phenomenologically consistent with the weak pairing of holes at a Fermi surface, a skyrmion superconductor is, in its simplest manifestation, formed from tightly-bound, bosonic skyrmion ``molecules.''
Interpolating between these limits, for instance through spin polarons or baby skyrmions, is an active area of research~\cite{babyskyrmions,Trions_TBG}.
In the case where both superconductors have $p$-wave pairing symmetry, such a transition could constitute a novel realization of the $p$-wave BCS-BEC phase transition~\cite{gurarie-pwave, crepel2023topological}.

Given that topology plays an essential role in a skyrmion superconductor by imbuing individual skyrmion textures with electric charge, it is important to understand the extent to which the weak-pairing superconductor presented here depends on the topology of its parent state.
For instance, one can imagine modifying the single-particle problem so that the low-energy bands are exponentially Wannierizable but the spread of these Wannier functions is sufficiently broad to favor a spontaneous---but topologically trivial---LAF insulating state~\cite{repellin_2020}.
One could determine whether superconductivity still manifests in this context and whether it is similarly robust.

Finally, we further remark on routes to experimentally realizing the double Hofstadter model, both in solid-state and cold-atom settings.
In the main text, we highlighted a precise mapping to the degrees of freedom of chiral TBG, but also commented on how these systems may differ energetically.
Given these differences, an interesting direction of future research would be to identify solid-state materials that realize double Hofstadter systems more faithfully, for instance in the family of moiré TMDs.
In the cold atom context, we presented a direct implementation of the double Hofstadter model using alkaline-earth atoms in an optical lattice.
While our study primarily sought to understand zero-temperature behaviors, directly accessing ground state physics in cold atom quantum simulators remains an outstanding challenge.
Namely, experiments typically either remove entropy by coupling to finite-temperature thermodynamic reservoirs~\cite{mazurenko_cold-atom_2017,yang_cooling_2020} or indirectly probe the ground state through quasi-adiabatic state preparation~\cite{BOHRDTReview}.
These obstacles motivate a systematic exploration of both the finite-temperature equilibrium phenomenology of the double Hofstadter model (e.g., applying the insights developed in Ref.~[\onlinecite{hoffman_fermionicQMC_2022}]), as well as non-equilibrium dynamical state preparation in this system, which has the potential to drastically influence the observed order~\cite{sahay2023quantum}.

\begin{acknowledgments}
We would like to thank Nick Bultinck, Junkai Dong, Ilya Esterlis, Lev Kendrick, Eslam Khalaf, Patrick Ledwith, Peter Littlewood, Francisco Machado, Ivar Martin, Andy Millis,  Alex Nikolaenko, Lokeshwar Prasad, Saran Prembabu, Daniel Shaffer, Senthil Todadri, Ruben Verresen, Pavel Volkov, and Yi-Zhuang You.
Our computations were run on the FASRC Cannon cluster supported by the FAS Division of Science Research Computing Group at Harvard University, on the Savio High Performance Computing cluster provided by the Berkeley Research Computing program at the University of California, Berkeley, and at the National Energy Research Scientific Computing Center, a DOE Office of Science User Facility supported by the Office of Science of the U.S. Department of Energy under Contract No. DE-AC02-05CH11231 using NERSC award BES-ERCAP0024721.

R.S. acknowledges support by the U.S. Department of
Energy, Office of Science, Office of Advanced Scientific
Computing Research, Department of Energy Computational Science Graduate Fellowship under Award Number
DESC0022158.
S.D. acknowledges support from the NSERC PGSD fellowship.
This research is funded in part by the Gordon and Betty
Moore Foundation’s EPiQS Initiative, Grant GBMF8683
to D.E.P.
T.S. was supported by a fellowship from Masason foundation, and by the U.S. Department of 
Energy, Office of Science, National Quantum Information Science Research Centers, Quantum 
Systems Accelerator.
S.A. was supported by the U.S. Department of Energy, Office of Science, Basic Energy Sciences, under Early Career Award No. DE-SC0022716.
J.H.’s research is part of the Munich Quantum Valley, which is supported by the Bavarian state government with funds from the Hightech Agenda Bayern Plus.
A.V. was supported by a Simons Investigator award and by the Simons Collaboration on Ultra-Quantum Matter, which is a grant from the Simons Foundation (651440, AV), and by NSF-DMR 2220703.
N.Y.Y acknowledges support via the AFOSR MURI program (via grant FA9550-21-1-0069), the ARO (via grant W911NF-21-1-0262) and the David and Lucile Packard Foundation. 
M.A. acknowledges support from the European Research Council (ERC) under the European Union’s Horizon 2020 research and innovation program (grant agreement No. 803047), from the Horizon Europe programme HORIZON-CL4-2022-QUANTUM-02-SGA via the project 101113690 (PASQuanS2.1) and from the Deutsche Forschungsgemeinschaft (DFG) via the Research Unit FOR 2414 under
Project No. 277974659 and under Germany’s Excellence Strategy –
EXC-2111 – 390814868.
S.C. acknowledges support from the ARO through the MURI program (grant number W911NF17-1-0323).
M.Z., S.C., and S.D. were supported by the U.S. Department of Energy, Office of Science, National Quantum Information Science Research Centers, Quantum Systems Accelerator (QSA).

\smallskip
\textit{Author Contributions}---S.D., R.S., D.P., and T.S. implemented the band-projected model.
R.S. and S.D. performed the numerical simulations and developed the numerical technique to extract the pair wavefunction and determine its symmetry properties.
S.D., R.S., D.P., and S.C. analyzed the data.
S.A. and J.H. built the multi-node parallel DMRG code.
R.S., N.Y. and M.A. developed the optical lattice implementation of the model.
S.D., R.S., D.P., S.C., A.V., and M.Z. clarified the connection to moir\'e materials and developed the theoretical analysis of the model at integer filling.
N.Y. and M.Z. conceived of the project.
D.P., A.V., S.C., N.Y., and M.Z. oversaw the project.
All authors discussed the results and wrote the manuscript.
\end{acknowledgments}

\bibliography{refs} 

\widetext
\newpage
\onecolumngrid

\definecolor{shadecolor}{gray}{0.9}

\onecolumngrid
\begin{center}
\textbf{\large Supplementary Material: ``Superconductivity in a Topological Lattice Model with Strong Repulsion''}
\end{center}

\setcounter{equation}{0}
\setcounter{figure}{0}
\setcounter{table}{0}
\makeatletter
\renewcommand{\thefigure}{S\arabic{figure}}
\renewcommand{\thesection}{S\Roman{section}}
\renewcommand{\thesubsection}{S\Roman{subsection}}
\renewcommand{\bibnumfmt}[1]{[S#1]}
\setcounter{section}{0}

\newcommand{\nocontentsline}[3]{}
\newcommand{\toclesslab}[3]{\bgroup\let\addcontentsline=\nocontentsline#1{#2\label{#3}}\egroup}


\newcommand{\Danfigscale}{0.75}
\newcommand{\otherfigscale}{0.5}

\appendix

\toclesslab\section{Organization of the Appendices}{}

Below, we provide a detailed report of the numerical evidence and analytics underlying the claims and methods used in the main text.
The appendices are organized in a similar order to the main text.
In particular, Appendix~\ref{app:micro_model} expounds upon details of the double Hofstadter model, which was introduced in Sec.~II of the main text.
Subsequently, Appendix~\ref{app:strong_coup}~and~\ref{app:NumericalDataUndoped} provides additional details regarding the results of Sec.~III. The former details the strong coupling theory of the double Hofstadter model, which we used to analytically argue for the presence of the LAF, and the latter provides numerical data supplementing that in the main text.
Following this, we provide additional numerical data for how we characterize the superconductor (Appendix~\ref{app:NumericalDataExistenceandRobustness}), as well as further theoretical details on the extraction of the pair wavefunction (Appendix~\ref{app:pairwave}) and the pairing symmetry (Appendix~\ref{app:PairingSymmetry}).
Finally, we provide details on the connection between the double Hofstadter model and contemporary experiments.
We provide a detailed exposition of our optical lattice implementation (Appendix~\ref{app:optical_lattice}) and our mapping to models of chiral twisted bilayer graphene (Appendix~\ref{app:Dictionary}).

\vspace{8mm}\toclesslab\section{Additional Details on the Double Hofstadter Model}{app:micro_model}

In the main text, we extensively studied the double Hofstadter model, namely its symmetries, non-interacting band structure, and ground states.
Here, we provide some additional details regarding the model's single-particle physics.
To set notation, let us recall the definition of the double Hofstadter model and then discuss the symmetries, band structure, and layer-polarized basis of the projected model.

Recall that the flux through the top/bottom layer is $\pm \Phi = \pm \Phi_0/q$ per square plaquette, where we fix $q=4$.
Our model has direct lattice vectors $\mathbf{a}_x = (qa,0)$ and $\mathbf{a}_y = (0,a)$, a basis for a rectangular lattice $\mathbb{L}$.
Consider a \textit{square} grid of sites $\widetilde{\mathbb{L}}$ with separation $a$ so that each unit cell $\mathbf{u} = m_1 \mathbf{a}_x + m_2 \mathbf{a}_y$ contains $q$ sites at sublattice positions
\begin{align} \label{eq-lattice-site}
    \mathbf{r} = \mathbf{u} + \v{\sigma} = \mathbf{u} + (\sigma a, 0)
\end{align}
for $0 \le \sigma < q$. At each site, we consider a row-vector of microscopic fermion creation operators
\begin{equation}
    \hat{\psi}^\dagger_{\sigma,\ell s}(\mathbf{u}) = \left(    \hat{\psi}^\dagger_{\sigma,\mathsf{T} \uparrow}(\mathbf{u}),
    \hat{\psi}^\dagger_{\sigma,\mathsf{B} \uparrow}(\mathbf{u}),
    \hat{\psi}^\dagger_{\sigma,\mathsf{T} \downarrow}(\mathbf{u}),
    \hat{\psi}^\dagger_{\sigma,\mathsf{B} \downarrow}(\mathbf{u})
    \right)_{\ell s},
\end{equation}
where layer $\ell \in \st{\mathsf{T}, \mathsf{B}} = \{+1,-1\}$ and spin $s \in \st{\uparrow,\downarrow} = \{+1,-1\}$ are on-site degrees of freedom, for a total of four flavors per site.
We often also use $\hat\psi^\dag_{\br \ell s}$ as a short-hand for the operator $\hat\psi^\dag_{\sigma, \ell s}(\bu)$ at the square lattice site $\br = \bu + \bm{\sigma}$. Crucially, the layers are vertically separated, with a magnetic flux of $+\Phi_0$ per unit cell for the top layer and $-\Phi_0$ per unit cell in the bottom layer. Adopting a Landau gauge and setting $a = 1$ henceforth, the Hamiltonian for the double Hofstadter model is
\begin{align}\label{eq:app_2hof_Hamiltonian}
    \hat{H} &= \hat{h} + \hat{V}\\
    \hat{h} &= -t \sum_{\mathbf{r} \in \widetilde{\mathbb{L}}} \left(\hat{\psi}^{\dagger}_{\mathbf{r} + \hat{y}} e^{\frac{2\pi i x}{qa} \ell^z} \hat{\psi}_{\mathbf{r}} + \hat{\psi}^{\dagger}_{\mathbf{r} + \hat{x}} \hat{\psi}_{\mathbf{r}} + \text{h.c.}\right) - g \hat{\psi}^{\dagger}_{\mathbf{r}} \ell^x \hat{\psi}_{\mathbf{r}} \\
    \hat{V} &= \frac{U_1}{2} \sum_{\mathbf{r} \in \widetilde{\mathbb{L}}} :\hat{n}_{\mathbf{r} \mathsf{T}}^2 + \hat{n}_{\mathbf{r} \mathsf{B}}^2: +\ U_2 \sum_{\mathbf{r} \in \widetilde{\mathbb{L}}} :\hat{n}_{\mathbf{r} \mathsf{T}} \hat{n}_{\mathbf{r} \mathsf{B}}: 
\end{align}
where $\hat{n}_{\br\ell} := \sum_s \hat{\psi}_{\br\ell s}^\dagger \hat{\psi}_{\br\ell s}$ and colons denotes normal ordering relative to the electronic vacuum. 
Each layer consists of a Harper-Hofstadter model with kinetic energy $t$, coupled by inter-layer tunneling $g$.
Moreover, the model is manifestly charge conserving and independent of spin and hence has continuous $U(1)_Q$ charge and $SU(2)$ spin symmetries. We consider separate in-layer repulsion $U_1 > 0$ and between-layer repulsion $U_2 > 0$.  We underscore that $\hat{H}$ is a purely repulsive model.

We remark that, in the main text, we find it convenient numerically to add a small field $B_z$ to the Hamiltonian:
\begin{equation}
    \hat{H} \to \hat{H} - B_z \sum_{\mathbf{r} \in \widetilde{\mathbb{L}}} \hat{\psi}^{\dagger}_{\mathbf{r}} s^z \ell^z \hat{\psi}_{\mathbf{r}}.
\end{equation}
Such a term breaks the $SU(2)$ symmetry down to $U(1)_S$ and is called a Zeeman term as it Zeeman shifts the electronic spin of the top (bottom) layer with $\pm B_z$ consistent with the physical picture of the double Hofstadter model (with opposite magnetic fields in opposite layers).

\vspace{6mm}\toclesslab\subsection{Space Group Representation in Momentum Space}{}

The following characterization of the symmetries of the double Hofstadter
model holds for any even integer $q,$ where $\Phi=\ell\Phi_{0}/q$
is the flux per plaquette in layer $\ell.$ In particular, this analysis
applies to the case $q=4$ considered in the main text. The full space
group is generated by the operations
\begin{align}
\hat{m}_x,\ \hat{m}_y,\ \hat{C}_{2x},\ \hat{C}_{2z},\ \hat{\mathcal{I}}\label{eq:symmetries}
\end{align}
defined in the main text. In particular, $\hat{T}_{1}=\hat{m}_x^2$
and $\hat{T}_{2}=\hat{m}_{y},$ that correspond to translations by
$qa\hat{x}$ and $a\hat{y}$, generate an abelian translation group.
On the other hand, the full group of magnetic translations is non-abelian
since $\hat{m}_y$ anti-commutes with $\hat{m}_y$
We therefore take $\hat{T}_{1}$ and $\hat{T}_{2}$ to define the minimal
unit cell. The corresponding primitive reciprocal lattice
vectors are:
\begin{align}
\mathbf{G}_{1}=\frac{2\pi}{qa}\hat{x},\qquad\mathbf{G}_{2}=\frac{2\pi}{a}\hat{y}.
\end{align}

Placing the system on a torus of size $qN_{x} \times N_{y}$ lattice sites, we define Fourier-transformed
fermion operators by:
\begin{align}
\hat{\psi}_{\sigma,\ell s}^{\dagger}(\mathbf{k})=\frac{1}{\sqrt{N}}\sum_{\mathbf{u}\in\mathbb{L}}e^{i\mathbf{k}\cdot(\mathbf{u}+\sigma\hat{\mathbf{x}})}\hat{\psi}_{\sigma,\ell s}^{\dagger}(\mathbf{u}),
\end{align}
where $\bm{\sigma}=\sigma\hat{\mathbf{x}}.$ Here, $N=N_{1}N_{2}$
is the total number of unit cells on the torus and $\mathbf{k}$ takes
values in the magnetic Brillouin zone. The actions of the symmetries Eq.~\eqref{eq:symmetries} on the
Fourier-transformed fermion operators are as follows:
\begin{alignat}{3}
 \Ad_{\mathcal{I}} \hat{\psi}^{\dagger}(\mathbf{k}) & =\hat{\mathcal{I}} \hat\psi^{\dagger}(\mathbf{k})\hat{\mathcal{I}}^{-1} &  & =\hat\psi^{\dagger}(\mathbf{k})R\ell^{x}s^{x} &  & \qquad\text{(anti-unitary)}\\
 \Ad_{C_{2z}} \hat\psi^{\dagger}(\mathbf{k}) & = \hat C_{2z}\hat\psi^{\dagger}(\mathbf{k})\hat C_{2z}^{-1} &  & =\hat\psi^{\dagger}(-\mathbf{k})R &  & \qquad\text{(unitary)}\\
 \Ad_{C_{2x}} \hat\psi^{\dagger}(\mathbf{k}) & = \hat C_{2x}\hat\psi^{\dagger}(\mathbf{k})\hat C_{2x}^{-1} &  & =\hat\psi^{\dagger}(C_{2x}\mathbf{k})\ell^{x}s^{x} &  & \qquad\text{(unitary)}\\
 \Ad_{m_x} \hat\psi^{\dagger}(\mathbf{k}) & = \hat{m}_x \hat\psi^{\dagger}(\mathbf{k})\hat{m}_x^{-1} &  & =\hat\psi^{\dagger}(\mathbf{k}+\pi\hat{y})Se^{-ik_{x}q/2} &  & \qquad\text{(unitary)}\\
 \Ad_{m_y} \hat\psi^{\dagger}(\mathbf{k}) & = \hat{m}_y \hat\psi^{\dagger}(\mathbf{k})\hat{m}_y^{-1} &  & =\hat\psi^{\dagger}(\mathbf{k})e^{-ik_{y}} &  & \qquad\text{(unitary)},
\end{alignat}
where for each symmetry $g$, we have defined an adjoint superoperator $\Ad_{g}\hat{\mcO} = \hat{g}\hat{\mcO}\hat{g}^{-1}$ that performs the action of conjugation.
Moreover, $R$ is a $q\times q$ permutation matrix that describes how
the sublattice degree of freedom transforms under a rotation about
an atom at sublattice $\sigma=0:$ 
\begin{align}
[R]_{\sigma'\sigma}=\delta_{(-\sigma'\,\text{mod}\,q),\sigma}.
\end{align}
Note that $R$ leaves the $\sigma=0$ sublattice equivalence class
invariant and satisfies $R=R^{-1}.$ Similarly, $S$ is a $q\times q$
permutation matrix that shifts the sublattices by $q/2$ sites in
the $\hat{x}$ direction, i.e. by half the unit cell: 
\begin{align}
[S]_{\sigma'\sigma}=\delta_{(\sigma'+q/2\,\text{mod}\,q),\sigma},
\end{align}
and likewise satisfies $S=S^{-1}.$ These sublattice matrices commute:
$[S,R]=0.$

Let us now comment on the algebraic relationships between the symmetry
generators.
Of relevance are the commutators of the adjoint
superoperators, which we show in Table~\ref{table-commutators}.
In particular, the entry of row $g$ and column $h$ is the action of the commutator $[\Ad_{g},\Ad_{h}]=\Ad^{-1}_{g}\circ\Ad^{-1}_{h}\circ\Ad_{g}\circ\Ad_{h}$
on the vector space of one fermion operators, which is spanned by the microscopic operators $\psi^{\dagger}(\mathbf{k}).$
Note that the diagonal of the table is the identity and that entries above the diagonal are related to those below via the identity
\begin{align}
[\Ad_h,\Ad_g] = \Ad^{-1}_h\circ\Ad^{-1}_g\circ\Ad_h\circ\Ad_g = \left(\Ad^{-1}_g\circ\Ad^{-1}_h\circ\Ad_g\circ\Ad_h\right)^{-1}=[\Ad_g,\Ad_h]^{-1},
\end{align}
which holds whether $g$ and $h$ are linear or anti-linear.
Also note that each commutator $[\Ad_g,\Ad_h]$ is always
a linear superoperator.
Computing the commutators explicitly, we find
that all the adjoint superoperators commute up to a momentum-dependent
phase.
In particular, these phases are such that they instead exactly \textit{commute} when acting on the vector space of charge-$2e$ operators, which is constructed in the usual way from the generators $\psi^{\dagger}(\mathbf{k})$ of the charge-$e$ fermionic Hilbert space space.
Moreover, all the symmetry superoperators square to the identity on this same vector space of charge-$2e$ operators.

\begin{table}
\begin{centering}
\begin{tabular}{|c|c|c|c|c|c|c|}
\hline 
\multicolumn{1}{|c}{} &  & \multicolumn{5}{c|}{$h$}\tabularnewline
\cline{3-7} \cline{4-7} \cline{5-7} \cline{6-7} \cline{7-7} 
\multicolumn{1}{|c}{} &  & $\mathcal{I}$  & $C_{2z}$  & $C_{2x}$  & $m_{x}$  & $m_{y}$\tabularnewline
\hline 
\multirow{5}{*}{$g$} & $\mathcal{I}$  & $1$  & $1$  & $1$  & $e^{-ik_{x}q}$  & $e^{-2ik_{y}}$\tabularnewline
\cline{2-7} \cline{3-7} \cline{4-7} \cline{5-7} \cline{6-7} \cline{7-7} 
 & $C_{2z}$  & $1$  & $1$  & $1$  & $e^{-ik_{x}q}$  & $e^{-2ik_{y}}$\tabularnewline
\cline{2-7} \cline{3-7} \cline{4-7} \cline{5-7} \cline{6-7} \cline{7-7} 
 & $C_{2x}$  & $1$  & $1$  & $1$  & $1$  & $e^{-2ik_{y}}$\tabularnewline
\cline{2-7} \cline{3-7} \cline{4-7} \cline{5-7} \cline{6-7} \cline{7-7}
 & $m_{x}$ & $e^{ik_{x}q}$  & $e^{ik_{x}q}$  & 1  & $1$  & $e^{i\pi}$\tabularnewline
\cline{2-7} \cline{3-7} \cline{4-7} \cline{5-7} \cline{6-7} \cline{7-7} 
 & $m_{y}$ & $e^{2ik_{y}}$  & $e^{2ik_{y}}$  & $e^{2ik_{y}}$  & $e^{-i\pi}$  & $1$\tabularnewline
\hline 
\end{tabular}
\par\end{centering}
\caption{The action of the commutator $[\Ad_{g},\Ad_{h}]=\Ad^{-1}_{g}\circ\Ad^{-1}_{h}\circ\Ad_{g}\circ\Ad_{h}$
on the vector space of one fermion (i.e. charge-$e$) operators spanned
by the microscopic operators $\mathcal{\psi}^{\dagger}(\bk).$
Here, $g$ and $h$ run over the generators of the magnetic symmetry
group. We remark that the adjoint superoperators corresponding to these generators commute up to a momentum-dependent phase.}
\label{table-commutators} 
\end{table}

\vspace{6mm}\toclesslab\subsection{Band Structure of the Double Hofstadter Model}{}

\begin{figure}
    \includegraphics[width=\textwidth]{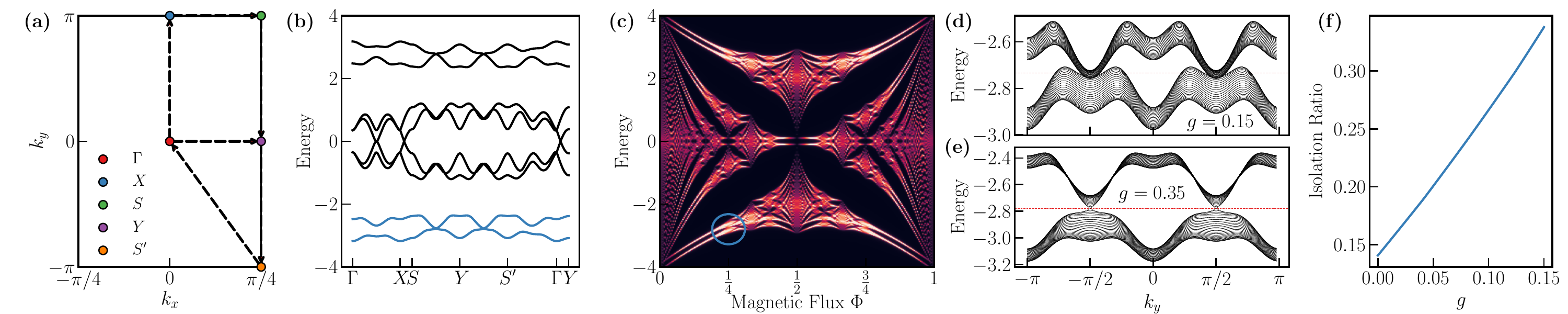}
    \caption{ \textbf{Band Structure of the Double Hofstadter Model.}
    (a) The Brillouin zone of the double Hofstadter model for $q=4$. Dashed lines show a high-symmetry path.
    (b) The spectrum of the double Hofstadter model for $t=1, g=0.35, q=4$ along the high-symmetry path. Each band is doubly-degenerate due to spin. We project down to the lower bands (blue). Note the presence of two Dirac cones between the two lowest bands at $\bk = (\pi/4,\pm \pi/2)$.
    (c) The ``double Hofstadter butterfly" spectrum for $t=1,g=0.1$ for $0 \le \Phi = p/q \le 1$. (Although we have taken $p=1$ in the discussion in the main text, one can use the standard generalization to rational fractions $p/q$ without changing the shape of the Brillouin zone.) Notice that the two lowest bands at $\Phi=1/4$ (blue circle) are continuously connected to the two lowest Landau levels in the limit $q \to \infty$.
    (d) Constant $k_x$ cuts of the bandstructure at $g=0.15$. Note that the Fermi level at half-filling (dashed red line) is above the Dirac point, yielding both electron and hole pockets. 
    (e) Constant $k_x$ cuts at $g=0.35$. The Fermi level now intersects the Dirac point, giving a Dirac semimetal.
    (f) Plot of the ``isolation ratio'' (defined as the ratio of the bandwidth of the low-energy manifold to the gap above it) vs. the interlayer tunneling $g$, for the range $0\leq g\leq 0.15$ considered in the superconducting analysis presented in the main text.
    }
    \label{fig:app:bandstructure}
\end{figure}

At a flux fraction of $\Phi = \Phi_0/q$ per square plaquette, the band structure of the double Hofstadter model features $2q$ bands per spin species. Moreover, since the model is spin rotationally-symmetric (in the $B_z=0$ limit), each of these bands is additionally two-fold degenerate in spin.
In this work, we focus on the lowest four bands, i.e. two bands times two spins.
For small values of the inter-layer tunneling $g$, these bands are narrow and energetically well-isolated from the rest of the spectrum, as shown in Fig.~\ref{fig:app:bandstructure}(b).
In this same figure, we observe that these low-energy bands host two Dirac cones of the same chirality in each spin sector. 
To clarify the situation, Fig.~\ref{fig:app:bandstructure}(e) shows the detailed view of the vicinity of the Dirac cones with Fermi levels at half occupancy marked by a dashed red line. At large $g=0.35$, this exactly intersects the Dirac cones, so bandstruture describes a Dirac semimetal.
At smaller values of the inter-layer tunneling [Fig.~\ref{fig:app:bandstructure}(d)], the Dirac cones are buried below the Fermi level, indicating a band metal with both electron and hole-like Fermi pockets. This is consistent with the behavior at $g=0$, where the bands are degenerate at each momentum and not entirely flat.

We now further comment on the structure and protection of the Dirac cones, assuming $g$ is large enough that the half-filled state is semi-metallic.
Whenever two Dirac cones are present in the spectrum, then symmetries are responsible for their topological protection and locations.
Specifically, $C_{2z}$ introduces an energetic degeneracy between $\bk$ and $-\bk$, whereas $C_{2x}$ relates $\bk$ to $C_{2x}\bk$, and the magnetic translation $m_x$ likewise makes $\bk$ and $\bk + \pi\hat{y}$ degenerate (assuming $q$ is even so that these momenta are both allowed).
This constrains the two Dirac cones to sit at only a finite number of isolated points in the Brillouin zone, namely $(k_x,k_y)\in\{0,\pm\pi/4,\pm\pi/2\} \times \{0,\pm\pi\}$. For energetic reasons, the Dirac cones of the double Hofstadter model incidentally reside at the particular locations $\bk = (-\pi/4, \pm \pi/2)$.
Provided that these Dirac cones exist, their topological protection can be guaranteed by the spacetime inversion symmetry $\mathcal{I}$ and spin $SU(2)$.
To see this, it is convenient to think in the layer-polarized basis that was introduced in the main text and will be expounded upon in more detail in Section~\ref{subsec-layerpolarized} of the supplemental material.
First, note that the spin $SU(2)$ of our model constrains the single particle Hamiltonian of the two lowest bands to be $h(\mathbf{k}) = \sum_{\alpha \in \{0, x, y, z\}} h_{\alpha}(\mathbf{k}) l^\alpha$.
Moreover, in an appropriate gauge, spacetime inversion acts by $h(\mathbf{k}) = l^x h^*(\mathbf{k}) l^x$, which precludes $l^z$ from the single-particle Hamiltonian.
Note that the gap between the two bands is only set by the $l^{x, y}$ components of $h$.
Since the Hamiltonian looks like $h(\mathbf{k}) = h_0(\mathbf{k}) l^0 + k_x l^x + k_y l^y + \mcO(\bk^2)$ in the vicinity of each Dirac cone, we conclude that symmetric single-particle perturbations may be able to move the Dirac cones below the Fermi energy but cannot get rid of these gap closings.
We conclude by noting that since our model is symmetric under spacetime inversion $\mathcal{I}$ and consists of two low-energy bands, the winding of polarization (as shown in Fig.~\ref{fig:wannier_data}d) implies $w_2=+1$ fragile topology~\cite{AhnPRX}.

We now brief comment on the energetic isolation of the lowest pair of bands. In this work, we focus on the case $q=4$ since it is the smallest $q$ for which the low-energy bands are well-isolated from the rest of the single-particle spectrum. In particular, as we show in Fig.~\ref{fig:app:bandstructure}(f), the ratio of bandwidth to bandgap ratio (or ``isolation ratio'') at $q=4$ and $g=0.0$ is $0.22/1.53 \approx 0.15 \ll 1$. Moreover, it remains relatively small for the values $g\leq 0.15$ considered in our superconducting analysis. In contrast, in the case of $q=3$ and $g=0.0$, this ratio is $0.73/1.27 \approx 0.57$, which is a factor of four larger. Since our projection to the lowest-two bands (per spin sector) is best justified when the isolation ratio is small, we restrict to $q=4$ to ensure our DMRG numerics are representative of the ground state of the full microscopic model.

\vspace{6mm}\toclesslab\subsection{Layer-Polarized Hybrid Wannier Computational Basis}{subsec-layerpolarized}

In the single-particle subspace of the low-lying, narrow bands, it
is desirable to construct a basis in which the Chern number is manifest.
Inspired by the $g=0$ limit, in which the Chern number is equivalent
to the layer index $\mathcal{C}=\pm\ell,$ we achieve this by constructing
a ``layer-polarized'' basis whose single-particle states are maximally-polarized
to one of the layers. Crucially, we use the gauge freedom within the
low-energy bands to construct a ``hybrid Wannier'' basis of orbitals
that are maximally exponentially-localized in the $x$ direction but
delocalized (namely, plane waves with momentum $k_{y}$) around the
circumference of the cylinder.

To achieve a basis with the aforementioned properties, we perform
1D Wannier localization on each set of Bloch states on each wire $k_{y}.$\footnote{Due to the finite circumference $L_{y}$ of the cylinder, $k_{y}$
takes $L_{y}$-many quantized values. We therefore refer to each constant-$k_{y}$
slice as a ``wire,'' along which $k_{x}$ is effectively continuous.} In doing so, we
obtain the polarizations $P_{x}(k_{y})\in[0,1)$ of the 1D exponentially-localized
orbitals. Within the Modern Theory of Polarization, these polarizations
are related to the eigenvalues of the Wilson loop along each wire
$k_{y}.$ In addition, we demand that the basis transforms in a prescribed
way under the anti-unitary symmetry $\mathcal{I},$ namely the following
purely off-diagonal form:
\begin{equation} \label{eq:I_action_constraint}
O_{\mathcal{I}}(\bk)=\langle v(\bk)|\hat{\mathcal{I}}|v(\bk)\rangle=e^{-2\pi i(P^{+}(k_{y})+P^{-}(k_{y}))k_{x}/N_{x}}s^{x}\sigma^{x},
\end{equation}
where $s^{x}$ and $\sigma^{x}$ are the usual Pauli matrix (in the spin and Wannier label), and the 1D Hybrid Wannier states are given by
\begin{align}
|v_{ls}(\bk)\rangle=\hat{\varphi}_{ls}^{\dagger}(\bk)|0\rangle=\sum_{\sigma,\ell}\hat{\psi}_{\sigma\ell s}^{\dagger}(\bk) |0\rangle \, v_{\sigma\ell,l}(\bk).
\end{align}
Since $\mathcal{I}$ anti-commutes with the position operator $\hat{\mathcal{I}}\hat{\mathbf{r}}=-\hat{\mathbf{r}}\hat{\mathcal{I}},$
then the two polarizations are opposite (modulo unity), which implies
that $P^{+}(k_{y})+P^{-}(k_{y})$ takes an integer value on each wire.

Having chosen the above off-diagonal action of $\mathcal{I}$, we
guarantee that our basis has several important and convenient properties.
Most crucially, the basis is a \textit{layer-polarized} basis.
As defined in the main text, such a basis (\textit{i}) diagonalizes the action of the (projected) layer operator
\begin{align} \label{eq-proj-ell-z}
O_{\ell^{z}}(\bk)=\langle v(\bk)|\hat{\ell}^{z}|v(\bk)\rangle=v^{\dagger}(\bk)\ell^{z}v(\bk),
\end{align}
with (\textit{ii}) eigenvalues $\lambda_l(\bk)$ that take a constant sign across the Brillouin zone.
To see this, we recall that $\hat{\mathcal{I}}\hat{\psi}^{\dagger}(\bk)\hat{\mathcal{I}}=\hat{\psi}^{\dagger}(\bk)\ell^{x}s^{x},$
and remark that Eq.~\eqref{eq:I_action_constraint} is then equivalent
to the constraint
\begin{equation} \label{eq:I_matrix_constraint}
v^{\dagger}(\bk)\ell^{x}v^{*}(\bk)=e^{i\gamma(\bk)}\sigma^{x},
\end{equation}
where the coefficient is the complex phase:
\begin{align}
e^{i\gamma(\bk)}=e^{-2\pi i(P^{+}(k_{y})+P^{-}(k_{y}))k_{x}/N_{x}}.
\end{align}
Invoking the constraint Eq.~\eqref{eq:I_matrix_constraint}, it is then straightforward to show that
\begin{align}
\sigma^{x}O_{\ell^{z}}^{*}(\bk)\sigma^{x}=-O_{\ell^{z}}(\bk).
\end{align}
Since $O_{\ell^{z}}(\bk)$ is a $2\times2$ Hermitian matrix, this expression implies that it is traceless and diagonal. Numerically, it takes the form $O_{\ell^{z}}(\bk)=\lambda(\bk)\sigma^{z}$
with $0.9 < \lambda(\bk) < 1$, for all parameters $g \in [0, 0.3]$ considered, so that the matrix is never singular.
This concludes the proof that the states $|v_{ls}(\bk)\rangle$ constitute a layer-polarized basis, justifying the use of the symbol $l$ (referred to in the main text as the ``dressed layer'' index) to label these states. Henceforth, we use $l^a$ rather than $\sigma^a$ to denote the Pauli matrices acting in the space of layer-polarized states.

\begin{figure*}
    \centering
    \includegraphics[width=\textwidth]{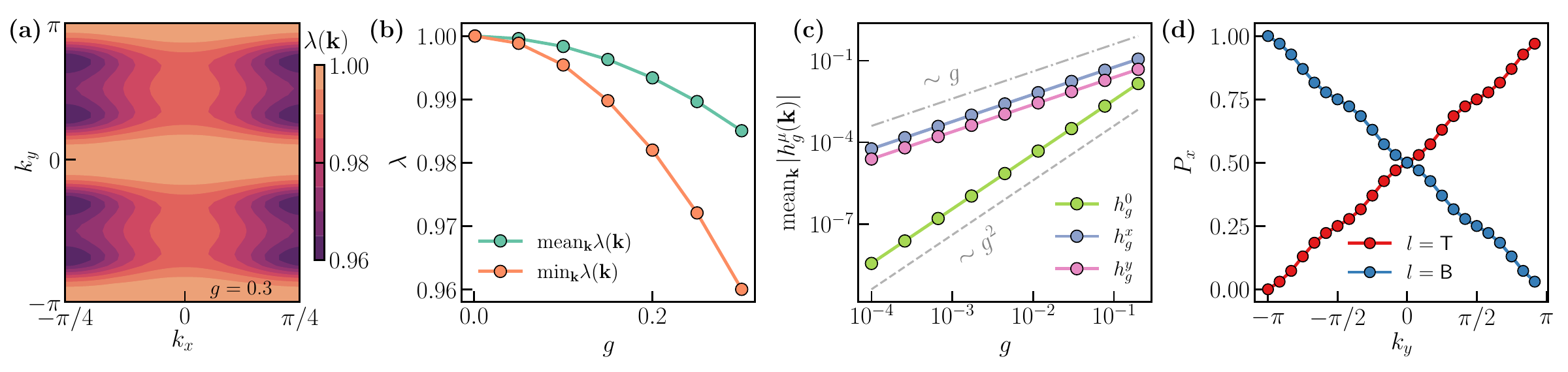}
    \caption{\textbf{Layer Polarized Basis of the Double Hofstadter Model.} (a) The positive eigenvalue $\lambda(\bk)$ of the projected layer operator in the layer-polarized basis at $g=0.3$.
    (b) The mean and minimum of this quantity, evaluated over the BZ, for a broad range of values $0 < g \leq 0.3$ of the inter-layer tunneling. Across this interval, $\lambda$ is bounded below by $0.96 \sim 1$, indicating near-maximal layer polarization.
    (c) The BZ-average of the components of the projected inter-layer tunneling operator $\hat\mcP \hat{h}_g \hat\mcP$ expressed in the layer-polarized basis. The dominant components $\mu=x,y$ scale linearly in $g$, whereas the $\mu=0$ component is quadratic and the $\mu=z$ component is made to vanish identically.
    (d) Polarizations $P^{\mathsf{T}/\mathsf{B}}(k_y)$ on cuts of constant $k_y$ at $g=0.3$. The Chern number of a band obeys $C = \int P(k_y) d k_y$. Here the two bands wind oppositely, with crossings at $k_y = -\pi,0$. This demonstrates the Chern number of the layer-polarized bands are $C=\pm 1$ for the top/bottom bands, and that these low-energy bands have a non-trivial fragile topological invariant~\cite{AhnPRX}, $w_2=+1$.
    }
    \label{fig:wannier_data}
\end{figure*}

Next, we derive the constraints imposed on the projected inter-layer tunneling in the layer-polarized basis. The matrix structure of $\hat\mcP \hat{h}_g \hat\mcP$ is related to the overlap $O_{\ell^{x}}(\bk)$, whose form is likewise constrained by Eq.~\eqref{eq:I_matrix_constraint} as follows:
\begin{align}
    l^{x}O_{\ell^{x}}^{*}(\bk)l^{x}=O_{\ell^{x}}(\bk).
\end{align}
Since $O_{\ell^{x}}(\bk)$ is a $2\times 2$ Hermitian matrix, this expression guarantees that it has no $l^z$ matrix component.
Though it may generally possess an identity component (i.e. a finite trace), we remark that this contribution is generally very small, namely of order $\mcO(g^2)$. In contrast, the off-diagonal $l^{x,y}$ components go as $\mcO(g)$ (see Fig.~\ref{fig:wannier_data}(c)).
Altogether, the projected tunneling operator takes the following form:
\begin{align}
    \hat\mcP \hat{h}_g \hat\mcP \approx \sum_\bk \hat\varphi^\dag_\bk \Big[ h_g^x(\bk) l^x + h_g^y(\bk) l^y \Big] \hat\varphi_\bk.
\end{align}

Applying similar arguments, it can be shown that the hybrid Wannier orbitals transform as:
\begin{alignat}{3} 
\Ad_{\mathcal{I}} \hat{\varphi}^{\dagger}(\mathbf{k}) & =\hat{\mathcal{I}}\hat{\varphi}^{\dagger}(\mathbf{k})\hat{\mathcal{I}}^{-1} &  & =\hat{\varphi}^{\dagger}(\mathbf{k})s^{x}l^{x}e^{i\gamma(\mathbf{k})} &  & \qquad\text{(anti-unitary)} \label{eq-wannier-Isym-operator}\\
\Ad_{C_{2z}} \hat{\varphi}^{\dagger}(\mathbf{k}) & =\hat{C}_{2z}\hat{\varphi}^{\dagger}(\mathbf{k}) \hat{C}_{2z}^{-1} &  & =\hat{\varphi}^{\dagger}(-\mathbf{k})e^{i\gamma(\mathbf{k})}s(k_{y}) &  & \qquad\text{(unitary)}\\
\Ad_{C_{2x}} \hat{\varphi}^{\dagger}(\mathbf{k}) & = \hat{C}_{2x}\hat{\varphi}^{\dagger}(\mathbf{k}) \hat{C}_{2x}^{-1} &  & =\hat{\varphi}^{\dagger}(C_{2x}\mathbf{k})s^{x}l^{x}s(k_{y}) &  & \qquad\text{(unitary)}\\
\Ad_{m_x} \hat{\varphi}^{\dagger}(\mathbf{k}) & = \hat{m}_x \hat{\varphi}^{\dagger}(\mathbf{k}) \hat{m}_x^{-1} &  & =\hat{\varphi}_{l}^{\dagger}(k_{x},k_{y}+\pi)e^{-ik_{x}q/2}e^{i\mu_{l}(\mathbf{k})} &  & \qquad\text{(unitary)} \label{eq-wannier-mx-operator} \\
\Ad_{m_{y}} \hat{\varphi}^{\dagger}(\mathbf{k}) & = \hat{m}_y \hat{\varphi}^{\dagger}(\mathbf{k}) \hat{m}_y^{-1} &  & =\hat{\varphi}^{\dagger}(\mathbf{k})e^{-ik_{y}} &  & \qquad\text{(unitary),}
\end{alignat}
where $s(k_{y})=\pm1$ is some wire-dependent sign factor and $e^{i\mu_{l}(\mathbf{k})}$ is some complex phase.
For our particular numerical implementation, we additionally have that
\begin{align}
    e^{i\gamma(-\bk)} = e^{-i\gamma(\bk)},\qquad s(-k_y)=s(k_y),
\end{align}
for all wires $k_y$. Moreover,
\begin{align}
    e^{i\mu_l(-\bk)} = e^{i\mu_l(\bk)}\qquad \mathrm{and} \qquad e^{i\mu_{+1}(\bk)}e^{i\mu_{-1}(\bk)} = 1,
\end{align}
so long as $k_y \not\in\{0,\pi\}.$
These identities are indispensable for our analysis of the superconducting pairing symmetry in Appendix~\ref{app:PairingSymmetry}. Finally, we remark that we assume our layer-polarized basis is periodic in the Brillouin zone:
\begin{align}
    \hat{\varphi}_{ls}^{\dagger}(\bk + \bG) = \hat{\varphi}_{ls}^{\dagger}(\bk),
\end{align}
or equivalently that
\begin{equation}
    v_{\sigma \ell, l}(\bk+\bG) = e^{-i \bG \cdot \v{\sigma}} v_{\sigma \ell, l}(\bk).
\end{equation}

\vspace{6mm}\toclesslab\subsection{Quantum Geometry}{}

To conclude our discussion of the single-particle aspects of the double Hofstadter model, we comment briefly on its quantum geometry. Quantum geometry studies the position operator projected into a few (often flat) bands. A key motivation for its study was determining when lattice systems were ``close enough" to the lowest Landau level that one should expect fractional quantum Hall physics to appear, but also appears in numerous other contexts including optics and studies of flat band superconductivity. Here we use it to be able to make basis-independent comparisons to other systems, notably the lowest Landau level and to the Bistritzer-MacDonald model.

\begin{figure}
    \includegraphics[width=\textwidth]{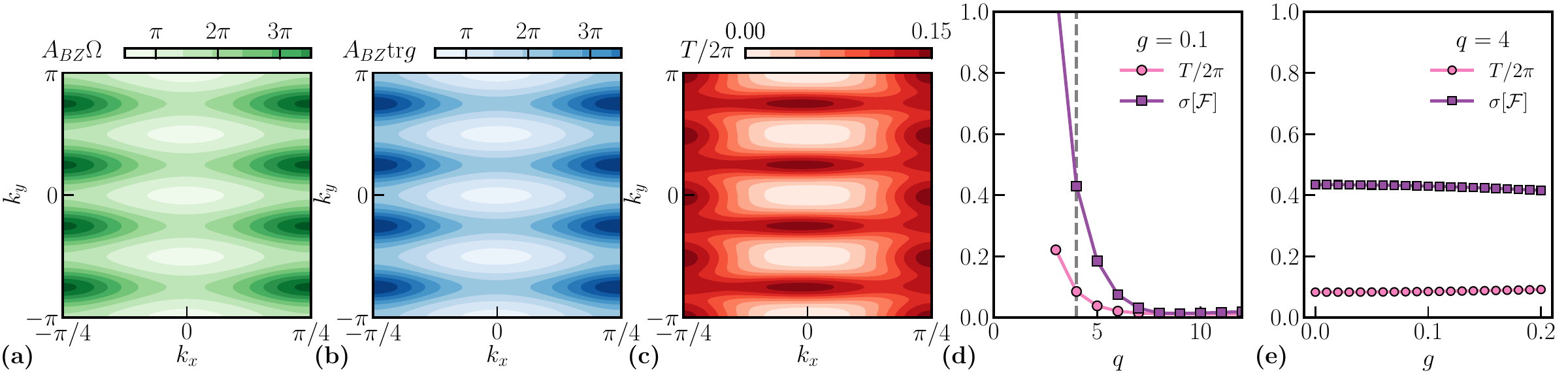}
    \caption{ \textbf{Quantum Geometry of the Double Hofstadter Model.} We work in the the top layer-polarized band at $q=4,g=0.1$ (unless otherwise noted).
    (a) Berry curvature $\Omega(\bk)$, normalized by the Brillouin zone area so that a uniform $C=1$ band would have value $2\pi$.
    (b) The trace of the Fubini-Study metric $g_{\mathrm{FS}}(\bk)$, defined in the text.
    (c) The violation of the trace inequality $T(\bk) = {\operatorname{Tr}} g_{\mathrm{FS}}(\bk) - \Omega(\bk)$ that vanishes for the lowest Landau level, or more generally for `ideal' Chern bands. 
    (d) The integrated trace violation $T/2\pi$ and standard deviation of the Berry curvature $\sigma[\mathcal{F}]$ as a function of flux $q$. These vanish as the band approaches the lowest Landau level limit at large $q$.
    (e) The same quantities evolve only weakly with $g$.
    }
    \label{fig:app_QGeometry}
\end{figure}

Consider a well-isolated band or set of bands whose Bloch periodic wavefunctions are $u_{\bk n}(\br)~=~e^{-i\bk\cdot\br} \psi_{\bk n}(\br)$ with associated projector $P_{\bk} = \sum_{n} \ket{u_{\bk n}}\bra{u_{\bk n}}$. Momentum space band geometry focuses on the Berry curvature $\mathcal{F}(\bk)$ and the Fubini-Study metric $g_{\mathrm{FS}}(\bk)$. These are defined by
\begin{equation}
    g_{\mathrm{FS}}^{\mu\nu}(\bk) = \mathrm{Re}[\eta^{\mu\nu}(\bk)],\quad
    \mathcal{F}^{\mu\nu}(\bk) = -\frac{1}{2}\mathrm{Im}[\eta^{\mu\nu}(\bk)],
\end{equation}
where 
\begin{equation}
\eta^{\mu\nu}(\bk) = \operatorname{Tr}[P_{\bk} \hat{r}^\mu Q_{\bk} \hat{r}^\nu P_{\bk}] = \sum_n \braket{\partial^{k_\mu} u_{\bk n}|Q_{\bk}| \partial^{k_\nu} u_{\bk n}}
\end{equation}
is the quantum geometry tensor and $Q_{\bk} = I - P_{\bk}$. The non-zero component of the curvature $\Omega = \mathcal{F}^{xy}$ integrates to the Chern number: $C = \frac{1}{2\pi} \int d\bk \Omega$. These are plotted for the double Hofstadter model for the top layer polarized band in Fig.~\ref{fig:app_QGeometry}. Both quantities are rather uniform, and both have the same peaks and troughs whose four-fold structure is due to the (weakly broken) magnetic translation symmetry in each layer. Since the metric is related to the interaction-induced Fock dispersion~\cite{abouelkomsan2023quantum}, the resemblance of these quantities to the LAF bandstructure (shown in the main text) is not accidental.

This data allows us to compare the double Hofstadter model to the lowest Landau level. To this end, we recall the \textit{trace inequality}\footnote{Comparing to the formula for the Chern number, the characteristic size of this quantity is $2\pi$. Hence we plot $T/2\pi$.}
\begin{equation}
    T = \int T(\bk) \, d^2\bk = \int {\operatorname{Tr}} g_{\mathrm{FS}}(\bk) - \Omega(\bk)\,  d^2{\bk} \ge 0.
\end{equation}
The lowest Landau level is entirely specified by its band geometry, in the sense that any $\n{C}=1$ Chern band where: (I) $T=0$, and (II) $\Omega(\bk)$ is constant is equivalent to the lowest Landau level \cite{Roy_2014}. Fig.~\ref{fig:app_QGeometry}(c) plots the $\bk$-resolved trace inequality, showing that the top layer-polarized band of the double Hofstader model is extremely close to the lowest Landau level. Indeed, in the limit $q \to \infty$, both $T\to 0$ and the standard deviation of the Berry curvature $\sigma[\mathcal{F}] \to 0$ [Fig.~\ref{fig:app_QGeometry}(d)]. The layer-polarized bands of the double Hofstadter model are therefore closely adiabatically connected to the lowest Landau levels. This can be seen visually in Fig.~\ref{fig:app:bandstructure} (c), where one can trace the adiabatic connection from $q=4$ (blue circle) to $q\to \infty$. Notably, this closeness of the top layer polarized band to the LLL (and the bottom layer polarized band to the LLL in opposite magnetic field) evolves only weakly with $g$ [see Fig.~\ref{fig:app:bandstructure} (e)]. Conceptually, one can therefore think of the physics of the double Hofstadter model as descending from that of pairs Landau levels in opposite magnetic fields. However, as we saw in the main text, the small deviations from uniformity across the Brillouin zone determine the locations of the Fermi surface and the character of any superconductivity.

Another system that can be thought of as two approximate Landau levels in opposite magnetic fields is twisted bilayer graphene \cite{TarnoChiral,ledwith2020fractional}. There, however, the deviations from the ideal \cite{wang2021exact} or vortexable limit \cite{ledwith2022vortexability} are concentrated near the $\Gamma$ point --- a time-reversal invariant momentum point --- whereas in the double Hofstadter model the trace inequality $T(\bk)$ is largest at non-high-symmetry points. The double Hofstadter model at $q=4,g=0.1$ has comparable trace inequality violation to the sublattice-polarized flat band of the Bistritzer-MacDonald model at chiral ratio $w_{0}/w_1 \approx 0.3$ \cite{ledwith2020fractional}. Given these quantitative comparisons to both Landau levels and twisted bilayer graphene, it is reasonable to expect phenomena from those systems to manifest in the double Hofstadter model.

\vspace{8mm}\toclesslab\section{Additional Details on the Strong-Coupling Theory of the LAF}{app:strong_coup}

This section provides a complete treatment of the strong-coupling theory presented in the main text, which gave analytical predictions for the ground state at $\nu = 2$ electrons per unit cell. This filling is equal to that of a band insulator occupying two of the four narrow bands.
Our treatment hews closely to related calculations in magic angle graphene, whose notation and conventions we largely follow \cite{bultinck2020ground,eslamskyrmeSC, pedagogical_patrick}. Provided that the interaction scale $U$ is smaller than the gap to the ``remote'' bands, (which is about $\sim 2t$ for $q=4$ at $g=0.1$), we expect the ground state at any filling $\nu \leq 4$ to be accurately captured by the flat-band projected Hamiltonian $\hat{\mcP}\hat{H}\hat{\mcP}.$
Below we carefully define the projected interaction and dispersion and show, in the symmetric $g=0$ limit, that the interaction selects out a manifold of Slater determinant ground states. We consider both the case $U_2=0$ where this manifold is comprised of bilayer quantum Hall ferromagnets (QHFMs) and the case $U_2 = U_1$ where these states compete with layer-polarized QAH states. Adding back $g>0$ perturbatively, we will see that both the LAF --- a particular family of anti-aligned bilayer QHFMs --- and QAH insulators are selected as ground states. Our strong-coupling analysis is heavily inspired a formalism developed for twisted bilayer graphene, introduced in \cite{kang2019strong,bultinck2020ground,BernevigTBGIV}.



\vspace{6mm}\toclesslab\subsection{Flat Band-Projected Interaction}{}

To make analytic progress on this Hamiltonian with strong-coupling theory, we must express the interaction in the appropriate form. We start from real space, transform to momentum space, then finally form an effective model for the flat band subspace. Consider a generic density-density interaction
\begin{align}
\hat{V} = \sum_{\ell,\ell'} \sum_{\br,\br'}V_{\ell\ell'}(\br - \br') :\hat{\rho}_{\br\ell} \hat{\rho}_{\br'\ell'}:\ , \qquad \hat{\rho}_{\br \ell} = \sum_{s} \hat{\psi}_{\br \ell s}^\dag \hat{\psi}_{\br \ell s},
\end{align}
where $V_{\mathsf{TB}}=V_{\mathsf{BT}}>0$ is the inter-layer interaction and $V_{\mathsf{TT}}=V_{\mathsf{BB}}>0$ is in-layer.
Next, writing the lattice site as $\br = \bu + \bm{\sigma}$ [see Eq.~\eqref{eq-lattice-site}], we consider the Fourier transformed fermion operators
\begin{equation}
    \hat{\psi}_{\sigma, \ell s}(\mathbf{u}) = \frac{1}{\sqrt{N}} \sum_b \sum_{\bk \in \text{BZ}} e^{i \bk\cdot(\mathbf{u} + \v{\sigma})} v_{\sigma \ell, b}(\bk) \hat{\varphi}_{b s}(\bk),
\end{equation}
where $\text{BZ}$ is the first Brillouin zone and $b$ is a ``generalized'' band index. One can think of the $v$'s as Bloch wavefunctions of the kinetic Hamiltonian defining a band basis. However, we use no specific properties of that basis; in fact, we will later specialize to the case where the $v$'s define the layer-polarized basis. In any case, we additionally require a periodic gauge:
\begin{equation} \label{eq-periodic-gauge}
    v_{\sigma \ell, b}(\bk+\bG) = e^{-i \bG \cdot \v{\sigma}} v_{\sigma \ell, b}(\bk). 
\end{equation}
This means the $v$'s are specified entirely by their values on momenta within the first Brillouin zone. The Fourier transformed density is naturally written in terms of \textit{layer-resolved form factors}, which are independent of spin:
\begin{equation}
    [\Lambda_{\bq \ell}(\bk)]_{bb'} = \left[v^\dag(\bk) {P}_{\ell} v(\bk+\bq)\right]_{bb'} = \sum_{\sigma} v_{\sigma \ell, b}(\bk)^* v_{\sigma \ell, b'}(\bk+\bq).
\end{equation}
Here ${P}_{\ell} = (\ell^0 + \ell\ell^z)/2$ is the projector onto the microscopic layer $\ell$.
We emphasize that the form factors are matrices that do not depend on spin; we therefore consider them as $2\times 2$ matrices in the generalized band index $b$. Our choice of periodic gauge implies the following useful identities:
\begin{equation}
    \Lambda_{-\bq \ell}(\bk+\bq) = \Lambda_{\bq \ell}(\bk)^\dagger, 
    \quad 
    \Lambda_{-\bG \ell}(\bk) = \Lambda_{\bG\ell}(\bk)^\dagger, \quad
    \Lambda_{\bq \ell}(\bk+\bG) = \Lambda_{\bq \ell}(\bk). 
    \label{eq:form_factor_identities}
\end{equation}
The Fourier transformed density operator may then be written as
\begin{equation} \label{eq:rho-qell}
    \hat\rho_{\bq \ell}
    = \frac{1}{\sqrt{qN}} \sum_{\br} e^{-i \bq \cdot \br} \hat\rho_{\br \ell} 
    = \frac{1}{\sqrt{qN}} \sum_{b,b',s} \sum_{\bk} \hat{\varphi}_{bs}^\dagger(\bk) [\Lambda_{\bq \ell}(\bk)]_{bb'} \hat{\varphi}_{b's}(\bk+\bq),
\end{equation}
where the second step follows from the Poisson summation formula and periodic gauge.
%
Using the Fourier transformed potential $V_{\ell\ell'}(\br) = \frac{1}{qN}\sum_{\bq\in\text{EBZ}}e^{-i\bq\cdot\br}V_{\ell\ell'}(\bq)$, the interaction then takes the form
\begin{equation}
   \hat{V}=  \sum_{\ell \ell'} \sum_{\bq\in\text{EBZ}}V_{\ell\ell'}(\bq):\hat\rho_{\bq \ell} \hat\rho_{-\bq \ell'}:
   \label{eq:interaction_momentum_space}
\end{equation}
where colons denote normal ordering with respect to the vacuum. Here, EBZ denotes the ``extended'' Brillouin zone associated
with the microscopic square lattice, to be distinguished from the BZ derived from
the $q\times1$ site unit cell.

Finally we make an effective model for the active bands. Since the active bands are energetically the lowest ones then, just as with the lowest Landau level, projecting into them amounts to restricting to a certain set of operators. Let $\mathcal{A} \cup \mathcal{R}$ be a partition of the generalized bands into ``active'' bands of the flat band subspace and all other ``remote'' bands.
The associated many-body projector is $\hat{\mcP}= \prod_{\bk\in\text{BZ}}\prod_{s}\prod_{b\in \mathcal{R}}(1-\hat{\varphi}_{bs}^{\dagger}(\bk) \hat{\varphi}_{bs}(\bk))$. Crucially, since the interaction is normal-ordered with respect to
the vacuum, and since the image of $\hat{\mcP}$ contains no states
that occupy the inactive bands $\mathcal{R},$ then $\hat{\mcP} \hat{V} \hat{\mcP}$
takes the same form as Eq.~\eqref{eq:interaction_momentum_space}, except that the band indices in the expression for the density operator are \textit{truncated} to $\mathcal{A}$. Therefore we have arrived at the flat band-projected interaction:
\begin{equation}
   \hat{\mcP} \hat{V} \hat{\mcP} = \sum_{\ell \ell'} \sum_{\bq\in\text{EBZ}} V_{\ell\ell'}(\bq): \hat{\mcP} \hat\rho_{\bq\ell} \hat{\mcP} \hat\rho_{-\bq\ell} \hat{\mcP}:.
   \label{eq:projected_interaction_momentum_space}
\end{equation}


\vspace{6mm}\toclesslab\subsection{Strong-Coupling Form of the Hamiltonian}{}

To apply strong-coupling theory, we must partition the full projected Hamiltonian into a \textit{positive-definite} interaction $\hat\mcV$ plus residual terms that are perturbatively small or act trivially within the ground state manifold of $\hat\mcV$.
In what follows, we discuss how this is performed in two distinct physical limits, namely $U_2=0$ and $U_2=U_1$.
In each case, strong-coupling theory requires that the interaction Hamiltonian be comprised of densities measured relative to a reference \textit{background charge density} at $\nu = 2$, rather than the $\nu=0$ vacuum.
We first focus on the limit $g=0$ where layer is a good quantum number, postponing a perturbative treatment of the inter-layer tunneling to the next section.

\subsubsection*{Case 1: Vanishing Inter-Layer Interactions}

In the limit $U_2 = 0$ where the inter-layer interactions vanish, it is natural to posit that this background charge density evenly occupies each layer and is translationally-invariant. A family of states whose charge densities satisfy these properties is the manifold of bilayer QHFMs, defined as
\begin{equation}
    \label{eq-BLQHFM}
    \ket{\v{n}_{\mathsf{T}},\v{n}_{\mathsf{B}}} = \prod_{\bk} \hat\varphi^\dag_{\mathsf{T},\bm{n}_\mathsf{T}}(\bk) \hat\varphi^\dag_{\mathsf{B},\bm{n}_\mathsf{B}}(\bk) \ket{0},
\end{equation}
where the Fermion operators, whose spins are aligned to the axes $\v{n}_l = (\cos \theta_l \sin \phi_l, \sin \theta _l\sin \phi_l, \cos \phi_l) \in S^2$, are defined in terms of the dressed layer orbitals (see Sec.~\ref{subsec-layerpolarized}) by
\begin{equation}
    \hat\varphi^\dag_{l,\bm{n}_l} = \cos(\theta_l/2) \hat\varphi^\dag_{l, \uparrow} + e^{i \phi_l} \sin(\theta_l/2) \hat\varphi^\dag_{l, \downarrow}.
\end{equation}
We now argue that the bilayer QHFMs Eq.~\eqref{eq-BLQHFM} are exact eigenstates of the projected density operator Eq.~\eqref{eq:rho-qell} in the $g=0$ limit, and use this fact to engineer a corresponding positive-definite ``strong-coupling'' interaction $\hat{\mcV}$ that exactly annihilates these states. This is pertinent to the original ground state problem because $\hat{\mcV}$ will be found to differ from the original projected interaction Eq.~\eqref{eq:projected_interaction_momentum_space} by a single-particle term that approximates a \textit{chemical potential}, up to perturbatively small momentum-dependent variations.

To show that the states $\ket{\v{n}_{\mathsf{T}},\v{n}_{\mathsf{B}}}$ are exact zero modes of the projected density, we use the enhanced symmetry at $g=0$. Note that the layer-projected form factors are greatly restricted by $U(1)$ layer conservation. Namely, since layer is a good quantum number, the dressed layer index $l$ is indistinguishable from the microscopic layer $\ell$, and correspondingly:
\begin{align} \label{eq-layerdiag-ffs}
    \left[\Lambda_{\bq \ell}(\bk)\right]_{ll'} \propto \delta_{l \ell} \delta_{l' \ell}.
\end{align}
The projected interactions therefore take the form
\begin{align}
    \hat{\mcP}\hat{\rho}_{\bq\ell}\hat{\mcP} \ket{\v{n}_{\mathsf{T}},\v{n}_{\mathsf{B}}} &= \frac{1}{\sqrt{qN}}\sum_{\bk\in\text{BZ}} \left[\Lambda_{\bq\ell}(\bk)\right]_{\ell \ell} \sum_{s}\hat{\varphi}_{\ell s}^{\dagger}(\bk) \hat{\varphi}_{\ell s}(\bk+\bq) \ket{\v{n}_{\mathsf{T}},\v{n}_{\mathsf{B}}}.
\end{align}
This expression vanishes due to Pauli exclusion unless $\bq = \bG$ is a reciprocal lattice vector, in which case $\hat{\varphi}(\bk + \bG) = \hat{\varphi}(\bk)$ due to the periodic gauge convention Eq.~\eqref{eq-periodic-gauge}. The Fermion bilinear then becomes a number density, and thus
\begin{align} \label{eq:rhobar1_eigenstates}
    \hat{\mcP}\hat{\rho}_{\bq\ell}\hat{\mcP} \ket{\v{n}_{\mathsf{T}},\v{n}_{\mathsf{B}}} &= \bar{\rho}_{\bq\ell} \ket{\v{n}_{\mathsf{T}},\v{n}_{\mathsf{B}}},
\end{align}
where
\begin{align} \label{eq:rhobar1_def}
    \bar{\rho}_{\bq\ell} = \frac{1}{\sqrt{qN}}\sum_{\bk\in\text{BZ}} \sum_{\bG} \delta_{\bq,\bG}\operatorname{tr}\left[\Lambda_{\bG\ell}(\bk)\right].
\end{align}
Here $\operatorname{tr}$ denotes a trace over $l$, which is trivial in the $g=0$ limit, but provides a definition for $\bar{\rho}$ that will later prove useful in the case where $g$ is perturbatively small. We have therefore demonstrated that the bilayer QHFM states are exact eigenstates of the density operator.

As forecasted above, this enables us to isolate a positive-definite portion of the projected interaction. As a preliminary step, define the \textit{relative density operator} by
\begin{align} \label{eq:projected_density_operator}
    \delta \hat{\rho}_{\bq\ell} = \hat{\mcP} \hat{\rho}_{\bq\ell}\hat{\mcP}- \bar{\rho}_{\bq\ell} \hat{I}.
\end{align}
By construction, this operator automatically annihilates each bilayer QHFM, defined above in Eq.~\eqref{eq-BLQHFM}.
Applying standard identities (e.g., see Ref.~\onlinecite{parker2021field}), one can then re-write the projected interaction in terms of these relative densities (without normal ordering) as
\begin{equation}
   \hat{\mcP} \hat{V} \hat{\mcP} = \sum_{\ell \ell'} \sum_{\bq\in\text{EBZ}} V_{\ell\ell'}(\bq) \delta \hat{\rho}_{\bq\ell} \delta \hat{\rho}_{-\bq\ell'} + \hat{h}_\mathrm{HF}[I/2] + \text{constant}.
\end{equation}
We remark that this identity holds even when $U_2 \geq 0$, in which case the off-diagonal layer components of $V(\bq)$ are non-zero.
In the above, $\hat{h}_\mathrm{HF}$ is the Hartree-Fock Hamiltonian and $I$ is the identity matrix in the spin and dressed layer indices. The Hartree-Fock Hamiltonian is a quadratic operator, defined relative to a correlation matrix $P(\bk)_{\alpha\beta} = \braket{\hat{\varphi}^\dagger_{\bk\beta}\hat{\varphi}_{\bk \alpha}}$, given by $\hat{h}_\mathrm{HF}[P] = \sum_{\bk, \alpha\beta} \hat{\varphi}_{\bk\alpha}^\dagger \left( h_\text{H}[P](\bk)_{\alpha \beta} + h_\text{F}[P](\bk)_{\alpha \beta} \right) \hat{\varphi}_{\bk\beta}$ with
\begin{subequations}
\label{eq:hartree-fock_equations}
\begin{align}
    h_\text{H}[P](\bk) &= \frac{2 s^0}{qN} \sum_{\ell\ell'} \sum_{\bG} V_{\ell\ell'}(\bG) \Lambda_{\bG\ell}(\bk) \operatorname{Tr}\left(\sum_{\bk'\in\mathrm{BZ}}P(\bk') \Lambda_{\bG\ell'}^{\dagger}(\bk') \right), \\
    h_\text{F}[P](\bk) &= -\frac{2}{qN} \sum_{\ell \ell'} \sum_{\bq \in \EBZ} V_{\ell\ell'}(\bq) \Lambda_{\bq\ell}(\bk) P(\bk+\bq) \Lambda_{\bq\ell'}(\bk)^\dagger.
\end{align}
\end{subequations}
Here, matrix multiplication is implicit, $\bG$ is restricted to reciprocal lattice vectors that lie in the first extended Brillouin zone (EBZ), $s^0$ is the identity matrix in spin, and $\operatorname{Tr}$ denotes a trace over \textit{both} bands and spin. 
The projection of the full Hamiltonian Eq.~\eqref{eq:app_2hof_Hamiltonian} may then be written in the ``strong-coupling'' form
\begin{equation}
    \hat{\mcP}\hat{H}\hat{\mcP} = \delta \hat{h} + \hat{\mcV} =  \left(\hat{\mcP} \hat{h} \hat{\mcP} + \hat{h}_{\mathrm{HF}}[I/2]
    \right)
    +
    \sum_{\ell \ell'} \sum_{\bq\in\text{EBZ}} V_{\ell\ell'}(\bq) \delta\hat{\rho}_{\bq\ell} \delta\hat{\rho}_{-\bq\ell'},
    \label{eq:Hamiltonian_drho_drho_form}
\end{equation}
up to an overall constant shift. Note that the last term is no longer normal ordered. Specializing again to the relevant case $U_2 = 0$, note that since $\delta\hat{\rho}_{-\bq\ell} = \delta\hat{\rho}^\dagger_{\bq\ell}$ and $V(\bq)\geq 0$, then the strong-coupling interaction $\hat{\mcV}$ is positive semi-definite, which implies that any states it annihilates are necessarily also its ground states. We therefore conclude that the bilayer QHFMs [Eq.~\eqref{eq-BLQHFM}] are all exact ground states of $\hat{\mcV}$.

In order to see that these states also constitute approximate ground states of the original projected Hamiltonian $\hat{\mcP}\hat{H}\hat{\mcP}$, it suffices to note that $\delta\hat{h}$, the quadratic term in the strong-coupling Hamiltonian, can be approximated by a chemical potential, which is constant at the fixed filling $\nu = 2$.
Indeed, not only do we find that $\delta h$ is proportional to the identity matrix $s^0 l^0$, but also that the momentum-dependent variations about its average are small.
In particular, they are dominated by the bare Hofstadter bandwidth, which is $\sim 0.1\, t$ (at our standard value $q=4$), as opposed to the contributions from $h_\mathrm{HF}[I/2]$.
These bandwidth fluctuations, in turn, are much smaller than the energy scale $U_1$ responsible for the ferromagnetic alignment in each layer, and in fact vanish identically in the Landau level limit $\Phi_0/q \to 0$ (see Fig.~\ref{fig:app:bandstructure}c).


\subsubsection*{Case 2: Equal In-Layer and Inter-Layer Interactions}

We now formulate the strong-coupling Hamiltonian for the limit $U_2 = U_1$, which is relevant to the mapping to chiral TBG, and more closely approximates the parameter regimes accessible in near-term optical lattice implementations.
Though the procedure will be almost identical to that above, we will discover that the family of approximate ground states is expanded, relative to the $U_2 = 0$ case, to include additional layer-polarized QAH insulating states.

As a first step, note that the interaction matrix takes a simple form in this limit:
\begin{align}
    V_{\ell\ell'}(\bq) =
    V(\bq)\begin{pmatrix}
        1 & 1\\
        1 & 1
    \end{pmatrix}_{\ell\ell'}.
\end{align}
In other words, all its components are equal. As a result, in the interaction Hamiltonian itself, it is no longer necessary to resolve the fermionic density operator by its layer-projected components. Define the all-layer density by
\begin{align}
    \hat\rho_{\bq} = \sum_\ell \hat\rho_{\bq\ell} = \frac{1}{\sqrt{qN}} \sum_{l,l',s} \sum_{\bk} \hat{\varphi}_{ls}^\dagger(\bk) [\Lambda_{\bq}(\bk)]_{ll'} \hat{\varphi}_{l's}(\bk+\bq),
\end{align}
where $\Lambda_{\bq}(\bk) = \sum_\ell \Lambda_{\bq\ell}(\bk)$ is the usual (layer-unresolved) form factor. This enables the projected interaction of Eq.~\eqref{eq:projected_interaction_momentum_space} to be written in a simple form:
\begin{equation}
   \hat{\mcP} \hat{V} \hat{\mcP} = \sum_{\bq\in\text{EBZ}} V(\bq) : \hat{\mcP} \hat\rho_{\bq} \hat{\mcP} \hat\rho_{-\bq} \hat{\mcP}:.
   \label{eq:projected_interaction_momentum_space_U2=U1}
\end{equation}
Moreover, since the bilayer QHFM states Eq.~\eqref{eq-BLQHFM} are eigenstates of the layer-resolved density operator [see Eq.~\eqref{eq:rhobar1_eigenstates}], then they are likewise eigenstates of the all-layer density operator:
\begin{align} \label{eq:rhobar2_eigenstates}
    \hat{\mcP}\hat{\rho}_{\bq}\hat{\mcP} \ket{\v{n}_{\mathsf{T}},\v{n}_{\mathsf{B}}} &= \bar{\rho}_{\bq} \ket{\v{n}_{\mathsf{T}},\v{n}_{\mathsf{B}}}, \qquad \bar{\rho}_{\bq} = \sum_\ell \bar{\rho}_{\bq\ell} = \frac{1}{\sqrt{qN}} \sum_{\bk\in\text{BZ}} \sum_{\bG} \delta_{\bq,\bG}\operatorname{tr}\left[\Lambda_{\bG}(\bk)\right].
\end{align}
Setting $U_2 = U_1$ and summing over $\ell$ and $\ell'$, Eq.~\eqref{eq:Hamiltonian_drho_drho_form} assumes precisely the same form, with the strong-coupling interaction term given explicitly by:
\begin{align} \label{eq:Hamiltonian_drho_drho_form2}
    \hat{\mcV} = \sum_{\bq\in\text{EBZ}} V(\bq) \delta\hat{\rho}_{\bq} \delta\hat{\rho}_{-\bq}.
\end{align}
The crucial difference compared to the $U_2 = 0$ case, however, is that this interaction admits a \textit{broader} family of ground states beyond the bilayer QHFMs. In particular, $\hat{\mcV}$ is readily found to annihilate the layer-polarized QAH states, defined as
\begin{align} \label{eq-QAHstates}
    \ket{\mathrm{QAH},l} = \prod_\bk \hat{\varphi}^\dag_{l,\uparrow}(\bk) \hat{\varphi}^\dag_{l,\downarrow}(\bk)\ket{0},
\end{align}
where $l = \mathsf{T}/\mathsf{B}$. Passing to the microscopic layer index $\ell$, which is indistinguishable from $l$ in the $g=0$ limit, explicitly acting the density operator on this state gives
\begin{align}
    \hat{\rho}_{\bq}|\mathrm{QAH},\ell\rangle &= \frac{1}{\sqrt{qN}}\sum_{l,l',s}\sum_{\bk\in\text{BZ}}\hat{\varphi}_{ls}^{\dagger}(\bk)\sum_{\ell'}[\Lambda_{\bq\ell'}(\bk)]_{ll'}\hat{\varphi}_{l's}(\bk+\bq)|\mathrm{QAH},\ell\rangle \\
    &= \frac{1}{\sqrt{qN}}\sum_{\bk\in\text{BZ}}[\Lambda_{\bq\ell}(\bk)]_{\ell\ell}\sum_{s}\hat{\varphi}_{\ell s}^{\dagger}(\bk)\hat{\varphi}_{\ell s}(\bk+\bq)|\mathrm{QAH},\ell\rangle \\
    &= \frac{2}{\sqrt{qN}}\sum_{\bk\in\text{BZ}}\sum_{\bG}\delta_{\bq,\bG}[\Lambda_{\bG\ell}(\bk)]_{\ell\ell}|\mathrm{QAH},\ell\rangle.
\end{align}
To see that this eigenvalue is precisely $\bar{\rho}_\bq$, we remark that the constraint of spacetime-inversion symmetry in our chosen periodic, layer-polarized gauge [see Eq.~\eqref{eq:I_matrix_constraint}] implies that the form factors must satisfy
\begin{align}
    \Lambda_{\bG}(\bk) = l^x \Lambda_{\bG}^{*}(\bk) l^x.
\end{align}
Since the form factors are layer-diagonal in the $g=0$ limit, as clarified in Eq.~\eqref{eq-layerdiag-ffs}, and the form factors at opposite reciprocal lattice vectors (which are both summed over) are related by $\Lambda_\bG(\bk)^\dag = \Lambda_{-\bG}(\bk)$ as per Eq.~\eqref{eq:form_factor_identities}, we then indeed have that
\begin{align}
    \delta\hat{\rho}_{\bq}|\mathrm{QAH},\ell\rangle = 0.
\end{align}
The QAH states and bilayer QHFMs are therefore both annihilated by the strong-coupling interaction Eq.~\eqref{eq:Hamiltonian_drho_drho_form2} in the $U_2=U_1$ limit. By arguments analogous to those presented in the $U_2=0$ subsection above, we reason that both of these families of states are approximate ground states of the full projected Hamiltonian.


\vspace{6mm}\toclesslab\subsection{Effects of Inter-Layer Tunneling}{}

We now restore a small amount of inter-layer tunneling $g$ and, working perturbatively within the ground state manifold of the strong-coupling interaction, find that both the LAF and QAH states are strong-coupling ground states in the $U_{2}=U_{1}$ limit. The choice of a layer-polarized basis, discussed and constructed in Sec.~\ref{subsec-layerpolarized}, is crucial to this analyis in the presence of finite $g$.
In particular, before embarking on our perturbative analysis, we establish how the action of symmetries in a layer-polarized basis constrains the structure of the form factors.

The most significant constraint comes from the anti-unitary spacetime inversion symmetry $\mathcal{I}$.
For the following analysis, it will be sufficient to work in the particular layer-polarized basis in which its action is as simple as possible. This is achieved by mapping the hybrid Wannier layer-polarized basis of Sec.~\ref{subsec-layerpolarized} to $\varphi^{\dagger}(\bk)\to\varphi^{\dagger}(\bk)e^{i\gamma(\bk)/2}$, so that Eq.~\eqref{eq-wannier-Isym-operator} becomes
\begin{align}
    \hat{\Isym} \hat{\varphi}^\dagger(\bk) \hat{\Isym}^{-1} = \hat{\varphi}^\dagger(\bk) s^x l^x.
\end{align}
Correspondingly, Eq.~\eqref{eq:I_matrix_constraint} becomes $v^\dag(\bk)\ell^x v^*(\bk) = \sigma^x$, which forgoes the exponential localization of the associated Wannier functions (or equivalently, smoothness of the gauge) while preserving the essential property of maximal layer polarization.
This forces the form factor to obey the following simple constraint:
\begin{align} \label{eq-general-ff-constraint}
    \Lambda_{\bq}(\bk) = l^{x}\Lambda_{\bq}^{*}(\bk)l^{x}.
\end{align}
Expanding the form factor as a linear combination of Pauli matrices $\Lambda_{\bq}(\bk)=\sum_{\mu=0,x,y,z}\Lambda_{\bq}^{\mu}(\bk)l^{\mu}$, then the above equation is equivalent to
\begin{align}
    \Lambda_{\bq}^{0}(\bk) = \Lambda_{\bq}^{0}(\bk)^{*},\quad 
    \Lambda_{\bq}^{x}(\bk) = \Lambda_{\bq}^{x}(\bk)^{*},\quad 
    \Lambda_{\bq}^{y}(\bk) = \Lambda_{\bq}^{y}(\bk)^{*},\quad 
    \Lambda_{\bq}^{z}(\bk) = -\Lambda_{\bq}^{z}(\bk)^{*}.
\end{align}
The form factor may therefore be written as the following sum of diagonal and off-diagonal terms, with $F_{\bq}^{S/A}(\bk),\Phi_{\bq}^{S/A}(\bk)\in\mathbb{R}:$
\begin{align} \label{eq-ff-decomp}
    \Lambda_{\bq}(\bk) = F_{\bq}^{S}(\bk)e^{i\Phi_{\bq}^{S}(\bk)l^{z}}+F_{\bq}^{A}(\bk)l^{x}e^{i\Phi_{\bq}^{A}(\bk)l^{z}} = \Lambda^S_{\bq}(\bk) + \Lambda^A_{\bq}(\bk),
\end{align}
which commute and anti-commute with $l^z$, respectively. 
This expression for the form factor may be verified, by explicit calculation, to satisfy the constraint Eq.~\eqref{eq-general-ff-constraint} above:
\begin{align}
    l^{x}\Lambda_{\bq}^{*}(\bk)l^{x} & = F_{\bq}^{S}(\bk)l^{x}e^{-i\Phi_{\bq}^{S}(\bk)l^{z}}l^{x}+F_{\bq}^{A}(\bk)l^{x}l^{x}e^{-i\Phi_{\bq}^{A}(\bk)l^{z}}l^{x} = \Lambda_{\bq}(\bk).
\end{align}
Taking the limit $g=0$ further constrains the form factors by restoring layer $U(1)$ symmetry, making them diagonal $(F^{A}=0)$. Otherwise, $F^{A}=\mathcal{O}(g)$ is of the same order as the inter-layer tunneling and must be treated with some care. We further remark the unitary discrete symmetries relate the form factors at symmetry-related momenta, which will not be relevant to our analysis.

These modifications to the form factors by the inter-layer tunneling --- namely the acquisition of off-diagonal components --- imprint themselves on both the strong-coupling Hamiltonian and its approximate ground states. This is fundamentally due to the fact finite $g$ breaks layer $U(1)$ symmetry, thereby distinguishing the microscopic layer index $\ell$ from the dressed layer index $l$ of our layer-polarized basis. The decomposition of the form factor in Eq.~\eqref{eq-ff-decomp} yields a similar partitioning of both the projected density and the associated reference charge densities:
\begin{align}
    \hat\rho^{S/A}_{\bq} = \frac{1}{\sqrt{qN}} \sum_{l,l',s} \sum_{\bk} \hat{\varphi}_{ls}^\dagger(\bk) [\Lambda^{S/A}_{\bq}(\bk)]_{ll'} \hat{\varphi}_{l's}(\bk+\bq),
    \qquad
    \bar{\rho}^{S/A}_{\bq} = \frac{1}{\sqrt{qN}} \sum_{\bk\in\text{BZ}} \sum_{\bG} \delta_{\bq,\bG}\operatorname{tr}\left[\Lambda^{S/A}_{\bG}(\bk)\right].
\end{align}
In particular, since $\Lambda^A$ is purely off-diagonal, then $\bar{\rho}^A$ vanishes identically.
In terms of the relative densities $\delta\hat\rho^{S/A}_{\bq} = \hat\rho^{S/A}_{\bq} - \bar\rho^{S/A}_{\bq}$, the strong-coupling interaction Eq.~\eqref{eq:Hamiltonian_drho_drho_form2} therefore reads
\begin{align}
    \hat{\mcV} = \hat{\mcV}^S + \hat{\mcV}^A = \sum_{\bq\in\text{EBZ}} V(\bq) \delta\hat{\rho}^S_{\bq} \delta\hat{\rho}^S_{-\bq} + \sum_{\bq\in\text{EBZ}} V(\bq) \left( \delta\hat{\rho}^A_{\bq} \delta\hat{\rho}^S_{-\bq} + \delta\hat{\rho}^S_{\bq} \delta\hat{\rho}^A_{-\bq} \right),
\end{align}
where $\hat\mcV^A = \mcO(g)$ since it contains one factor $\Lambda^A$, and 
we dropped the $AA$ term since it is $\mcO(g^2)$.

The bilayer QHFMs Eq.~\eqref{eq-BLQHFM} and QAH states Eq.~\eqref{eq-QAHstates} are annihilated by $\delta\hat\rho^{S}_{\bq}$ but not by $\delta\hat\rho^{A}_{\bq}$. This motivates performing degenerate perturbation theory in the ground state manifold of $\hat\mcV^S$, which consists of precisely these states.
The perturbations include not only the anti-symmetric term $\hat\mcV^A$, but also certain terms in the single-particle Hamiltonian $\delta \hat{h}$, which includes corrections from the re-organization of the interaction, analogous to Eq.~\eqref{eq:Hamiltonian_drho_drho_form}. As in that case, the $l^0$ component of $\delta h$ is small relative to the interaction scale and moreover acts diagonally in (and does not energetically distinguish) the states in the degenerate manifold. This leaves the off-diagonal components of the total single-particle Hamiltonian, which we collect altogether:
\begin{align}
    \delta \hat{h}_g = \sum_\bk \hat\varphi^\dag_\bk \Big[ h_g^x(\bk) l^x + h_g^y(\bk) l^y \Big] \hat\varphi_\bk = \mcO(g).
\end{align}
Since both this term and $\delta\hat\rho^{A}_{\bq}$ change the relative dressed-layer occupation of the degenerate ground states, they therefore do not modify their energies at first order. Nonetheless, they do facilitate \textit{second-order} virtual processes in which fermions hop between the two dressed layers and hop back.
Since an $\mcO(U)$ energetic penalty is incurred in the virtual state, assuming it is not Pauli blocked, then some of the degenerate states will enjoy an $\mcO(g^2/U)$ energetic benefit and therefore be singled out as the preferred ground states in the presence of finite $g$.
This intuition can be made more quantitatively precise by computing second-order perturbative energetic corrections using the Schrieffer-Wolff formalism~\cite{coleman2015introduction}, analogous to that detailed in Refs.~\cite{bultinck2020ground, pedagogical_patrick} in the context of twisted bilayer graphene. In particular, both ``layer-antiferromagnetic'' (LAF) and QAH states are energetically favorable over bilayer ferromagnetic states. 
As in the main text, LAF states are defined as $\ket{\bm{n},-\bm{n}}$, the subset of the states in Eq.~\eqref{eq-BLQHFM} that have opposite spin polarization in opposite Chern sectors.

\begin{figure}
    \centering
    \includegraphics[width=\textwidth]{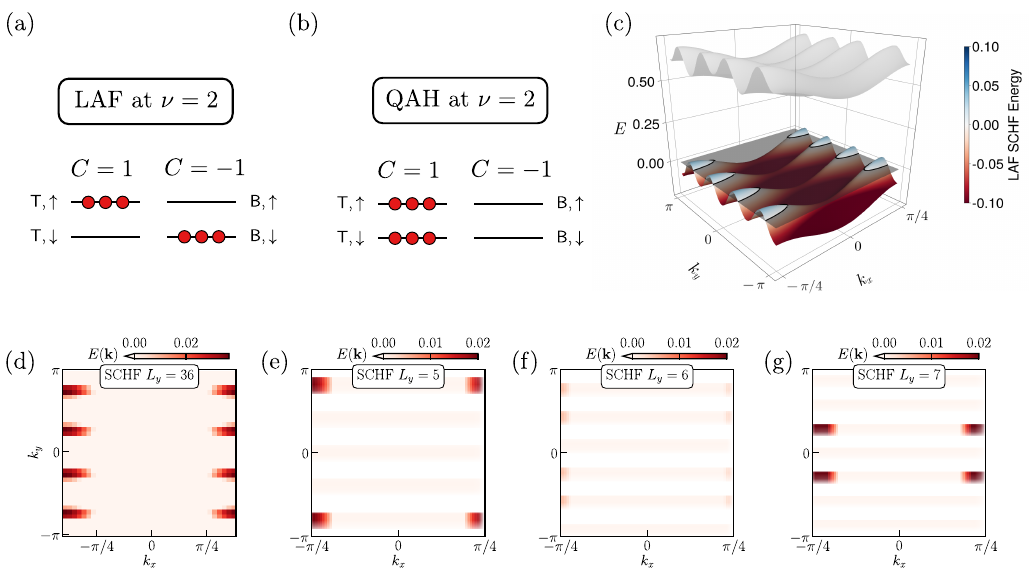}
    \caption{\textbf{Strong Coupling Ground States}. (a) One possible LAF state that fills the $(\mathsf{T},\uparrow)$ and $(\mathsf{B},\downarrow)$ orbitals for each $\bk$. (b) One possible quantized anomalous Hall state filling both $\mathsf{T}$ orbitals, with a total Chern number of $C=2$. (c) The bandstructure of the LAF state within self-consistent Hartree Fock at filling $\nu=1.8$ and $(g,U_1,U_2) = (0.15,1.5,0.0)$, matching the large-bond dimension superconductivity data. The horizontal plane at $E=0$ is the chemical potential, and the top gray band is entirely unoccupied. Each band is doubly degenerate. Computed on $36\times 36$ $\bk$-points and linearly interpolated for display.
    (d) The lower band of the LAF band structure (c) without interpolation and only showing data above the Fermi level. (e,f,g) Same as (d) except with $L_y=5,6,7$ and $\nu=2-1/L_y$, yielding Fermi wavevectors $k_F^{L_y=5} = 0.59/a,\,  k_F^{L_y=6} = 0.68/a,\,  k_F^{L_y=7} = 0.59/a$.
    }
    \label{fig:app_SCHF_LAF_bandstructure}
\end{figure}







\vspace{8mm}\toclesslab\section{Additional Numerical Data for Phase Diagram at $\nu = 2$}{app:NumericalDataUndoped}




    


\begin{figure}
    \centering
    \includegraphics[width=\textwidth]{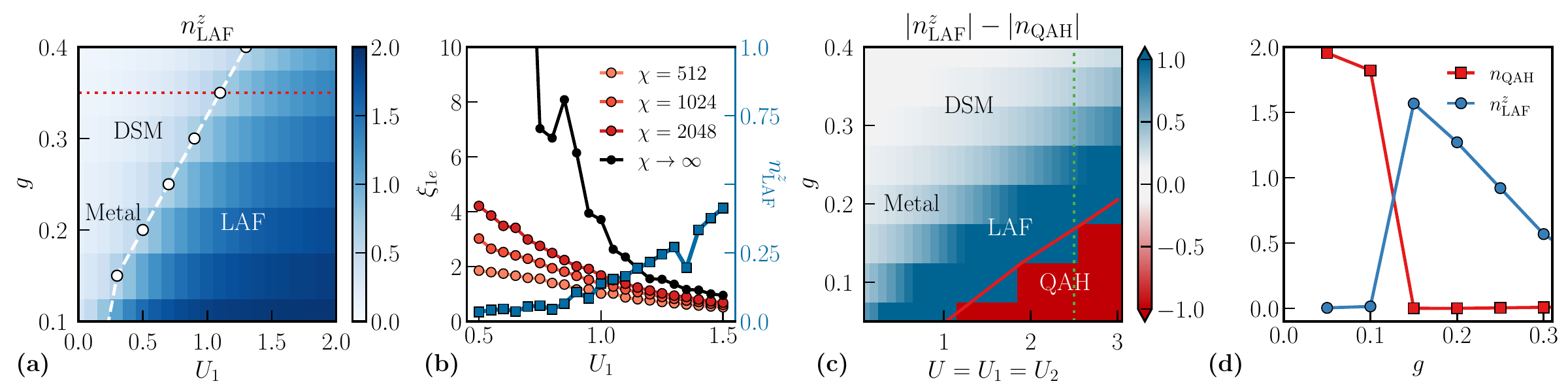}
    \caption{\textbf{Additional Details for the Undoped Phase Diagram at $\nu=2$}.
    (a)~LAF polarization as a function of $g$ and $U_1$ at $U_2=0$, $B_z = 10^{-2}$.
    (b)~DSM-to-LAF phase transition at $g=0.25$ [dotted red line in (a)]. The electron-electron correlation length $\xi_{1e}$ is extrapolated to $\chi=\infty$ as described in the text. The LAF polarization (right axis) becomes significant near $U_1 = 1$, matching the divergence in $\xi_{1e}$. Here we take a smaller value of the Zeeman pinning field, namely $B_z = 10^{-4}$, to sharpen the transition.
    (c)~Phase diagram as a function of $g$ and $U = U_1 = U_2$ with $B_z = 10^{-2}$.
    (d)~QAH-to-LAF transition as a function of $g$ at fixed $U_1 = U_2 = 2.5$ with $B_z= 10^{-2}$ [dotted green line in (c)].
    }
    \label{fig:app_undoped_phase_detail}
\end{figure}

In this Appendix we give additional details of the DMRG phase identification at the ``undoped'' filling of $\nu=2$, which corresponds to filling two of the four low-lying narrow bands of the double Hofstadter model.
Fig.~\ref{fig:app_undoped_phase_detail}(a) recapitulates the phase diagram shown in the main text. At weak interactions, the half-filled narrow bands form Dirac semimetal (DSM) at large $g$ and a metal at lower $g$, characterized by small electron/hole pockets.
Meanwhile at high $U_1$, the LAF polarization
\begin{equation}
    n^z_\mathrm{LAF} = \sum_{\bk} \langle \hat{\varphi}_{\bk}^\dagger \ell^z s^z \hat{\varphi}_{\bk}\rangle
\end{equation}
increase steadily towards 2, its maximal possible value, suggesting the state is the LAF insulator predicted by strong-coupling theory.
To verify this, Fig.~\ref{fig:app_undoped_phase_detail}(b) examines the phase diagram as a function of $U_1$ at fixed $g$. Since the LAF is strongly gapped, one expects electron-electron correlations to behave as
\begin{equation}
    \langle \hat{\varphi}_{x, k_y}^\dagger \hat{\varphi}_{0,k_y}\rangle_{\mathrm{LAF}} \sim e^{-x/\xi_{1e}(k_y)},
\end{equation}
where the correlation length $\xi_{1e}(k_y)$ should be finite.
By contrast, in the gapless metallic phase, one expects power-law correlations of $\langle \hat{\varphi}_{x, k_y}^\dagger \hat{\varphi}_{0,k_y}\rangle$, leading to diverging $\xi_{1e} \to \infty$ as $\chi \to \infty$. However, even in a gapped phase, $\xi_{1e}$ may increase slowly with $\chi$ before converging to a finite value. Therefore, further analysis is needed to determine if its extrapolated value at infinite bond dimension is finite or likely infinite.

To this end, we employ a recently-developed method for extrapolating correlation lengths in DMRG \cite{rams2018precise,vanhecke2019scaling}.
Namely, we compute several of the largest eigenvalues of the transfer matrix $T X_i = \lambda_i X_i$ in the electric charge $Q_E = 1$ sector for each $k_y$, and select those from the same ``excitation branch'' \cite{rams2018precise}.
These are related to the correlation lengths as $\epsilon_i = 1/\xi_i = -\log \n{\lambda_i}$ with $\epsilon_i \le \epsilon_{i+1}$, where we report $\xi$ in units of the unit cell length $u=4a$ (i.e., $q=4$ square lattice sites). Correspondingly, $\epsilon$ is reported in units of $u^{-1}$.
For a gapless system with a power-law correlation in charge sector $Q$, both $\epsilon_1^Q, \epsilon_2^Q \to 0$ as $\chi\to \infty$. As a result, the quantity
\begin{equation}
    \delta^Q = \epsilon_2^Q - \epsilon_1^Q
    \label{eq:scaling_delta}
\end{equation}
also vanishes at $\chi\to \infty$, serving as a good ``scaling variable'' that measures the deviation from the gapless state \cite{rams2018precise,vanhecke2019scaling}.
To estimate $\xi(\chi \to \infty)$, we therefore compute $\epsilon(\delta)$ at each available $\epsilon(\chi),\delta(\chi)$ and extrapolate to $\delta = 0$ with a linear fit. 
This extrapolation is shown in Fig.~\ref{fig:app_undoped_phase_detail}(b) for $\xi_{1e}$ using numerical data at bond dimensions $\chi=512,1024,2048$.
At $U_1 \lesssim 1$, the extrapolated correlation length $\xi_{1e}(\chi \to \infty)$ apparently diverges---in concord with a metallic phase---while at larger $U_1$ the extrapolated value is finite and on the scale of a few unit cells. As a technical point, we note that taking a finite $B_z$ broadens the transition region.
To sharpen the transition (while still avoiding Hohenberg-Mermin-Wagner issues), we take $B_z = 10^{-4}$ in panel (b).

Fig.~\ref{fig:app_undoped_phase_detail}(c,d) show the competition between the LAF and QAH phases when $U_1 = U_2$. At these parameters, the two phases are degenerate within strong coupling theory. Surprisingly, they are also exactly degenerate within self-consistent Hartree-Fock. DMRG, however, shows a clear transition near $g=0.13$ at $U = 2.5$, with simultaneous and opposite jumps in $n^z_{\mathrm{LAF}}$ and the QAH polarization, namely
\begin{equation}
    n_{\mathrm{QAH}} = \sum_{\bk} \langle \hat{\varphi}_{\bk}^\dagger \ell^z s^0 \hat{\varphi}_{\bk}\rangle.
\end{equation}
The transition therefore appears to be first order.

\vspace{8mm}\toclesslab\section{Additional Numerical Data for the Existence and Robustness of Superconductor}{app:NumericalDataExistenceandRobustness}

In the main text, we provided conclusive numerical evidence for the existence of superconductivity at $L_y = 5$ by performing a large bond dimension scaling analysis, and summarized the extent of evidence for this phase at higher circumferences $L_y = 6, 7$.
We subsequently demonstrated that the superconductivity at $L_y = 5$ belongs to a robust superconducting phase that persists even upon including additional repulsive interactions.
In this appendix, we provide additional numerical data and details supporting both of these claims.

In Subsection~\ref{appsubsec-ScalingAnalysis}, we provide a detailed account of our scaling analysis of the charge-$2e$ and charge-$1e$ correlations as a function of bond dimension and cylinder circumference.
This includes a depiction of the superconducting correlation functions for each channel, a description of our procedure for obtaining a scaling collapse and the exponents of algebraic decay, a depiction of the $1e$ and $2e$ correlation lengths as a function of bond dimension, and our procedure for extrapolating the correlation lengths to infinite bond dimension.
Crucially, we provide data for both the case of zero and non-zero interlayer repulsion $U_2$.
Subsequently, in Subsection~\ref{appsubsec-Robustness}, we provide the raw numerical data supporting the $L_y=5$ superconducting phase diagrams shown in Fig.~5 of the main text.


\vspace{6mm}\toclesslab\subsection{Scaling Analysis of Correlations}{appsubsec-ScalingAnalysis}{}

\begin{figure}
    \centering
    \includegraphics[width = 0.95\textwidth]{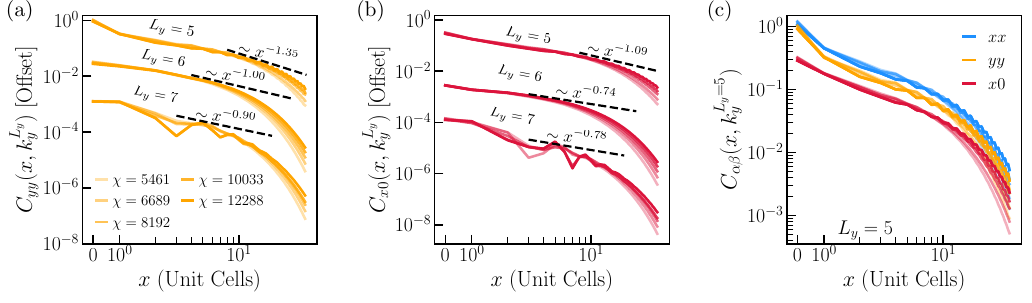}
    \caption{\textbf{Additional Superconducting Correlations Functions.} In panels (a) and (b), we depict the superconducting correlation function for the $yy$ and $x0$ channels, respectively, at circumferences $L_y = 5, 6, 7$. Similar to the the $xx$ channel shown in the main text, clear signatures of algebraic decay are observed in these two channels. Likewise, an arbitrary vertical offset is added to $L_y = 6,7$ for ease of viewing. In panel (c), the correlation functions at $L_y = 5$ for all three channels are shown without any vertical offset, allowing for the direct comparison of their magnitudes. While $xx$ and $yy$ are relatively close, especially at short distances, the $x0$ correlation function is notably smaller, a pattern that holds at all circumferences.}   \label{fig:supp_correlation_lengths}
\end{figure}

In the main text, we claimed that we saw algebraic decay of the superconducting correlation function
\begin{equation} \label{eq-appcorrfunc}
    C_{\alpha\beta}(x, k_y) = N_{2h}^{-1} \langle \hat\Delta_{\alpha\beta}^{\dagger}(x; 0, k_y) \hat\Delta_{\alpha\beta}(0; 0, k_y) \rangle
\end{equation}
in the channels $\alpha \beta = xx, yy, \text{ and } x0$.
There, in Fig.~4(a), we showed the superconducting correlation functions in the $xx$ channel.
In Fig.~\ref{fig:supp_correlation_lengths}(a, b), we show that the $yy$ and $x0$ likewise exhibit clear signatures of algebraic decay, as claimed in the main text.
%
%
In Fig.~\ref{fig:supp_correlation_lengths}(c), we compare the magnitudes of the correlation functions for the three different channels at $L_y = 5$.
Since $C_{\alpha\beta}$ is normalized by the density of total hole pairs $N_{2h}$, then its value at $x=0$ roughly corresponds to the average \textit{fraction} of Cooper pairs occupying the $\alpha \beta$ channel.
%
With this in mind, at $x=0$ we find that the $xx$ and $yy$ channels have nearly equal magnitude, while $x0$ is roughly three times smaller, signaling a lower number density of Cooper pairs in the latter channel.
This justifies our treatment of the $x0$ channel as subdominant.

\subsubsection{Collapse to Algebraic Scaling Hypothesis}

\begin{figure}
    \centering
    \includegraphics[width = 0.7\textwidth]{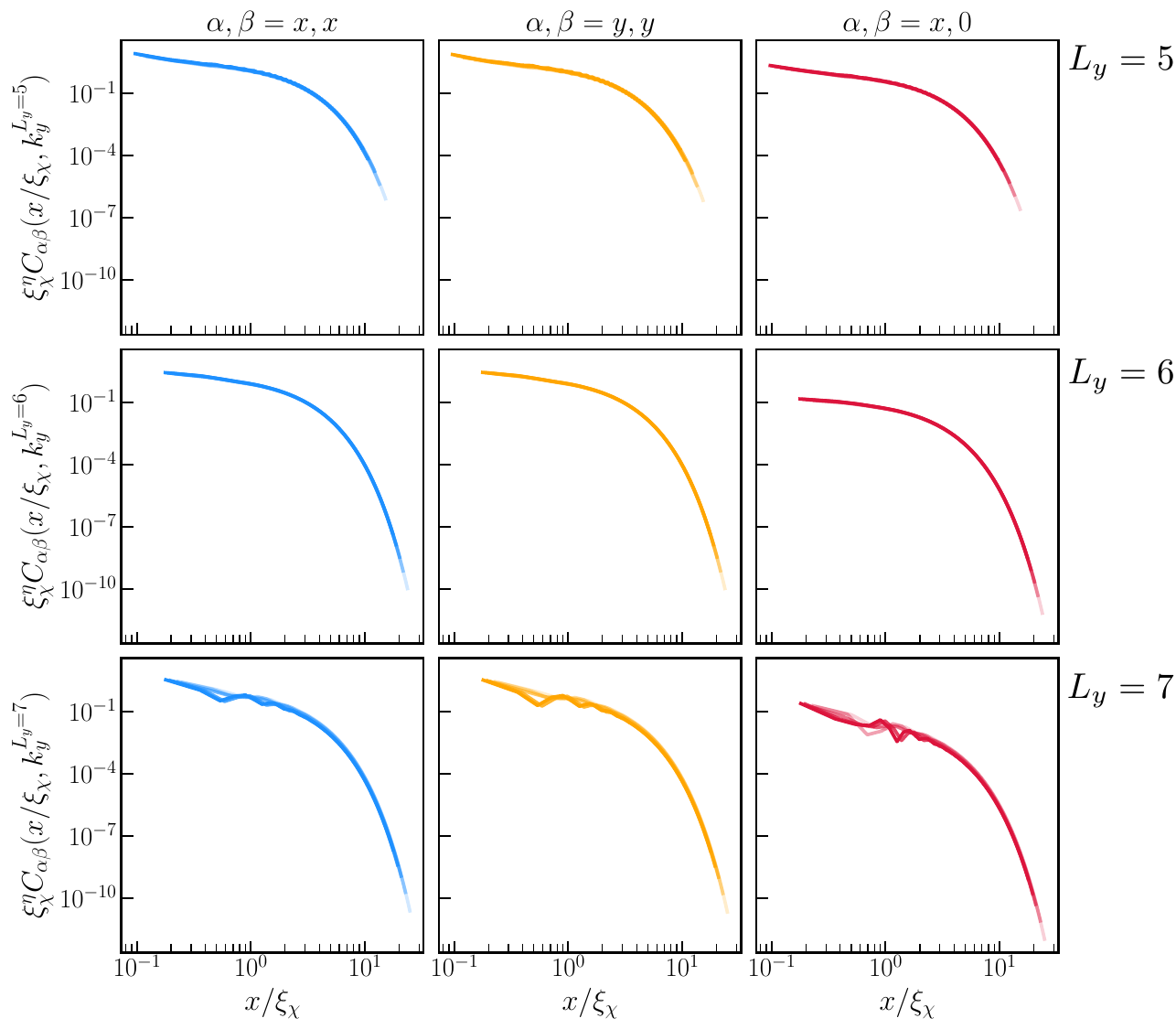}
    \caption{\textbf{Additional Scaling Collapses for Superconducting Correlation Functions.} Here, we depict the results of the finite entanglement scaling collapse for each circumference ($L_y = 5, 6, 7$ for the top, middle, and bottom row) and condensed channel ($x,x$, $y, y$, and $x, 0$; labeling the columns). We find an excellent scaling collapse for each correlation function shown and the exponents $\eta$ are shown in Fig.~4(c) of the main text. Data is shown for $\chi = 5461,6689,8192,10033, 12288$.}
    \label{fig:supp_scaling_collapses}
\end{figure}

In the main text, we performed a finite-entanglement scaling collapse to quantitatively demonstrate the approach to algebraic decay.
To this end, we adopted the scaling ansatz:
\begin{equation}
\label{eq:SC_scaling_collapse}
    C_{\alpha \beta}^{(\chi)}(x, k_y) = \xi^{-\eta} g_{\alpha \beta}(x/\xi_{2e, \chi}, k_y).
\end{equation}
We now detail our precise quantitative procedure for performing this collapse and provide additional numerical data in support of our claims.
In our numerics, the data for the superconducting correlation function is organized in the form of tuples $\{(x_i, C^{\chi}(x_i, k_y)\}$ for each simulated bond dimension $\chi$ and each position $x_i$.
To perform the finite entanglement scaling collapse, the $x$-axis is first rescaled by the correlation length, yielding a data set $\{(x_i/\xi_{\chi}, C^{(\chi)}(x_i/\chi, k_y)\}$.
Linearly interpolating this data, we obtain a continuous function $\widetilde{C}^{(\chi)}(x/\xi_{\chi}, k_y)$, which we can then scale and sample from.
Namely, given this interpolation and Eq.~\eqref{eq:SC_scaling_collapse}, we posit the following cost function:
\begin{equation}
    \mathcal{Q}(\eta) =  \frac{1}{N_{\chi} (N_{\chi} - 1)}\sum_{\chi \neq \chi'} \int_{y_0}^{y_1} dy\ \frac{1}{\xi_{\chi}^{\eta}\xi_{\chi'}^{\eta}}\left|\widetilde{C}^{(\chi)}(y, k_y) \xi_{\chi}^{\eta} - \widetilde{C}^{(\chi')}(y, k_y) \xi_{\chi'}^{\eta}\right|^2.
\end{equation}
It is chosen based on the following principles: (\textit{i}) it is positive semi-definite, (\textit{ii}) it vanishes if and only if Eq.~\eqref{eq:SC_scaling_collapse} holds, and (\textit{iii}) its prefactor $(\xi_{\chi}^{\eta}\xi_{\chi'}^{\eta})^{-1}$ makes the integrand dimensionless. More heuristically, it is more stable (compared to several considered alternatives) to the removal of some data from the fit, which suggests that it is more robust.
A finite entanglement scaling collapse is achieved for the value of $\eta$ that minimizes the cost function shown above, while the error in the optimal value $\eta = \eta^{*}$ is estimated from the inverse Hessian as
\begin{equation}
    \delta \eta = \sqrt{Q(\eta^*)\left| \frac{\partial^2 Q}{\partial \eta^2} \right|_{\eta = \eta^{*}}^{-1}}.
\end{equation}
We remark that $y = x/\xi_{\chi}, y_0$ and $y_1$ are chosen to isolate the algebraic ``shelf'' of the correlation function --- clearly visible in Fig.~\ref{fig:supp_correlation_lengths} as the region over which multiple curves coincide with the algebraic fit --- and to ensure that the function $Q(\eta)$ is convex at the obtained minimum.
We remark that, when performing the optimization procedure for $L_y = 6, 7$, we discarded certain low bond dimension correlation functions, as including them inhibited the discovery of a convex solution.
%
This suggests that these bond dimensions were not yet in the `scaling regime' at $\chi \gtrsim e^{O(L_y)}$ and could therefore not be collapsed appropriately under the scaling hypothesis.
In particular, for $L_y = 6$, the bond dimensions taken were $\chi \geq 8192$ and for $L_y = 7$, only the last two bond dimensions could be used $\chi \geq 10033$.
%
%
With these choices of $\chi$, we have ensured that this optimization problem is convex, and we have further ensured that our method is robust to changing the values of $y_0, y_1$.
%
Fig.~\ref{fig:supp_scaling_collapses} shows the scaling collapses for all channels. The collapses have generally high quality, both in terms of $Q(\eta)$ and to the eye. This constitutes strong evidence that superconducting correlations are gapless in all three channels $xx,yy,x0$ at all systems sizes $L_y =5,6,7$ studied.
The extracted exponents are also tabulated in Table~\ref{tb:exp_no_repulsion}.

\begin{table}
\begin{center}
\begin{tabular}{||c c c c||} 
 \hline
  $L_y \backslash \alpha \beta$ & $xx$ & $yy$ & $x0$\\ [0.5ex] 
 \hline\hline
 5 & $1.3 \pm 0.2$ & $1.4 \pm 0.3$ & $1.1 \pm 0.2$ \\ 
 \hline
 6 & $1.02 \pm 0.04$ & $1.00 \pm 0.07$ & $0.74 \pm 0.07$ \\
 \hline
 7 & $0.9 \pm 0.3$ & $0.9 \pm 0.3$ & $0.7 \pm 0.4$ \\ [1ex] 
 \hline
\end{tabular}
\end{center}
\caption{\textbf{Extracted Power Law Exponents from Scaling Collapse.} The table shows the exponents $\eta(L_y)$ extracted for each channel $\alpha \beta$ by collapsing under the finite entanglement scaling hypothesis of Eq.~\eqref{eq:SC_scaling_collapse}. The exponents here were obtained using the data at $\nu = 2 - 1/L_y$ and $(g, U_1, U_2, d) = (0.15, 1.8, 0, 0)$.}
\label{tb:exp_no_repulsion}
\end{table}

%

\subsubsection{Correlation Lengths and Extrapolation} \label{sec-zerodU2-xis}

\begin{figure}
    \centering
    \includegraphics[scale=\otherfigscale]{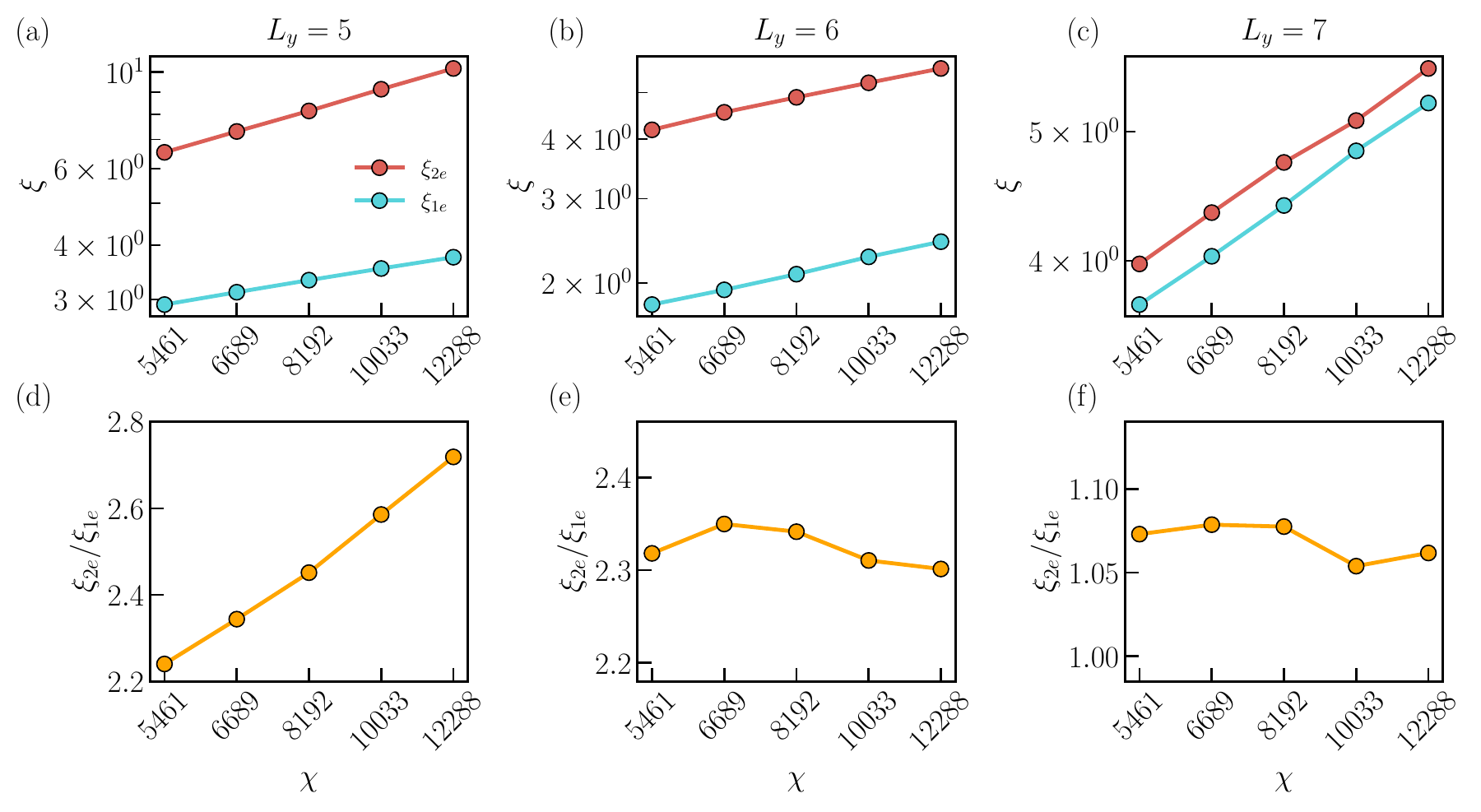}
    \caption{\textbf{Bond Dimension Scaling of Correlation Lengths.} In panels (a, b, c), we depict the $2e$ and $1e$ correlation lengths as a function of bond dimension at a hole doping $\nu = 2 - 1/L_y$ for the parameter point $(g, U_1, U_2, d) = (0.15, 1.8, 0, 0)$ and circumferences $L_y = 5, 6, 7$. We find that for all circumferences, both correlation lengths grow linearly with $\chi$. For $L_y = 5, 6$ [panels (a, b)], the $2e$ correlation lengths are substantially larger in magnitude suggesting non-trivial pair correlations in these states, while for $L_y = 7$, the difference in magnitude is not substantial. In panels (d, e, f), we check the ratio of these correlation lengths, which distinguishes a Luttinger Liquid from a Luther-Emery liquid based on the theory of finite entanglement scaling (See Section V.A of the main text). We find that for $L_y = 5$ [panel (d)], the ratio of the correlation lengths increases as a function of bond dimension strongly suggesting a Luther-Emery liquid. For $L_y = 6, 7$, this ratio does not increase suggesting that we cannot rule out the possibility of a Luttinger liquid at these circumferences.  
    }
    \label{fig:SC_xi_scaling_zero_dU2}
\end{figure}

\begin{figure}
    \centering
    \includegraphics[scale=\otherfigscale]{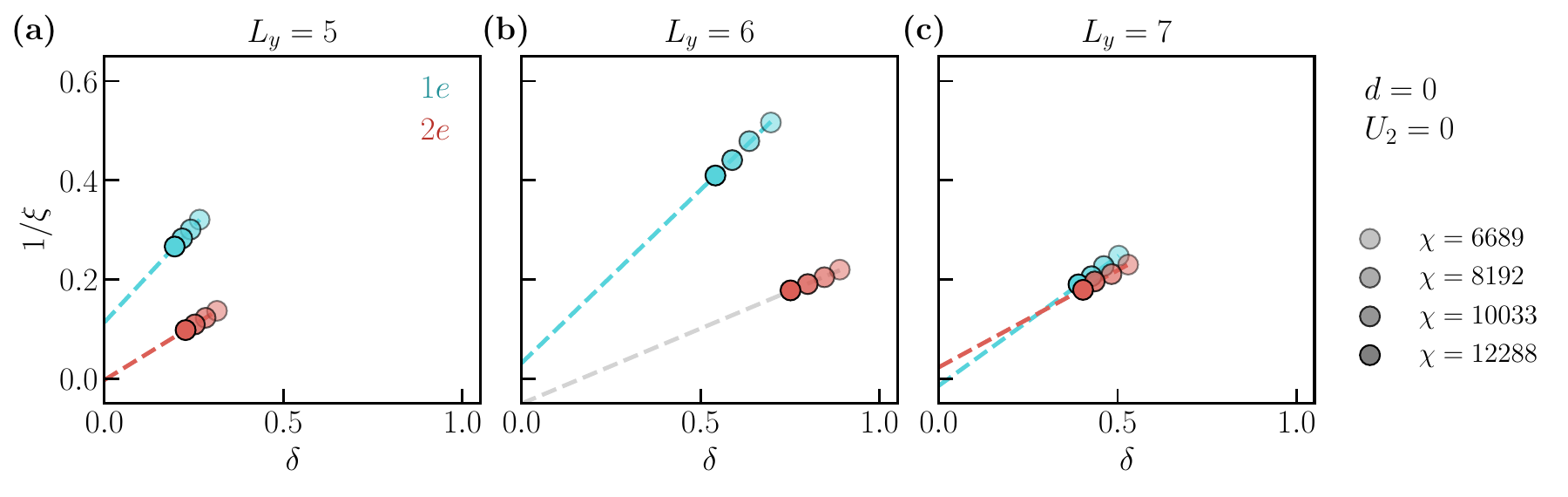}
    \caption{\textbf{Extrapolation of Inverse Correlation Lengths.}
    Attempted extrapolation of the inverse correlation lengths $1/\xi_Q$ in the $Q = 1e$ and $2e$ charge sectors. (a) At circumference $L_y=5$, we find that $\xi_{1e}$ extrapolates to a finite value while $\xi_{2e}\to\infty,$ the hallmark of a Luther-Emery liquid.
    (b) Though we find that $\xi_{2e}$ substantially exceeds $\xi_{1e}$ at $L_y=6$, the former extrapolates to a negative value (therefore marked in gray), signaling the state is too far from convergence to confidently identify it as a Luther-Emery liquid or Luttinger liquid.
    (c) At $L_y=7$, we find that the $1e$ and $2e$ excitations extrapolate close to zero with the former slightly negative and the latter appearing gapped. While these observations could arise from a Luttinger liquid, the changing behavior of the correlation functions (Fig.~\ref{fig:supp_correlation_lengths} and Fig.~4(a) in the main text) as a function of bond dimension (e.g., short-distance behavior changing non-monotonically in $\chi$) suggest that this state is also too far from convergence to derive any sharp conclusions.
    }
    \label{fig:SC_xi_extrapolation_zero_dU2}
\end{figure}

We now study the correlation lengths as a function of bond dimension at each cylinder circumference. By performing a quantitative extrapolation in bond dimension, we attempt to ascertain if the system is gapped or gapless in each sector. It is worthwhile to review the possible options. If the system is a Luttinger liquid, then one expects gapless excitations in both the one-electron ($1e$) and two-electron ($2e$) sectors, whose corresponding correlation lengths $\xi$ would diverge as $\chi \to \infty$. Equivalently, $1/\xi_{1e}$ and $1/\xi_{2e}$ would vanish. By contrast, for a Luther-Emery (i.e. a 1D superconductor), there should be a gap to single-electron excitations, so $1/\xi_{1e}$ would remain positive as $\chi \to \infty$. 

Clearly it is impossible to actually take this limit, so we are obliged to extrapolate in $\chi$. Such a procedure is always fraught, due to the presence of finite $\chi$ phase transitions where the properties of the state change dramatically as a function of bond dimension. And indeed we often observe such a transition at $L_y=5$ from a LAF-metal state at low $\chi$ to a Luther-Emery liquid at high $\chi$. If one \textit{assumes} that no such transition will occur beyond available bond dimensions, and further assumes that the available bond dimension is ``close'' to $\chi \to \infty$, then one may apply the scaling techniques pioneered in \cite{rams2018precise, vanhecke2019scaling}, and described in App. \ref{app:NumericalDataUndoped} above. We recall that describing a gapped state generically requires bond dimension $\chi \sim e^{O(L_y)}$, so describing a gapless state likely requires bond dimensions growing at least this fast. This suggests that studying extrapolation at fixed maximal bond dimension --- a constraint imposed upon us by limited numerical resources --- grows less reliable as $L_y$ increases.

Nevertheless, we extrapolate using all available data in Fig.~\ref{fig:SC_xi_extrapolation_zero_dU2}, where we report $\xi$ in units of the size of the unit cell, namely $u=4a$, with $\delta$ accordingly being reported in units of $u^{-1}$. The data is clearest at $L_y=5$. There, both $\xi_{1e}$ and $\xi_{2e}$ are increasing with $\chi$. Extrapolating the inverse correlation lengths versus $\delta$ [see Eq.~\eqref{eq:scaling_delta}] by a linear fit implies that $1/\xi_{2e}$ is consistent with zero in the $\delta\to 0$ limit, whereas $1/\xi_{1e}$ is strictly positive. Taken in conjunction with the clear power law behavior for superconducting correlations discussed above, we thus identify the $L_y=5$ state as a Luther-Emery liquid. For $L_y=6$, the extrapolated $1/\xi_{2e}$ inverse is negative --- a non-physical result since $\xi_{2e} > 0$. This fitting difficulty suggests the correlations are still changing non-linearly and may not be in the ``almost infinite $\chi$" regime required by scaling theory. So although the extrapolated value of $1/\xi_{1e}$ is slightly positive, consistent with an electronic gap for a Luther-Emery liquid, we cannot confidently rule out a Luttinger liquid at $L_y=6$. Finally at $L_y=7$ the extrapolated values of both inverse correlation lengths are close to zero (with $1e$ slightly negative). This behavior is consistent with a Luttinger liquid.

Let us make a few observations on the distance to the ``scaling regime" at asymptotically large $\chi$. It is instructive to consider how changing $L_y$ changes the `normal state' --- the non-superconducting metal obtained by doping the LAF bandstructure. This is shown in Fig.~\ref{fig:app_SCHF_LAF_bandstructure}(d-g). There one can see that at asymptotically large $L_y$ there are four Fermi pockets (wrapping around the right to left edges of the Brillouin zone). At $L_y=5,6,7$ the wires intersect only some of these cuts, resulting in $2,4$ and $2$ Fermi pockets, respectively. It is striking that the minimal $\delta$ at fixed bond dimension is roughly proportional to the number of Fermi pockets, in accord with the idea that it measures the ``distance to infinite bond dimension". Moreover, these Fermi pockets have slightly different Fermi momenta, namely $k_F^{L_y=5} = 0.15 u^{-1}, k_F^{L_y=6} = 0.17 u^{-1}, k_F^{L_y=7} = 0.15 u^{-1}$ (see Fig.~\ref{fig:app_SCHF_LAF_bandstructure}), which provide a natural inverse length scale against which to evaluate the smallness of $\delta$. In particular, we find that $\delta$ greatly exceeds $k_F^{L_y}$ for each of the larger circumferences. Taken with the arguments above, this suggests that $L_y=6,7$ remain far from convergence and that vastly larger bond dimensions may be required to resolve the nature of their ground states definitively.


\vspace{6mm}\toclesslab\subsection{Scaling Analysis of Correlations With Additional Repulsion}{}

In the main text and above, we detailed our scaling analysis of the charge-$2e$ and charge-$1e$ correlations as a function of bond dimension and cylinder circumference at the fixed parameters $(g,U_1) = (0.15,1.5)$. In particular, this quantitative analysis was performed in the absence of additional repulsion $d,U_2$.
To demonstrate that the superconductor that appears is not a consequence of a vanishingly small attractive interaction generated between the layers, we now examine the effect of adding additional long-range interactions in inter-layer channels:
\begin{equation}
	\hat{V} = \sum_{\br} e^{-\frac{\n{\br-\br'}^2}{2d^2}} :\left[
		\frac{U_1}{2} (\hat n_{\br\Top}^2 + \hat n_{\br\Bot}^2)
		+ U_2 \hat n_{\br\Top} \hat n_{\br\Bot}
	\right]:
\end{equation}
where $U_2$ controls the strength of interlayer repulsion and $d$ controls their range.
%
We perform the same scaling analysis as before at parameters $(g, U_1, U_2, d) = (0.1, 1.4, 0.2, 0.358)$. This point is marked with a star in Fig.~5(c) of the main text and crucially has inter-layer repulsion.

\begin{figure}
    \centering
    \includegraphics[width = \textwidth]{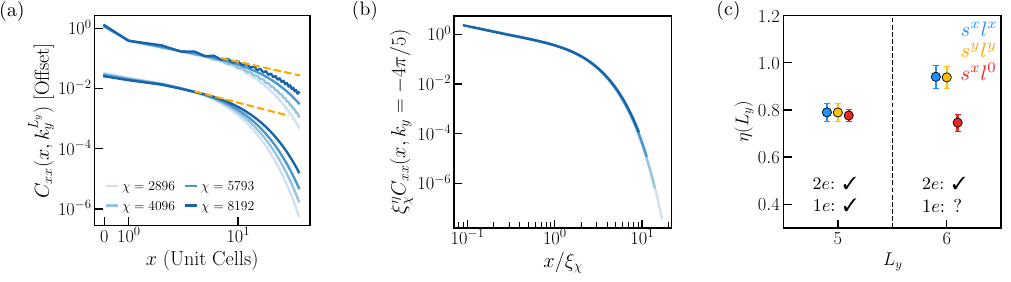}
    \caption{\textbf{Superconducting Correlations at Finite Repulsion.} We perform a scaling analysis on the superconducting correlations in the ground state at $(g, U_1, U_2, d) = (0.1, 1.4, 0.2, 0.358)$ and doping $\nu = 2 - 1/L_y$. In panel (a), we show the correlation functions for the $xx$ channel at circumferences $L_y = 5, 6$, finding a clear tendency to algebraic decay. We remark that an arbitrary vertical offset has been added to circumference $L_y = 6$ for ease of viewing. In panel (b), we verify the tendency to algebraic decay quantitatively by performing a finite entanglement collapse to the form Eq.~\eqref{eq:SC_scaling_collapse}. Finally, in panel (c), we report the exponent of algebraic decay extracted for $L_y = 5, 6$ for each condensed channel $\alpha \beta = xx, yy, \text{ and } x0$. At the bottom of the panel, we summarize our data indicating if the $2e$ correlations are gapless and $1e$ correlations are gapped at each circumference with a \cmark. These panels demonstrate that our results from the main text are qualitatively unchanged as we introduce finite inter-layer repulsion $U_2$ and finite interaction range. }
    \label{fig:repulsive_SC_scaling}
\end{figure}

Similar to the analysis in the main text, let us start by examining the structure of the superconducting (charge-$2e$) correlation functions with increasing bond dimension and cylinder circumference.
In particular, in Fig.~\ref{fig:repulsive_SC_scaling}(a), we show the superconducting correlation function for $L_y = 5, 6$, finding that they both show a clear tendency to algebraic decay,  similar to the case at $d=U_2=0$.
We demonstrate this quantitatively once more via a finite entanglement scaling collapse in Fig.~\ref{fig:repulsive_SC_scaling}(b) and show the extracted exponents for the two $L_y$'s in Fig.~\ref{fig:repulsive_SC_scaling}(c).
Our results indicate that at finite repulsion $(U_2, d) = (0.2, 0.358)$, our system has algebraically decaying $2e$ correlations at $L_y = 5, 6$ but, unlike the case without repulsion, its exponents of algebraic decay do not clearly decrease with cylinder circumference.
We largely suspect that this is due to the fact that the fillings and locations of Fermi pockets are qualitatively different at the two circumferences, preventing a direct comparison of the two circumferences.

\begin{figure}
    \centering
    \includegraphics[scale=\otherfigscale]{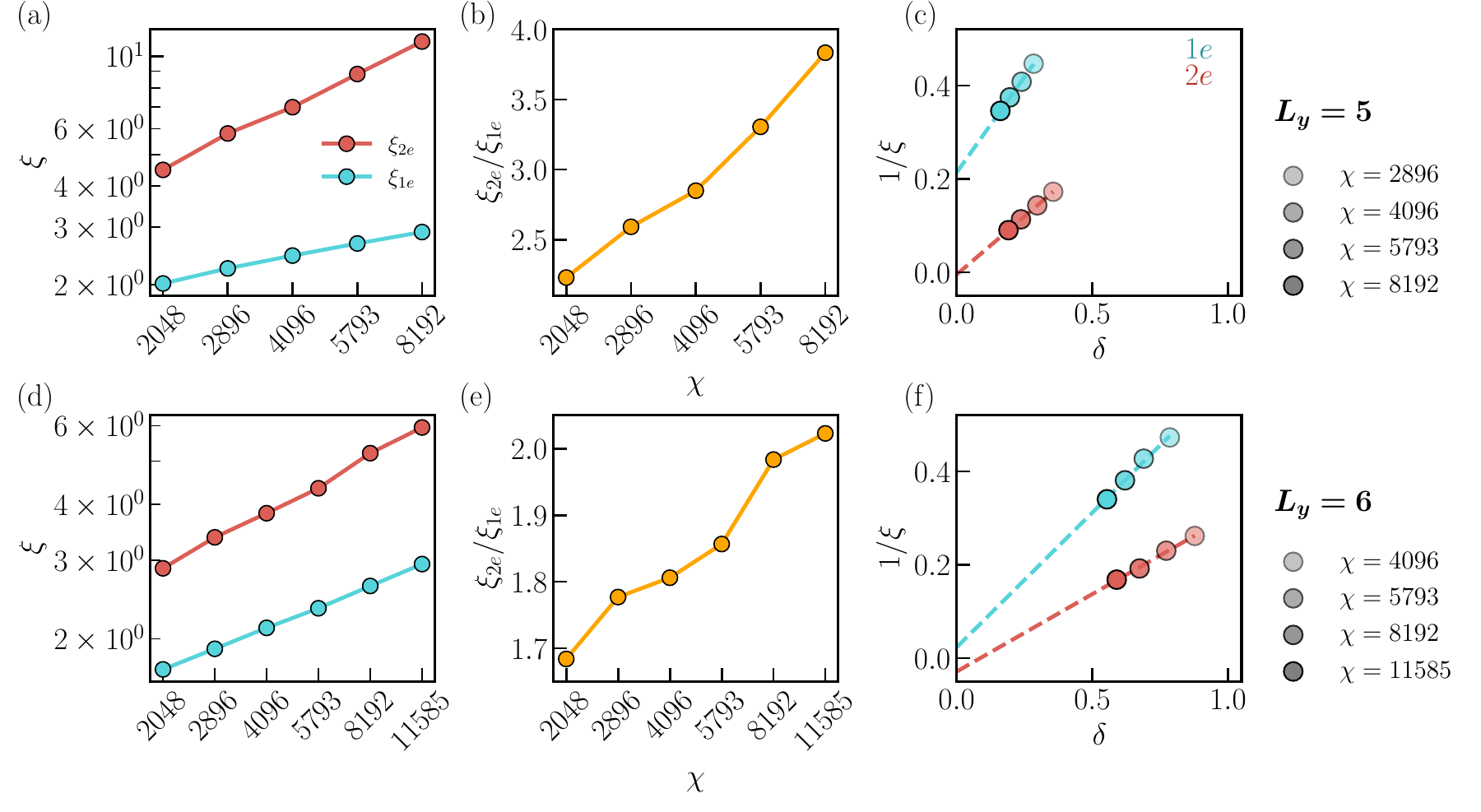}
    \caption{\textbf{Bond Dimension Scaling of Correlation Lengths at finite $(d,U_2)$.} In panels (a) and (d), we show the one-electron and two-electron correlation lengths as a function of bond dimension for $L_y = 5, 6$ at a parameter point $(g, U_1, U_2, d) = (0.1, 1.4, 0.2, 0.358)$ with finite inter-layer repulsion. We find that both appear to be growing linearly but the one-electron correlation lengths generally grow much slower. This is easiest to see from panels (b) and (e) where the ratio of $\xi_{2e}/\xi_{1e}$ is found to be growing with bond dimensions for both $L_y = 5, 6$. From the discussion of Section~V.A of the main text, this suggests that both parameter points are a Luther-Emery liquid. For $L_y = 5$, this is corroborated in panel (c) by performing an extrapolation to infinite bond dimension as we find a gapped $1e$ correlations and gapless $2e$ correlations. However, for $L_y = 6$, panel (e) yields negative correlation lengths for $\xi_{2e, \infty}^{-1}$ and $\xi_{1e, \infty}^{-1}$ close to zero suggesting that we are too far from a ``scaling regime'' for $L_y = 6$ to distinguish a Luther-Emery liquid from a Luttinger Liquid. 
    }
    \label{fig:SC_xi_extrapolation_finite_dU2}
\end{figure}

We next turn to the analysis of the correlation lengths. The raw correlation lengths are given in Fig. \ref{fig:SC_xi_extrapolation_finite_dU2}(a,b). At $L_y=5$, both $\xi_{1e}$ and $\xi_{2e}$ are increasing with $\chi$, but their ratio grows rapidly as a function of bond dimension, apparently without bound [see Fig. \ref{fig:SC_xi_extrapolation_finite_dU2}(d)]. In particular, it greatly exceeds the value $1/2$ expected from Wick's theorem in the limit of a weakly-interacting metallic state.
This is suggestive, but does not conclusively show on its own, that there is a gap to electronic excitations. The situation is similar at $L_y=6$ but here the ratio, although above two, is not as large at the bond dimensions we can access.

To gain greater clarity about the nature of the state, e.g. the existence of a charge gap, we again attempt to extrapolate both the $2e$ and $1e$ correlation lengths. Fig.~\ref{fig:SC_xi_extrapolation_finite_dU2}(c,f) shows linear extrapolations of $1/\xi$ versus $\delta$ for $L_y =5,6$. At $L_y = 5$, $1/\xi_{2e}$ extrapolates to zero, in accord with the power law superconducting correlations above, where as $1/\xi_{1e}$ is positive. This is consistent with a charge gap, and we may again confidently identify the $L_y=5$ at finite $(d,U_2)$ as a Luther-Emery liquid, i.e. a superconductor. In fact, the indications of superconductivity here, such as $1/\xi_{1e}$, appear to be larger or more clear than the $(d,U_2) = (0,0)$ case studied above. Unfortunately this clarity does not extent to the $L_y=6$ case, where both extrapolations are consistent with zero. Given the ratio $\xi_{2e}/\xi_{1e} \gg 1/2$, ordinarily a good heuristic indication of pairing, and that the smallest attainable $\delta$ remains much larger than the Fermi wavevector (see Sec.~\ref{sec-zerodU2-xis}), this evidence does not yet yield a firm conclusion for $L_y=6$. As usual, higher bond dimensions could resolve this issue. Due to computational expense, $L_y=7$ was not studied in detail here.

In summary, adding finite $(d,U_2)$ does not have a clear effect at $L_y=6$, but appears to further stabilize superconductivity at $L_y=5$. This success at $L_y=5$ invites us to determine the boundaries of the superconducting phase at this circumference. This is the topic of the next section.

\vspace{6mm}\toclesslab\subsection{Robustness of Superconductivity at $L_y=5$}{appsubsec-Robustness}

\begin{figure}
    \centering
    \includegraphics[scale=\otherfigscale]{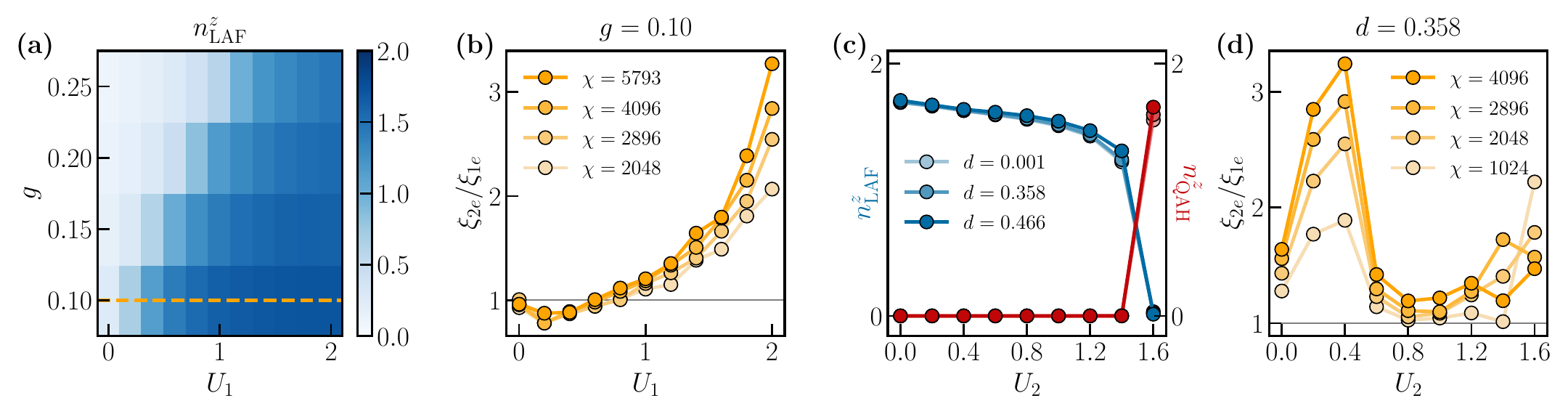}
    \caption{\textbf{LAF Polarization of the Superconductor and Robustness to Additional Repulsion.} (a) LAF polarization of the doped state, which exhibits a widespread LAF phase in the strong coupling regime at fixed $d=U_2=0$.
    To diagnose superconducting order, we take a cut at fixed $g$ (orange line) and, in (b), plot the ratio $\xi_{2e}/\xi_{1e}$. Its growth with bond dimension and substantial magnitude (i.e., in excess of unity at the largest available bond dimension) suggests the presence of superconductivity at $U_1 \gtrsim 0.6$.
    (c) Plot of the LAF and QAH (i.e. layer) polarization as a function of $U_2$ at fixed $(g,U_1) = (0.1,1.4)$. For each value of the interaction range $d$, the doped state remains substantially LAF-polarized until $U_2$ exceeds $U_1$, at which the phase abruptly transitions to QAH.
    Fixing $d=0.358$, in (d) we plot the superconducting diagnostic $\xi_{2e}/\xi_{1e}$, finding evidence for superconducting correlations at $U_2 \lesssim 1.2$.
    All plots at $L_y=5$ and $\nu = 2 - 1/5$.
    }
    \label{fig:robustness_gU1_dU2}
\end{figure}

We now focus on circumference $L_y=5$ cylinders and study the extent of the superconducting phase in terms of several parameters: (i) interlayer tunneling $g$, (ii) in-layer interactions $U_1$, (iii) interlayer interactions $U_2$, (iv) interaction range $d$, (v) electron filling $\nu$, and (vi) spin-layer Zeeman field $B_z$. We will show that the superconducting phase is at least perturbatively stable to each of these parameters, and has a large phase extent in most parameters. This justifies our referring to the superconductor as ``robust" in the main text.

To start, we examine the effect of $g$ and $U_1$. Fig.~\ref{fig:robustness_gU1_dU2}(a) shows the LAF polarization in the $(g,U_1)$ plane from DMRG at filling $\nu=1.8$.
Comparing to Fig.~\ref{fig:app_undoped_phase_detail}(a) above, the LAF polarization appears in essentially the same region of parameter space; doping the LAF does not destroy the order. Fig.~5(a) of the main text showed that the superconducting phase persists over almost the whole region. To corroborate this claim, Fig \ref{fig:robustness_gU1_dU2}(b) plots the ratio $\xi_{2e}/\xi_{1e}$ at fixed $g=0.1$ as a function of $U_1$. We adopt the heuristic that the state is superconducting if $\xi_{2e}/\xi_{1e}$ is growing and exceeds unity at the largest available bond dimension. In the line cut, one can see this clearly holds at large $U_1$, but there is a small region near $U_1 = 0.2$ that is most LAF polarized but does not have significant superconducting correlations. At these bond dimensions, these states appear to be metallic and we refer to them as a LAF metal. However, larger bond dimensions tend to increase the superconducting correlations in a finite $\chi$ transition, albeit at larger $\chi$ at parameters close to the edge of the LAF-polarized region.
We therefore cannot determine if the LAF metal is a finite bond dimension artifact or a genuine phase. Either way, the vast majority of the LAF polarized region is clearly adiabatically connected to the state at $(g,U_1) = (0.15,1.5)$ that we previously determined to be a Luther-Emery liquid.

Next, we examine the effect of additional repulsive interactions $(d,U_2)$, corresponding to Fig.~5(c) of the main text. Fig.~\ref{fig:robustness_gU1_dU2}(c) shows the LAF and quantum anomalous Hall (QAH) polarizations as a function of $U_2$ for each $d$ at $\nu=2-1/5$. As before, the polarizations hew closely to their respective values at $\nu=2$ (not shown), with a transition from the LAF insulator to a Chern-polarized QAH insulator when $U_2$ exceeds $U_1$.
Fig.~\ref{fig:robustness_gU1_dU2}(d) shows the ratio $\xi_{2e}/\xi_{1e}$ on the same axes, which exceeds unity and grows with bond dimension for $U_2 \lesssim 1.2$. By our heuristic, this suggests all these points lie in the superconducting phase, and in fact the largest superconducting correlations are at $U_2 \sim 0.4$ at the largest available bond dimension. However, the ratio drops off precipitously around $U_1 \approx 0.6$. This might signal a drop in the superconducting gap (perhaps even to zero) that would only be apparent at much larger bond dimensions, or might simply be a regime where more entanglement is required to adequately describe the state.
Altogether, this demonstrates both that the superconductor is stable to significant additional repulsion, and that such interactions may accelerate the convergence of the superconducting state in bond dimension.

\begin{figure}
    \centering
    \label{fig:robustness_nu_Bz}
    \includegraphics[scale=\otherfigscale]{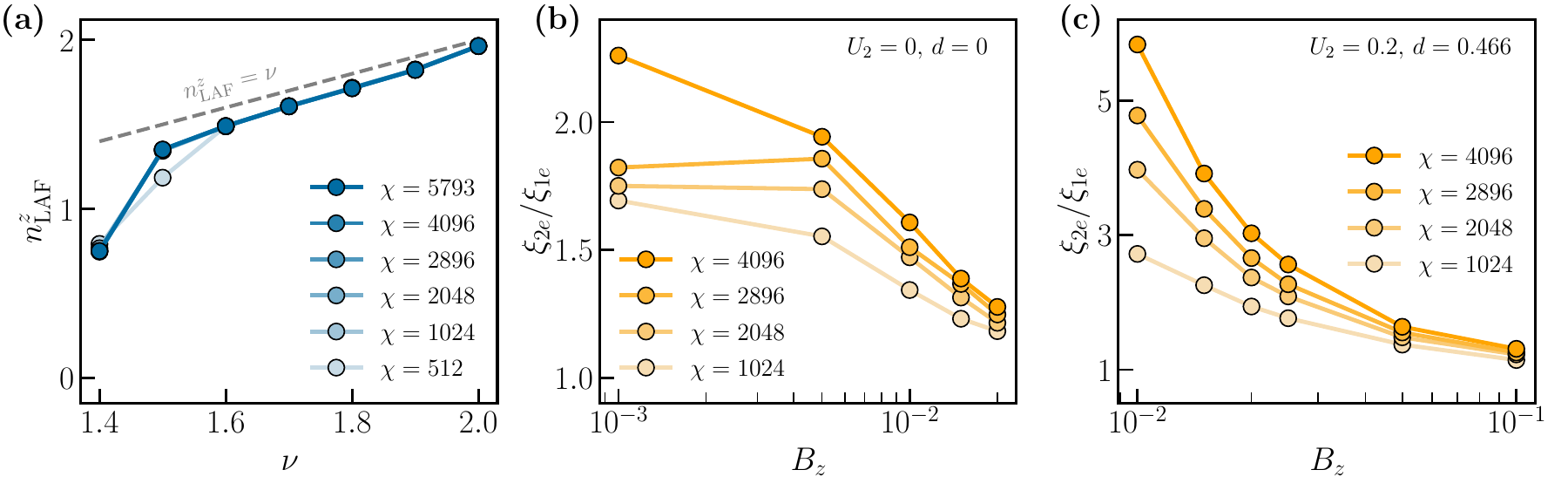}
    \caption{\textbf{Additional data for the $\nu$ and $B_z$ scans.} (a) The LAF polarization vs. the total electronic density at fixed $(g,U_1)=(0.1,1.8)$ and $d=U_2=0$. Over a significant range of densities that subsumes the putative superconducting region $1.7\lesssim \nu < 2$, the LAF polarization is near-maximal, decreasing approximately linearly with the density of hole dopants (i.e. $2-\nu$).
    (b) Plot of $\xi_{2e}/\xi_{1e}$ as a function of the Zeeman pinning field at $d=U_2=0$ and $(g,U_1)=(0.1,1.4)$. Its large magnitude and continued growth with bond dimension signals the prevalence of superconducting correlations down to very small values, namely $B_z = 10^{-3}$ (an order of magnitude smaller than $10^{-2}$, the value taken in the main text).
    (c) Plot of the same quantity at a finite amount of additional repulsion, enabling the numerical stabilization of DMRG simulations up to $B_z = 10^{-1}$, at which the state may still be superconducting.
    All plots are at $L_y=5$.
    }
\end{figure}

The superconductivity is also stable across a range of electronic densities below $\nu=2$. Fig.~\ref{fig:robustness_nu_Bz}(a) shows that the LAF polarization is stable to significant hole doping, decreasingly almost linearly with filling down to $\nu=1.5$, whereupon the behavior becomes more irregular. We remark that the states at $\nu \leq 1.5$ may be translation-breaking, though careful consideration of different circumferences would be necessary to draw any firm conclusions.
Comparing to Fig.~5(b) of the main text, we see that superconductivity appears to persist down to $\nu=1.6$.
In particular, these data suggest that the superconductor is resilient to significant changes in the Fermi velocity of the underlying LAF band structure, which increases with decreasing density --- recall from Fig.~\ref{fig:app_SCHF_LAF_bandstructure} the structure of the LAF Fermi surface.

Finally we discuss the subtle role of $B_z$. Recall from Appendix~\ref{app:micro_model} that we have added a small field $-B_z \sum_{\br} \hat{\psi}^\dagger_{\br} s^z \ell^z \hat{\psi}_{\br}$, with a typical value of $B_z =10^{-2}$ in our numerics. Briefly, the reason for this is that Hohenberg-Mermin-Wagner considerations forbid spontaneous symmetry breaking of a continuous symmetry in quasi-1D, a prerequisite for the formation of the LAF. Without this pinning field, one would expect a quantum disordered phase with extremely long but ultimately decaying LAF fluctuations, just as in the Haldane model~\cite{AffleckReview}. To be able to realize a truly polarized ground state without a huge entanglement cost, we add this $B_z$ field at an energy scale significantly below the other microscopic scales in the problem. Another choice would have been to add the square of the LAF order parameter, which has the advantage that it does not pick out a particular quantization axis.

To confirm that $B_z$ is innocuous to the superconducting physics, we examine the effect of tuning it. Fig.~\ref{fig:robustness_nu_Bz}(b) shows the ratio $\xi_{2e}/\xi_{1e}$ as a function of $B_z$. This quantity appears to increase without bound, all the way from $2\cdot 10^{-2}$ (on the order of the value $B_z = 10^{-2}$ fixed in the main text), down to $10^{-3}$, with a decrease in $B_z$ appearing to accelerate convergence at fixed bond dimension.
This strongly suggests that the value of $B_z$ we used here only serves to stabilize the LAF order, but does not play a critical role in pairing on top of it. Working at non-zero $(d,U_2)$ to circumvent convergence issues, in Fig.~\ref{fig:robustness_nu_Bz}(c) we show that superconductivity apparently persists to at least moderately larger $B_z \approx 0.1$, though a suppression in $\xi_{2e}/\xi_{1e}$ at fixed bond dimension suggests that the larger-$B_z$ states are more difficult to converge.
In fact, the longer-range interactions appear necessary to avoid norm errors at these larger $B_z$ values. We conclude that the particular value of $B_z$ is incidental to the superconducting phase at $L_y=5$, and are therefore justified in treating it as a small perturbation used for numerical convenience.

In summary, we have found that the Luther-Emery liquid at $L_y=5$ is resilient to six independent perturbations. This underscores the theme that doping the LAF insulator leads to a robust superconducting phase at this circumference.

\vspace{8mm}\toclesslab\section{Method for Extracting the Pair Wavefunction}{app:pairwave}

In the main text, we laid out a technique for extracting the pair wavefunction in charge-conserving cylinder iDMRG up to an overall momentum-independent constant of proportionality.
Here, we provide additional details about the arguments and assumptions that go into the technique.
%
To start, recall that the pair wavefunction is defined as
\begin{equation} \label{eq-pwfunction}
    \Delta_{\alpha \beta}(\mathbf{k}) \equiv \langle \hat\phii^{T} (\mathbf{k}) s^\alpha \ell^{\beta} \hat\phii(-\mathbf{k}) \rangle \equiv \langle \hat{\Delta}^{\dagger}_{\alpha \beta}(\mathbf{k}) \rangle.
\end{equation}
The magnitude and sign structure (or more generally, phase) of the pair wavefunction over momentum space are central to the character of a weak-pairing superconducting phase.
As such, $\Delta(\mathbf{k})$ is an invaluable tool for analyzing not only the symmetry properties of the superconducting state, but also the gap of its Bogoliubov quasiparticles via the relative magnitude of $\Delta(\mathbf{k})$ near the Fermi surface.

Extracting this quantity in charge-conserving iDMRG is non-trivial.
In particular, the matrix product state used in our iDMRG is restricted to have a definite $U(1)_Q$ charge.
As a result, when the algorithm attempts to capture a superconducting state that spontaneously breaks the $U(1)_Q$ symmetry, the resulting matrix product state will be similar in spirit to a ``cat state'' or the GHZ ground state of the Ising symmetry breaking phase found when one enforces being in a definite parity sector.
The practical consequence of this is that the matrix product state ansatz cannot have a non-zero one-point expectation value for any $U(1)_Q$ charged operator.
To see this explicitly, let $\ket{\psi}$ be the matrix product state and suppose that $e^{i \theta \hat{Q}} \ket{\psi} = e^{i \theta q} \ket{\psi}$.
Then, for \textit{any} linear operator $\hat\mcO$ with $U(1)$ charge $\alpha \neq 0$, we have that
\begin{equation}
    \langle \hat\mcO \rangle_{\psi} = \langle e^{-i \theta \hat{Q}} \hat\mcO e^{i \theta \hat{Q}} \rangle_{\psi} = e^{i \alpha \theta} \langle \hat\mcO \rangle_{\psi} \ \implies \ \langle \hat\mcO \rangle_{\psi} = 0,
\end{equation}
where $\langle \cdots \rangle_{\psi} = \bra{\psi} \cdots \ket{\psi}$. As such, the pair wavefunction, which encodes the expectation value of the Cooper pair operator, will trivially vanish for any state in our charge-conserving numerics, owing to the definite charge of each such state.
%
Our goal will therefore be to use the state \textit{obtained in numerics} to extract the pair wavefunction computed, in principle, from another state in the superconducting ground state manifold that does \textit{not} have definite charge.
With this goal in mind, this Appendix is organized into the following subsections:
\begin{enumerate}
    \item First, working directly in the 2D $(L_y \to \infty)$ and infinite bond dimension limit, we will show how one would extract the pair wavefunction above (up to a $\mathbf{k}$-independent constant of proportionality).

    \item Subsequently, we will discuss the case of a finite cylinder (still at infinite bond dimension) and will explicate our expectations regarding the approach to the 2D limit. Namely, we show how the 2D pair wavefunction can be approximately probed using data at finite, but sufficiently large, circumferences.
\end{enumerate}

\vspace{6mm}\toclesslab\subsection{Extracting the Pair Wavefunction in the 2D Limit}{appsec-extractingpwfin2D}

Let us first consider how to extract the pair wavefunction in the two-dimensional limit.
Throughout this section, let $\ket{\psi}$ be a wavefunction that: (\textit{i}) lives in the Hilbert space of a $(2 + 1)$d quantum system, (\textit{ii}) carries a definite $U(1)_Q$ charge and lives in the Hilbert space of a $(2 + 1)$d quantum system, and (\textit{iii}) is superconducting.
Such a state will fail to satisfy a physical property, typically assumed in most quantum systems, called \textit{cluster decomposition}.
Namely, not all long-distance (disconnected) two-point functions will decompose into products of one-point functions.
This is because:
\begin{equation}\label{eq-cluster}
    \lim_{\mathbf{r} \to \infty} \left[ \langle \hat\Delta^{\dagger}(\mathbf{r}) \hat\Delta(0) \rangle_{\psi} - \langle \hat\Delta^{\dagger}(\mathbf{r}) \rangle_{\psi} \langle \hat\Delta(0) \rangle_{\psi} \right] \neq 0,
\end{equation}
where $\hat\Delta(\mathbf{r})$ is any charged operator centered at location $\mathbf{r} \in \mathbb{R}^2$ that has long-range correlations in the superconducting state (i.e., $\langle \hat\Delta^\dag(\mathbf{r}) \hat\Delta(0) \rangle_{\psi} \to \text{constant}$ as $\mathbf{r} \to \infty$).\footnote{We take the vanishing of the left-hand side of Eq.~\eqref{eq-cluster}, for all operators $\hat\mcO$ in place of $\hat\Delta$, to be the \textit{definition} of a state that satisfies cluster decomposition.}
Before explicating our method for extracting the pair wavefunction numerically, we note two key assumptions and provide justification for their applicability.

\begin{shaded}
\noindent
    \textbf{Assumption 1:} If the system we are probing is infinite in at least one direction, then for any states $\ket{\psi}$ and $\ket{\phii}$ degenerate with one another in the manifold of superconducting ground states, we have that
    \begin{equation}
        \langle \hat\Delta^{\dagger}(\mathbf{r}) \hat\Delta(0) \rangle_{\psi} = \langle \hat\Delta^{\dagger}(\mathbf{r}) \hat\Delta(0) \rangle_{\phii} .
    \end{equation}
    As above, $\hat\Delta(\mathbf{r})$ is defined as any \textit{local} charged operator having long-range correlations in the superconducting state.
\end{shaded}

\textit{Rationale.} We provide a heuristic argument for this assumption based on the stability of a symmetry-breaking phase to perturbations that don't explicitly break the symmetries of the Hamiltonian.
Suppose that $\ket{\psi}$ and $\ket{\phii}$ are two states that are degenerate with one another and both live in the superconducting ground state manifold of a $U(1)_Q$-symmetric Hamiltonian $\hat{H}$.
Furthermore, suppose that these superconducting ground states are part of a stable phase of matter, such that any $U(1)_Q$-symmetric perturbation to the Hamiltonian---provided it is sufficiently small---does not drive a transition out of this superconducting phase.
%
In particular, any such perturbation should not be able to energetically split the degenerate ground state manifold of the superconductor.
Let us now suppose, for the sake of contradiction, that there exists a local, charge-neutral, and bounded (to exclude objects like the total charge) operator $\hat{\mathcal{A}}(\mathbf{r})$ such that $\langle \hat{\mathcal{A}} (\mathbf{r}) \rangle_{\psi} \neq \langle 
\hat{\mathcal{A}} (\mathbf{r}) \rangle_{\phii}$.
In this case, we could add the following $U(1)_Q$ symmetry-preserving perturbation to our Hamiltonian: 
\begin{equation} \label{eq-perturbedH}
    \hat{H} \to \hat{H} + \varepsilon \sum_{\mathbf{r}} 
 \left[\hat{\mathcal{A}} (\mathbf{r}) + \hat{\mathcal{A}}^{\dagger}(\mathbf{r}) \right].
\end{equation}
For sufficiently small $\varepsilon$, the energies of $\ket{\psi}$ and $\ket{\phii}$, at first order in perturbation theory, would then change to
\begin{equation}
    E[\ket{\psi}] - E[\ket{\phi}] \approx \varepsilon \left(  \langle \hat{\mathcal{A}} (\mathbf{r})  + \hat{\mathcal{A}}^{\dagger} (\mathbf{r})  \rangle_{\psi} - \langle \hat{\mathcal{A}} (\mathbf{r}) + \hat{\mathcal{A}}^{\dagger}(\mathbf{r}) \rangle_{\phii} \right),
\end{equation}
which is finite, assuming that $\langle \hat{\mathcal{A}}(\mathbf{r}) \rangle_{\psi} - \langle \hat{\mathcal{A}}(\mathbf{r}) \rangle_{\phii}$ has a finite real part.\footnote{If it is purely imaginary, we can instead add the anti-Hermitian part of $\hat\mcA$ to the Hamiltonian in Eq.~\eqref{eq-perturbedH}.}
This contradicts the assumption that the superconductivity in the ground state is part of a stable phase of matter.
As such, $\langle \hat{\mathcal{A}} (\mathbf{r}) \rangle_{\psi}  = \langle \hat{\mathcal{A}} (\mathbf{r}) \rangle_{\phii}$.
As a corollary, because $\hat{\Delta}^{\dagger}(\mathbf{r}) \hat{\Delta}(0)$ is charge neutral, then for any finite $\mathbf{r}$:
\begin{equation}
 \langle \hat{\Delta}^{\dagger}(\mathbf{r})  \hat{\Delta}(0) \rangle_{\psi} = \langle \hat{\Delta}^{\dagger}(\mathbf{r}) \hat{\Delta}(0) \rangle_{\phii}. 
\end{equation}

\vspace{3 mm}

\begin{shaded}
\noindent
    \textbf{Assumption 2:} The second observation is that if If $H$ is a Hamiltonian that realizes a superconducting ground state then, in its manifold of degenerate ground states, there exist states $\ket{\phii}$ that satisfy cluster decomposition. 
\end{shaded}

\textit{Rationale.} This assumption is physically motivated on the grounds of locality: for physical states in a system, distant local operators should not be able to influence one another.
Indeed, the cluster decomposition property is sometimes assumed in formal definitions of quantum field theory.

\vspace{2 mm}

With these two assumptions, we can now understand how to extract the pair wavefunction up to an overall constant of proportionality.
Once again, let $\ket{\psi}$ be a superconducting ground state that carries a definite $U(1)_Q$ charge and lives in the Hilbert space of a two-dimensional system.
Now, for fixed $x$ and $p_y$, consider the following function:
\begin{equation} \label{eq-proxygap}
    \wilde{\Delta}^{\alpha \beta}_{x, p_y; \psi}(\mathbf{k}) = \langle \hat{\Delta}^{\dagger}_{\alpha \beta}(0;\mathbf{k}) \hat{\Delta}_{\alpha \beta} (x; 0,p_y) \rangle_{\psi}, 
\end{equation}
where
\begin{equation}
    \hat{\Delta}^{\dagger}_{\alpha \beta}(x; \mathbf{k}) = \sum_{\delta x} \hat{\Delta}^{\dagger}_{\alpha \beta}(x; \delta x, k_y) e^{i k_x \delta x}
\end{equation}
and
\begin{equation}
 \hat{\Delta}^\dag_{\alpha\beta}(x; \delta x,k_{y}) \equiv
    \begin{cases}
        \hat{\varphi}^T \left(x + \frac{\delta x}{2},k_{y}\right)\cdot s^{\alpha} l^{\beta} \cdot \hat{\varphi}\left(x -\frac{\delta x}{2},-k_{y}\right) & \text{(even $\delta x$)} \\
        \hat{\varphi}^T \left(x + \frac{\delta x + 1}{2}, k_{y}\right) \cdot s^{\alpha} l^{\beta} \cdot \hat{\varphi}\left(x -\frac{\delta x - 1}{2}, -k_{y}\right) & \text{(odd $\delta x$)}
    \end{cases}.
\end{equation}
In the above, $x$ denotes the approximate center of mass of the fermion pair, whereas $\delta x$ denotes their relative position. The two operators in Eq.~\eqref{eq-proxygap} are therefore separated by a distance $x$, which we will take to be large.
As a preliminary remark, we note that
\begin{equation} \label{eq-Delta-translationinv}
\begin{aligned}
    \langle \hat{\Delta}_{\alpha \beta}^{\dagger}(x; \mathbf{k}) \rangle_{\phii} &= \sum_{\delta x} \langle \hat{\Delta}_{\alpha \beta}^{\dagger}(x; \delta x, k_y) \rangle_{\phii} e^{i k_x \delta x}\\
    &= \frac{1}{N_x} \sum_{x, \delta x} \langle \hat{\Delta}_{\alpha \beta}^{\dagger}(x; \delta x, k_y) \rangle_{\phii} e^{ik_x \delta x} \equiv  \langle \hat{\Delta}^{\dagger}_{\alpha \beta}(\mathbf{k}) \rangle_{\phii}  =  \langle \hat{\phii}^{T}(\mathbf{k}) s^\alpha l^\beta \hat{\phii}(-\mathbf{k})\rangle_{\phii}
\end{aligned}
\end{equation}
for any cluster decomposition-satisfying state $\ket{\phii}$ that is also invariant under translations.
We are now prepared to make the central claim of this section.


\begin{shaded}
\noindent
\textbf{Claim:} The object defined in Eq.~\eqref{eq-proxygap}, evaluated for the translation-invariant state $\ket{\psi}$ with definite $U(1)_Q$ charge, is proportional to the pair wavefunction computed in a cluster-decomposition satisfying state $\ket{\phii}$. In equations:
 \begin{equation}
         \wilde{\Delta}^{\alpha \beta}_{x, p_y; \psi}(\mathbf{k}) = C^{\alpha \beta}_{x, p_y} \langle \hat{\Delta}^{\dagger}_{\alpha \beta} (\mathbf{k}) \rangle_{\phii} + \mathcal{O}(e^{-|x - \zeta_{\text{SC}}|/ \xi})
\end{equation}
where $\xi$ is the connected correlation length of the superconducting state and $\zeta_{\text{SC}}$ is the ``size'' of the Cooper pair (which coincides with its coherence length in certain contexts). 
\end{shaded}

\textit{Argument.} Suppose that in our superconductor, Cooper pairs have a ``finite size'' $\zeta_{\text{SC}}$ for which
\begin{equation}
    \hat{\Delta}_{\alpha \beta}(x; \delta x, k_y) \ket{\psi'} = 0,  \quad \text{ if }\quad \delta x > \zeta_{\text{SC}},
\end{equation}
for any superconducting state $\ket{\psi'}$ in the ground state manifold. It then immediately follows that
\begin{align}
    \langle \hat{\Delta}^{\dagger}_{\alpha \beta}(0;\mathbf{k}) \hat{\Delta}_{\alpha \beta} (x; 0,p_y) \rangle_{\psi} &= \sum_{\delta x}  \langle \hat{\Delta}^{\dagger}_{\alpha \beta}(0; \delta x, k_y) \hat{\Delta}_{\alpha \beta} (x; 0,p_y) \rangle_{\psi}\ e^{ik_x \delta x}\\ &= \sum_{\delta x = -\zeta_{\text{SC}}}^{\zeta_{\text{SC}}}  \langle \hat{\Delta}^{\dagger}_{\alpha \beta}(0; \delta x, k_y) \hat{\Delta}_{\alpha \beta} (x; 0,p_y) \rangle_{\psi} e^{i k_x \delta x},
\end{align}
%
Now, by Assumptions 1 and 2 above, there exists 
a state $\ket{\phii}$ in the superconducting ground state manifold that satisfies cluster decomposition.
As such:
\begin{align}
\langle \hat{\Delta}^{\dagger}_{\alpha \beta}(0; \mathbf{k}) \hat{\Delta}_{\alpha \beta} (x; 0,p_y) \rangle_{\psi} &= \sum_{\delta x = - \zeta_{\text{SC}}}^{\zeta_{\text{SC}}} \langle \hat{\Delta}^{\dagger}_{\alpha \beta}(0; \delta x, k_y) \hat{\Delta}_{\alpha \beta} (x; 0,p_y) \rangle_{\phii} e^{ik_x \delta x}\\
&= \sum_{x = - \zeta_{\text{SC}}}^{\zeta_{\text{SC}}} \langle \hat{\Delta}^{\dagger}_{\alpha \beta}(0; \delta x, k_y) \rangle_{\phii}  \langle \hat{\Delta}_{\alpha \beta} (x; 0,p_y) \rangle_{\phii} e^{ik_x \delta x} + \mathcal{O}\left(e^{- |x - \zeta_{\text{SC}}|/\xi}\right)\\
&= \langle \hat{\Delta}^{\dagger}_{\alpha \beta}(0; \mathbf{k}) \rangle_{\phii} \langle \hat{\Delta}_{\alpha \beta}(x; 0, p_y) \rangle_{\phii} + \mathcal{O}\left(e^{- |x - \zeta_{\text{SC}}|/\xi}\right) \\
&= C_{x, p_y}^{\alpha \beta} \langle \hat{\Delta}^{\dagger}_{\alpha \beta}(\mathbf{k}) \rangle_{\phii} + \mathcal{O}\left(e^{- |x - \zeta_{\text{SC}}|/\xi}\right),
\end{align}
where in the second step we used cluster decomposition and in the last step we used Eq.~\eqref{eq-Delta-translationinv}. 
This completes our argument.

\vspace{6mm}\toclesslab\subsection{Extracting the Pair Wavefunction at Large Circumference}{}

We now discuss how the above discussion is modified on a cylinder of infinite length and ``large'' circumference. 
In particular, consider a wavefunction $\ket{\psi(L_y)}$ that lives in the Hilbert space defined on this geometry and assume that it has a definite $U(1)_Q$ charge.
Such as state should be thought of as the $\chi \to \infty$ limit of a sequence of wavefunctions obtained numerically in iDMRG.
If the wavefunction is in the charge-$2e$ superconducting phase, it will have algebraic long-range order of the following form:
\begin{equation}
    \langle \hat{\Delta}_{\alpha \beta}^{\dagger}(0; 0, k_y) \hat{\Delta}_{\alpha \beta}(x; 0, p_y) \rangle_{\psi(L_y)} \sim \frac{1}{|x|^{\eta(L_y)}},
\end{equation}
for some spin and layer channel $\alpha\beta$. More precisely, provided that $\delta x\ll x$, we can say that
\begin{equation}
    \langle \hat{\Delta}^{\dagger}(0; \delta x, k_y) \hat{\Delta}(x; 0, p_y) \rangle_{\psi(L_y)} \approx \frac{A_{p_y}(\delta x, k_y; L_y) }{|x|^{\eta(L_y)}}  + B_{\delta x, p_y}(L_y) e^{-x/\xi(L_y)},
\end{equation}
where $\xi(L_y)$ is a residual finite $L_y$ ``connected'' correlation length. Given that $\eta(L_y)\to 0$ as $L_y \to \infty$, then in the 2D limit the above relation becomes
\begin{equation}
    \lim_{L_y \to \infty} \langle \hat{\Delta}^{\dagger}(0; \delta x, k_y) \hat{\Delta}(x; 0, p_y) \rangle_{\psi(L_y)} = A_{p_y}(\delta x, k_y; \infty) + B_{\delta x, p_y}(\infty) e^{-x/\xi},
\end{equation}
where $\xi = \lim_{L_y\to\infty} \xi(L_y)$ is the 2D connected correlation length and
\begin{equation} \label{eq-2Dcorrelation_from_1D}
    A_{p_y}(\delta x, k_y; \infty) = \lim_{x \to \infty} \langle \hat\Delta^{\dagger}(0; \delta x, k_y) \hat\Delta(x; 0, k_y) \rangle_{\psi}.
\end{equation}
In this expression, $\ket{\psi}$ is a wavefunction that carries a definite $U(1)_Q$ charge and lives in the many-body Hilbert space of the 2D system.
Note that from Section~\ref{appsec-extractingpwfin2D} above, we know how to extract the pair wavefunction given knowledge of Eq.~\eqref{eq-2Dcorrelation_from_1D}.
Thus, the problem of extracting the pair wavefunction reduces to computing the latter expression up to an overall $\delta x$-independent constant of proportionality.
Doing so is simple if we assume that $A_{p_y}(\delta x, k_y; L_y) \approx A_{p_y}(\delta x, k_y; \infty)$ at the $L_y$'s accessible to us in numerics.
Then, if $x \gg \xi(L_y)$: 
\begin{equation}
    \langle \hat{\Delta}^{\dagger}(0; \delta x, k_y) \hat{\Delta}(x; 0, p_y) \rangle_{\psi(L_y)} \approx \frac{1}{|x|^{\eta(L_y)}}A_{p_y}(\delta x, k_y;\infty) \propto A_{p_y}(\delta x, k_y; \infty).
\end{equation}
Using this result, we can therefore approximately probe the pair wavefunction of the 2D system using data from our finite circumference simulations.
Though it is difficult to prove rigorously that our extraction method yields a function proportional to the pair wavefunction, we remark that the sensibility of our numerical results (e.g. its peaking at the Hartree Fock mean field Fermi surface and our obtaining a gap function $\delta(\mathbf{k})$ for which $\delta^\dagger(\mathbf{k}) \delta(\mathbf{k}) \sim P_{\text{LAF}}$) give us confidence in its merit as a useful tool for the characterization of superconducting states.

\vspace{8mm}\toclesslab\section{Additional Details of Pairing Symmetry Analysis}{app:PairingSymmetry}

In this appendix, we provide the theoretical details underlying our characterization of the pairing symmetry.
To do so, we provide a brief reminder regarding how one formally characterizes pairing symmetry in a superconducting state.
In particular, we describe how the pairing symmetry, in systems with both unitary and anti-unitary symmetries, is characterized by the induced co-representation of the symmetry group on the pair wavefunction.
Subsequently, we use this formalism to determine the pairing symmetry of the double Hofstadter model.

\vspace{6mm}\toclesslab\subsection{Characterizing Pairing Symmetry: Induced Co-Representations on the Pair Wavefunction}{}

In this subsection, we detail systematically how to characterize the pairing symmetry of a superconducting ground state. 
In particular, the pairing symmetry refers to the representation of the magnetic space group $G$ under which the pair wavefunction $\Delta_{\alpha \beta}(\mathbf{k})$ transforms.
To be a bit more precise, let us start by noting that the pair wavefunction $\Delta_{\alpha \beta}(\mathbf{k})$ lives in the vector space $\mathbb{C}^{4 \times 4} \otimes \mathcal{F}_{\text{BZ}}$ where the $4 \times 4$ comes from the different channels $\alpha \beta$ and $\mathcal{F}_{\text{BZ}}$ is the space of functions over the Brillouin zone.
Our goal will be to first determine an action of the magnetic space group on the pair wavefunction such that it transforms under a \textit{co-representation} of the group--- the generalization of the notion of a representation to include both unitary and anti-unitary symmetries.\footnote{For a short but lucid review see Appendix~B of Ref.~\onlinecite{ChenGuWen_SPTCohomology}, or Ref.~\onlinecite{cracknell} for a more comprehensive treatment.}
We recall the definition for convenience. Let $G$ be a group with group elements $g \in G$ that are either linear or anti-linear.
Following Ref.~\onlinecite{ChenGuWen_SPTCohomology}, define a function $s$ such that $s(g) = \pm 1$ for linear (anti-linear) group elements. We write $F^{s(g) = +1} = F$ and $F^{s(g) = -1} = F^{*}$ for complex numbers $F$.
Then, a co-representation of $G$ is a homomorphism from the group to the space of operators on a vector space $V$ defined over the complex numbers:
\begin{equation}
    g \to \begin{cases} \label{eq-nug}
    \nu_g &  s(g) = +1 \\
    \nu_g \mathsf{K} &  s(g) = -1, \\
    \end{cases}
\end{equation}
where $\nu_g \in \mathsf{GL}(V, \mathbb{C})$ (i.e. the group of automorphisms of $V$), $\mathsf{K}$ indicates complex conjugation, and where $\nu_g$ matrices must satisfy a characteristic composition property:
\begin{equation} \label{eq-corepalgebra}
    \nu_{gh} = \nu_g \nu_{h}^{s(g)}
\end{equation}
%
Below we will show the action of our symmetry group on the pair wavefunction and demonstrate that it forms a co-representation. We will then organize the vector space $\mathbb{C}^{4 \times 4} \otimes \mathcal{F}_{\text{BZ}}$ into a direct sum of irreducible co-representations. The subspaces the pair wavefunction we find lives in determines its pairing symmetry. The irrep that the pair wavefunction belongs to determines its pairing symmetry.

To define the action of the symmetry group $G$ on the pair wavefunction, let us take $\hat{g}$ to be the (anti-)unitary operator corresponding to group element $g \in G$.
Let us make the assumption (which will be relevant to the case at hand; see Subsection~\ref{subapp:-CooperPairAction}) that \textit{generators} of our magnetic space group $\hat{g}$ act on our Cooper pair operators $\Delta_{\alpha \beta}(\mathbf{k})$ by acting on the momentum label only:
\begin{equation}
    \hat{g}^{\dagger}\hat{\Delta}^{\dagger}(\mathbf{k}) \hat{g} = \hat{\Delta}^{\dagger}(g^{-1}\mathbf{k}),
\end{equation}
where we have suppressed the $\alpha \beta$ indices, and $g^{-1}\mathbf{k}$ is the action of the symmetry on the momentum label.
Define the action of the element on the pair wavefunction $\Delta(\mathbf{k}) \equiv \langle \psi, \hat{\Delta}^{\dagger}(\mathbf{k}) \psi \rangle$ as:
\begin{equation}
     (g \cdot \Delta)(\mathbf{k})  = \langle \hat{g} \psi, \hat{\Delta}^{\dagger}(\mathbf{k})  \hat{g}\psi \rangle = \langle \psi, \hat{g}^{\dagger}\hat{\Delta}^{\dagger}(\mathbf{k})  \hat{g}\psi \rangle^{s(g)} =  \Delta^{s(g)}(g^{-1}\mathbf{k}),
\end{equation}
where we have used the notation $\langle a, b \rangle = \langle a|b\rangle$ to denote the inner product as it will be more convenient for working with anti-unitary operators.
While this action is derived from the quantum mechanical data, we can now think of $\Delta(\mathbf{k})$ as an element of the space of functions on the Brillouin zone $\mathcal{F}_{\mathrm{Bz}}$, on which the action of the group is $(g \cdot F)(\mathbf{k}) = F^{s(g)}(g^{-1} \mathbf{k})$.
%
%
On the vector space of such functions, this map is manifestly a group homomorphism and is anti-linear.
To demonstrate that this action indeed does form a co-representation, it is helpful to decompose functions $F(\mathbf{k})$ into ``irreducible'' functions that are real and transform definitely under the action of $\mathbf{k} \to g^{-1} \mathbf{k}$.
Explicitly, we have that:
\begin{equation}
F(\mathbf{k}) = \sum_{\Gamma \in \text{Irrep}(G)} \sum_{\mu = 1}^{\text{dim}(\Gamma)} \eta^{\Gamma, \mu}_F  Y_{\Gamma,\mu} (\mathbf{k}),
\end{equation}
where $Y_{\Gamma,\mu}(\mathbf{k})\in \mathbb{R}$ and $Y_{\Gamma,\mu}(g^{-1}\mathbf{k}) = \mathbb{V}^{\Gamma, g}_{\mu \nu} Y_{\Gamma, \nu} (\mathbf{k})$.
In this basis, the action of our group can be expressed as: 
\begin{equation}
    (g \cdot F)(\mathbf{k}) = F^{s(g)}(g^{-1} \mathbf{k}) =  \nu_g \mathsf{K}^{s(g)} F(\mathbf{k}),
\end{equation}
where $\nu_g$ is a block diagonal matrix whose rows and columns are labeled by $\Gamma$ and $\mu$ that rotates the coefficients $\eta^{\Gamma, \mu}$ and $\mathsf{K}^{s(g) = +1} = 1$ and $\mathsf{K}^{s(g) = -1} = \mathsf{K}$. 
Its blocks are precisely the matrices $\mathbb{V}^{\Gamma, g}_{\mu \nu}$ labeled above.
We now demonstrate that the $\nu_g$'s above form a co-representation.
In particular, note that $(gh \cdot \Delta)(\mathbf{k}) = \nu_{gh} \mathsf{K}^{s(gh)} \Delta(\mathbf{k})$, by definition.
However,
\begin{equation}
    (gh \cdot F)(\mathbf{k}) = (g \cdot ( h \cdot F))(\mathbf{k}) = \nu_g \mathsf{K}^{s(g)} (h \cdot F)(\mathbf{k}) = \nu_g \mathsf{K}^{s(g)} \nu_h \mathsf{K}^{s(h)} F(\mathbf{k}) = \nu_g \nu_h^{s(g)} \mathsf{K}^{s(gh)} F(\mathbf{k}).
\end{equation}
As such, the $\nu$ matrices satisfy the characteristic algebra of a co-representation $\nu_{gh} = \nu_{g} \nu_h^{s(h)}$.
As a remark, a change of basis $Y(\mathbf{k}) \to \mathbb{W}^{\dagger} Y(\mathbf{k})$, generates similarity transformations on the $\nu$'s:
\begin{equation}
    \nu_g \to \mathbb{W}^{\dagger} \nu_g \mathbb{W}^{s(g)}.
\end{equation}
Two co-representations are said to be \textit{equivalent} if they are related by such a similarity transformation.

\vspace{6mm}\toclesslab\subsection{Pairing Symmetry}{subapp:-CooperPairAction}

Generalities out of the way, we now focus on the specific case at hand.
Recall that the magnetic space group of our model is generated by:
\begin{equation}
    m_x \qquad m_y \qquad C_{2z} \qquad C_{2x} \qquad \mathcal{I}
\end{equation}
as discussed in Appendix~\ref{app:micro_model}.
Note that, because our Cooper pairs carry zero momentum and charge-$2e$, the action of $m_y$ on such objects is trival and the remaining generators mutually commute (c.f. Table~\ref{table-commutators}) and square to the identity.
As such, the group acts on the pair wavefunction as $(\mathbb{Z}_2)^3 \times \mathbb{Z}_2^\mathsf{K}$, where $\mathbb{Z}_2^\mathsf{K}$ is anti-unitary.
This fact greatly simplifies the possible representations that the pair wavefunction can live in.
Since the group is abelian, the $\nu_g$'s are constrained to be numbers and since they square to the identity, this implies that the index $\nu^g$ defined in Eq.~\eqref{eq-nug} satisfies:
\begin{equation}
    1 = \begin{cases}\nu_g^2 = 1 & s(g) = +1 \\ |\nu_g|^2 = 1 & s(g) = -1,
    \end{cases}
\end{equation}
which follows from the algebra of the co-representation Eq.~\eqref{eq-corepalgebra}.
This constrains:
\begin{equation}
    \nu_g = \begin{cases}
        \nu_g = \pm 1 & s(g) = +1 \\
        \nu_g = e^{i \theta} & s(g) = -1.
    \end{cases}
\end{equation}
Finally, note that under basis transformations $\mathbb{W} = e^{i \phi/2}$, 
\begin{equation}
    \nu_g \to \begin{cases} \nu_g & s(g) = +1 \\ e^{-i\phi} \nu_g & s(g) = -1 
    \end{cases}
\end{equation}
This implies that there is only one possible regular co-representation of the pair wavefunction under the $\mathbb{Z}_2^K$ as they are all related by a similarity transformation.
Therefore, the pairing symmetry is characterized by three ``symmetry indices'' (or characters) $\nu_g$ for $g \in \{C_{2z}, C_{2x}, m_x\}$.
To determine $\nu_g$ for these generators, let us tabulate the action of the symmetry operators on the Cooper pair operators $\Delta_{\alpha \beta}^{\dagger}(\mathbf{k})$ for the dominant condensed channels $\alpha \beta = xx$ and $yy$, which follows from Eqs.~\eqref{eq-wannier-Isym-operator}-\eqref{eq-wannier-mx-operator}:
\begin{align}
    \hat{C}_{2z} \hat{\Delta}^{\dagger}_{\alpha \beta}(\mathbf{k}) \hat{C}^{-1}_{2z} &= \hat{\Delta}^{\dagger}_{\alpha \beta}(C_{2z}\mathbf{k}) \\ 
    \hat{C}_{2x} \hat{\Delta}^{\dagger}_{\alpha \beta}(\mathbf{k}) \hat{C}^{-1}_{2x} &= \hat{\Delta}^{\dagger}_{\alpha \beta}(C_{2x}\mathbf{k}) \\ 
    \hat{m}_x \hat{\Delta}^{\dagger}_{\alpha \beta}(\mathbf{k}) \hat{m}^{-1}_{x} &= \hat{\Delta}^{\dagger}_{\alpha \beta}(\mathbf{k} + \bG_2/2) \\
    \hat{\Isym} \hat{\Delta}^{\dagger}_{\alpha \beta}(\mathbf{k}) \hat{\Isym}^{-1} &= \hat{\Delta}^{\dagger}_{\alpha \beta}(\mathbf{k}).
\end{align}
Symmetry actions in hand, we can extract the $\nu_g$'s directly from the pair wavefunction.
From the data shown in Fig.~\ref{fig:pairwf_data_zerodU2} below, we can extract the symmetry indices directly, as shown Table~\ref{tb:pairing_symmetry}. Note that $x0$ has a different symmetry index under the magnetic translation $\hat{m}_x$, which is consistent with it having a different algebraic exponent $\eta_{L_y}$ (see Fig.~\ref{fig:repulsive_SC_scaling}c).
\begin{table}[H]
\begin{center}
\begin{tabular}{||c c c c||} 
 \hline
  $L_y \backslash \alpha \beta$ & $xx$ & $yy$ & $x0$ \\ [0.5ex] 
 \hline\hline
 5 & $(-1, -1, \text{\xmark})$ & $(-1, -1, \text{\xmark})$ & $(-1, -1, \text{\xmark})$ \\ 
 \hline
 6 & $(-1, -1, -1)$ & $(-1, -1, -1)$ & $(-1, -1, +1)$ \\
 \hline
 7 & $(-1, -1, \text{\xmark})$ & $(-1, -1, \text{\xmark})$ & $(-1, -1, \text{\xmark})$ \\ [1ex] 
 \hline
\end{tabular}
\end{center}
\caption{\textbf{Pairing Symmetry of the Superconductor.} Symmetry Indices $(\nu_{C_{2z}}, \nu_{C_{2x}}, \nu_{m_x})$ for the different condensed channels and circumferences. An \xmark$\ $indicates that $m_x$ is not a symmetry for odd circumferences, so that $\nu_{m_x}$ is not well defined.}
\label{tb:pairing_symmetry}
\end{table}

\begin{figure}
    \centering
    \includegraphics[scale=\otherfigscale]{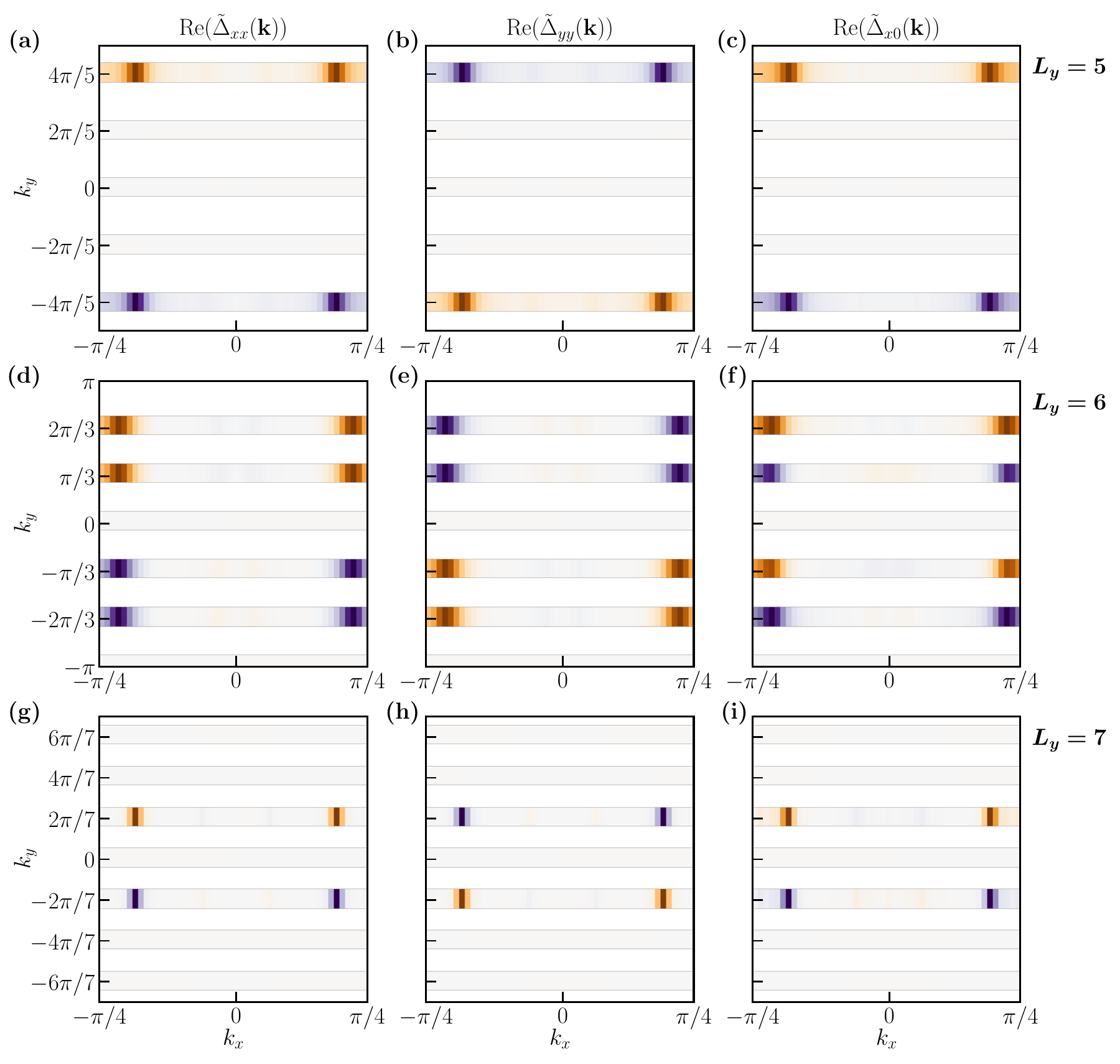}
    \caption{\textbf{Momentum Space Structure of the Pair Wavefunction.} Plot of the proxy pair wavefunction for the two dominant superconducting channels $\alpha\beta = xx,yy$ at all three circumferences $L_y = 5,6,7$, extracted from charge-conserving iDMRG using the methods described in the text.
    Each function is manifestly odd under $C_{2x}$ and $C_{2z}$, with the latter indicating $p$-wave pairing symmetry.
    At $L_y=6$, where the magnetic translation $m_x$ is a symmetry of the finite-circumference system, the pair wavefunction in the $xx$ and $yy$ channels is odd under shifts $k_y\to \pi$, where it is even in the sub-dominant $x0$ channel.
    We remark that the overall normalization of the pair wavefunction---which we have argued is proportional to the true pair wavefunction---is set arbitrarily. In each case, its imaginary part is smaller by orders of magnitude and is therefore omitted. Each plot is computed at $\chi = 12288$.
    }
    \label{fig:pairwf_data_zerodU2}
\end{figure}

\vspace{8mm}\toclesslab\section{Additional Details on the Optical Lattice Implementation}{app:optical_lattice}

In the main text, we outlined a concrete experiment blueprint for realizing the double Hofstadter model using fermionic alkaline-earth atoms in an optical lattice.
In this section of the appendix, we provide additional details of this blueprint.
Namely, in Section~\ref{appsubsec:alkaline-earth}, we start by discussing some general properties of alkaline-earth atoms and how they interact in an optical lattice.
In doing so, we will re-iterate how we envision encoding the degrees of freedom of the double Hofstadter model in the atomic levels of these atoms.
Given this encoding, we will provide a survey of how different species of alkaline-earth atoms interact and present an optical scheme for driving some of these species' native interactions to a regime most favorable for observing the LAF and superconductivity.
%
After this, in Section~\ref{appsubsec-Engineering SingleParticleTerms}, we will describe the laser driving protocol that enables the engineering of the complex hopping of the double Hofstadter model.
To do so, it will be convenient to summarize previous proposals for engineering the single-layer Hofstadter model, which will naturally suggest a generalization to our model.
%

\vspace{6mm}\toclesslab\subsection{Alkaline-Earth Atoms and Their Interactions in an Optical Lattice}{appsubsec:alkaline-earth}

Let us start by briefly reviewing some of the physics of fermionic alkaline-earth atoms.
Such atoms have two valence electrons in their outer $s$-orbital, endowing them with rich atomic properties that can be exploited for the use of quantum simulation.
In particular, they have two low-energy orbital ``clock'' states: the atomic ground state $\ket{g} =\, ^{1}S_0$ as well as a long-lived metastable excited state $\ket{e} =\, ^{3}P_0$.
In these two clock states, the nuclear spin $I$ almost perfectly decouples from the electronic angular momentum because $J = 0$.
Since the strength of the van-der-Waals interactions between such atoms is dominantly controlled by their electronic states~\cite{GorkovSUN}, this decoupling implies that interactions between fermionic alkaline-earth atoms are largely independent of the nuclear spin, aside from constraints due to Pauli exclusion.
This leads to an $SU(N)$ symmetry (with $N = 2I + 1$) in the interacting Hamiltonian of such atoms, constraining certain coupling constants to be equal to one another without the need for fine-tuning.
This motivates our use of such atoms for simulating the double Hofstadter model.

To be more quantitative, consider an alkaline-earth atom whose nuclear spin states are labeled by $\ket{a}$ where $a \in -I, \cdots, I$.
Owing to the aforementioned $SU(N)$ symmetry, to specify the two-body interactions between these atoms, we need only consider them in the following internal states: 
\begin{align}
    \ket{gg}_{ab} &\propto\ket{gg} \otimes \left( \ket{ab} -\ket{ba} \right) \quad \quad \quad \ \qquad \ket{ee}_{ab} \propto\ket{ee} \otimes \left( \ket{ab} -\ket{ba} \right)\\
    \ket{+}_{ab} &\propto (\ket{ge} + \ket{eg}) \otimes \left( \ket{ab} -\ket{ba} \right) \qquad \ket{-}_{ab} \propto (\ket{ge} - \ket{eg}) \otimes \left( \ket{ab} +\ket{ba} \right)
\end{align}
where $a\neq b$ are arbitrary.
The interactions between atoms in these states is characterized by four $s$-wave scattering lengths independent of the specific internal states $a, b$:\footnote{The $p$-wave scattering channel is negligible because it is suppressed at low-temperature.}
\begin{equation}
    a_{gg}, \qquad a_{ee}, \qquad a_{eg}^-, \qquad a_{eg}^+.
\end{equation}
The scattering lengths of different atomic species are specified in Table \ref{tab:scattering_lengths} below. 
\begin{table}[H]
    \centering
    \begin{tabular}{|c||r|r|r|r|}
        \hline
        Species / Scattering Length & $a_{gg}$ & $a_{ee}$ & $a_{eg}^-$ & $a_{eg}^+$  \\ [0.5ex]
        \hline\hline
         $^{173}\text{Yb}\ (I = 5/2)$ \cite{hofer2015observation, scazza2014observation} & $\ 199.4 a_0\ $ & $\ 306.2 a_0\ $ & $\ 219.7 a_0\ $ & $\> 1800 a_0\ $ \\ \hline
         $^{171}\text{Yb}\ (I = 1/2)$ \cite{bettermann2020clock} & $\ -2.8(3.6) a_0\ $ & $\ 104(7) a_0 \ $ & $\ 389(4) a_0\ $ & $\ 240(4) a_0\ $ \\
         \hline
         $^{87}\text{Sr}\ (I = 9/2)$ \cite{Sr87, goban2018emergence} & $\ 96.2(0.1) a_0\ $ & $\ 176.3(9.5) a_0 \ $ & $\ 69.1 a_0\ $ & $\ 161.3 a_0\ $ \\
         \hline
    \end{tabular}
    \caption{Scattering Lengths for Three Different Fermionic Alkaline-Earth Species}
    \label{tab:scattering_lengths}
\end{table}

\subsubsection{Alkaline Earth Atom Hamiltonian}

Scattering lengths in hand, the bare Hamiltonian for alkaline-earth atoms in an optical lattice is given by~\cite{GorkovSUN}:
\begin{align}
    \hat{\mathcal{H}} &= \sum_{\alpha s} \int d^3 \mathbf{r}\, \hat{\psi}_{\alpha s}^{\dagger}(\mathbf{r}) \left( -\frac{\hbar^2}{2M} \nabla^2 + V_{\alpha}(\mathbf{r})\right) \hat{\psi}_{\alpha s}(\mathbf{r})  + \hbar \omega_0 \int d^3 \mathbf{r}\, (\hat{\rho}_e(\mathbf{r}) - \hat{\rho}_g(\mathbf{r})) \\
    &\qquad + \frac{(g_{eg}^+ + g_{eg}^-)}{2} \int d^3 \mathbf{r}\,  \hat{\rho}_{e}(\mathbf{r}) \hat{\rho}_g(\mathbf{r}) + \sum_{\alpha, s < s'} g_{\alpha \alpha} \int d^3\mathbf{r}\, \hat{\rho}_{ \alpha s}(\mathbf{r}) \hat{\rho}_{ \alpha s'}(\mathbf{r})\\
    &\qquad + \frac{g_{eg}^+ - g_{eg}^-}{2} \sum_{s, s'} \int d^3 \mathbf{r}\, \hat{\psi}_{g, s}^{\dagger}(\mathbf{r}) \hat{\psi}_{e, s'}^{\dagger}(\mathbf{r}) \hat{\psi}_{g, s'}(\mathbf{r})\hat{\psi}_{e, s}(\mathbf{r}),
\end{align}
where $\hat{\psi}^{\dagger}_{\alpha s}(\mathbf{r})$ is the fermionic creation operator for atoms in internal states $\ket{\alpha s}$, $\alpha = g, e$ labels the electronic clock states, and $s = -I, \dots, I$ denotes one of the $N = 2I + 1$ nuclear Zeeman states.
Moreover, $\omega_0$ is the bare splitting between the clock states and $V_{\alpha}(\mathbf{r})$ describes an external trapping potential, which could be clock state-dependent, but is assumed to be independent of the nuclear spin.
Here, we have introduced
\begin{equation}
    g_X = \frac{4 \pi \hbar^2 a_X}{M},
\end{equation}
where $M$ is the atomic mass and $a_X$ is the $s$-wave scattering length.
Now, let us assume that $V_{\alpha}(\mathbf{r})$ is periodic with period $a$ in the $xy$-plane and is confining in the $z$-direction.
Then, the spectrum of the first term in the above equation is organized into bands.
Let us further assume that our atoms only occupy the lowest available Bloch band of this potential.
As such, we can project our fermion operators into the lowest band:
\begin{equation}
    \hat{\mathcal{P}} \hat{\psi}_{\alpha s}^{\dagger}(\mathbf{r}) \hat{\mathcal{P}} = \sum_j  \hat{c}^{\dagger}_{j \alpha s} w_{\alpha}^{*}(\mathbf{r} - \mathbf{r}_j),
\end{equation}
where $w_{\alpha}$ are the Wannier functions of the band of $\alpha$ clock states and $\mathbf{r}_j$ is the locations of the lattice sites.
%
The Hamiltonian projected to the lowest band is then
\begin{align}
    \hat{H} &= \hat{\mathcal{P}}\hat{\mathcal{H}}\hat{\mathcal{P}} = \hat{H}_0 + \hat{H}_{\omega} + \hat{H}_{\text{int}},
\end{align}
where we have split up the single-particle terms and the interaction terms each defined as:
\begin{align} \label{eq-bareprojHamOL}
     \hat{H}_0 + \hat{H}_{\omega} &=- \sum_{\langle n, m\rangle, \alpha, s} t_{\alpha} (\hat{c}_{n \alpha s}^{\dagger}\hat{c}_{m \alpha s}  + \text{h.c.}) + \hbar \omega_0 \sum_m (\hat{n}_{m e} - \hat{n}_{m g}), \\
     \hat{H}_{\text{int}} &= \sum_{m, \alpha} \frac{U^{\alpha}_1}{2} \hat{n}_{m \alpha} (\hat{n}_{m \alpha} -1)  + U_2 \sum_m \hat{n}_{me} \hat{n}_{mg} + J_{\text{ex}} \sum_{m, s, s'} \hat{c}_{mgs}^{\dagger} \hat{c}_{mes'}^{\dagger} \hat{c}_{mgs'} \hat{c}_{mes}.
\end{align}
Our goal will be to engineer the above Hamiltonian into the double Hofstadter model with parameters favorable for realizing the LAF and potentially the superconductor.
We will start with engineering the interactions in Sec.~\ref{appsec:engineeringinteractions} and then move to the single-particle terms in Sec.~\ref{appsubsec-Engineering SingleParticleTerms}.
Before starting, let us remind ourselves of the encoding mentioned in the main text:
\begin{equation} \label{eq-nucspinencoding}
    \ket{\mathsf{T} \uparrow} = \ket{g, a} \qquad \ket{\mathsf{T} \downarrow} = \ket{g, b} \qquad \ket{\mathsf{B} \uparrow} = \ket{e, a} \qquad \ket{\mathsf{B} \downarrow} = \ket{e, b}
\end{equation}
where $a \neq b$ are fixed and arbitrary nuclear spin states in $\{-I, \cdots I\}$.
Then, $U_1^{\mathsf{T}/\mathsf{B}}$ will be the in-layer interaction strength, $U_2$ will be the between-layer interaction, and $J_{\text{ex}}$ is an $SU(2)$-symmetric ``Heisenberg'' exchange interaction.
Here, the coupling constants are given by:
\begin{equation}
    t_{\alpha} = -\int d^3 \mathbf{r}\, w^*_{\alpha}(\mathbf{r}) \left[ -\frac{\hbar^2}{2M} \nabla^2 + V_{\alpha}(\mathbf{r}) \right]w_{\alpha}(\mathbf{r} - \mathbf{r}_0) \qquad U_1^{\alpha} = g_{\alpha\alpha} \int d^3\mathbf{r}\, |w_{\alpha}(\mathbf{r})|^4,
\end{equation}
and also $U_{2} = (U_{eg}^+ + U_{eg}^-)/2$ and $J_{\text{ex}} = (U_{eg}^+ - U_{eg}^-)/2$, with
\begin{equation}
    U_{eg}^{\pm} = g_{eg}^{\pm} \int d^3\mathbf{r}\, |w_{e}(\mathbf{r})|^2 |w_{g}(\mathbf{r})|^2.
\end{equation}

\subsubsection{Optically Engineering Interactions Between Alkaline Earth Atoms} \label{appsec:engineeringinteractions}

If we were to not spatially separate the $e$ and $g$ atoms, so that $w_e = w_g$, then the interaction strengths for all atomic species shown in Table~\ref{tab:scattering_lengths} would be in a regime not favorable for seeing the LAF insulator and, therefore, the LAF superconductor.
In particular, the species that have the most favorable scattering lengths are $^{173}\text{Yb}$ and $^{87}\text{Sr}$, owing to the fact that $a_{gg}$ and $a_{ee}$ are comparable.
However, even these species would not be ideal for seeing superconductivity because $a_{eg}^{\pm}$ for both are such that $U_2$ and $J_{\text{ex}}$ are too large and the latter is ferromagnetic, disfavoring anti-ferromagnetism between the layers necessary for LAF order.
Here, we outline a scheme for suppressing $U_2$ and $J_{\text{ex}}$.

\begin{figure}[H]
    \centering
    \includegraphics[width =0.55\textwidth]{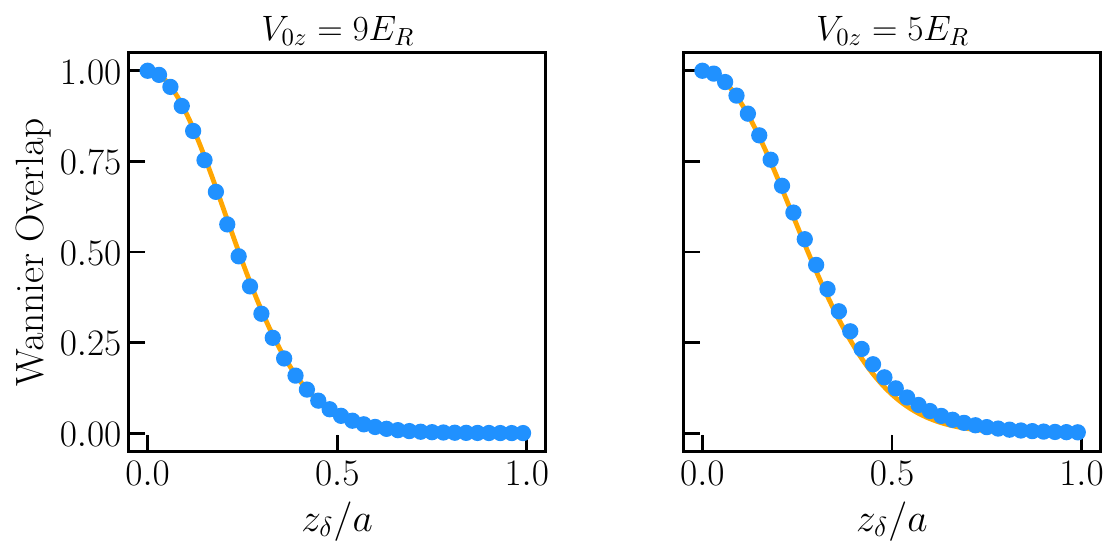}
    \caption{\textbf{Numerically Obtained and Predicted Overlap Integrals Between Clock State Wannier Functions.} In the two panels above, we compare the numerical value of the Wannier overlaps  $\int d^3\mathbf{r}\, |w_{e}(\mathbf{r})|^2 |w_{g}(\mathbf{r})|^2$ with the value predicted by approximating the Wannier function as the lowest harmonic oscillator wavefunction.
    We do so by plotting the overlap integral as a function of the separation $z_{\delta}$ between the centers of the Wannier functions of the two clock states and for two values of the $z$-trap depth $V_{0z}$. 
    Here, the circular dots are obtained numerically and the solid lines are the harmonic oscillator prediction, showing excellent quantitative agreement.}
    \label{fig:overlapsuppression}
\end{figure}

This scheme relies on using a state-dependent optical potential such that $w_e(\mathbf{r})$ and $w_g(\mathbf{r})$ are spatially separated in the $\hat{z}$ direction by $z_{\delta}$.
Then, $U_1^{\mathsf{T}/\mathsf{B}}$ are unaffected but $U_2$ and $J_{\text{ex}}$ decrease with the decreasing overlap between their Wannier functions.
Let us now obtain semi-quantitative estimates on the size of these overlap integrals.
Following a seminal paper by Jacksh et al.~\cite{Jaksch_bosonic_1998}, let us make a harmonic approximation for the potential wells of $V_{\alpha}(\mathbf{r})$, assume that the lattice spacing of the system is $2a = \lambda = 2\pi/k$, and further assume that the depth of the potential is $V_{0i}$ in the direction $i = x, y, z$.
Since our atoms are confined to a 2D plane, we assume that $V_{0z} \gg V_{0, x/y}$.
The frequency of the oscillator is then $\nu_i = \sqrt{4 E_R V_{0i}}/\hbar$, where $E_R = \hbar^2k^2/2M$ is the recoil energy~\cite{Jaksch_bosonic_1998}.
Moreover, the localization length of the harmonic oscillator wavefunction in direction $i$ is
\begin{equation}
    \xi_{i} = \sqrt{\frac{\hbar}{M \nu_i}}  = \frac{1}{\sqrt{2M}}  \left(\frac{2M\hbar^2}{V_{0 i} k^2} \right)^{1/4} = \sqrt{\frac{a}{2\pi}} \left(\frac{2\hbar^2}{V_{0i} M} \right)^{1/4}.
\end{equation}
We henceforth assume that $\xi_x = \xi_y \equiv \xi_{xy}$ and $\xi_z \ll \xi_{xy}$, such that the harmonic wave functions look like
\begin{equation}
    w(x, y, z)  = \frac{1}{\sqrt{ \pi^3 \xi_{xy}^2 \xi_{z}}} e^{-(x^2 + y^2)/(2\xi_{xy}^2)} e^{-z^2/(2\xi_z^2)}.
\end{equation}
Then, the relevant Wannier overlap is given by:
\begin{equation}
    U_{eg}^{\pm} \sim g_{eg}^{\pm}\frac{1}{\xi^4_{xy} \xi^2_z}\int d^3\mathbf{r} \, e^{-|\boldsymbol{\rho}|^2/\xi_{xy}^2} e^{-(z - z_\delta)^2/\xi^2} e^{-z^2/\xi^2}  \sim g_{eg}^{\pm} \frac{1}{\xi^2_{xy}\xi_z} e^{-z_{\delta}^2/(\xi^2_z)}, 
\end{equation}
where $z_{\delta}$ is the separation between the $e$ and $g$ atomic sites and we took $\xi_e = \xi_g$ for simplicity.
This suggests that if we separate the $e$ and $g$ atoms by an amount $z_{\delta}$, we will suppress $U_{eg}^{\pm}$ by an amount that falls off rapidly as a Gaussian in the separation.
The rate of the fall-off can be controlled by increasing the trap depth in the $z$-direction, $V_{0 z}$.
For reference, in Fig.~\ref{fig:overlapsuppression}, we compare the above prediction for the Wannier overlap $ \int d^3\mathbf{r}\, |w_{e}(\mathbf{r})|^2 |w_{g}(\mathbf{r})|^2$ with the true Wannier overlap of numerically-obtained Wannier functions for two values of $V_{0z}$  and as a function of $z_{\delta}$.

Assuming $J_{\text{ex}}$ and $U_2$ are wholly suppressed, the parameters of the resulting model if we take $^{87}\text{Sr}$ atoms are
\begin{equation}
    U_1^{\mathsf{T}} = (96.2\, a_0) u_0, \qquad U_1^{\mathsf{B}} = (176.3\, a_0) u_0 \qquad (^{87}\text{Sr}). 
\end{equation}
Alternatively, taking $^{173}\text{Yb}$ atoms, the parameters are then
\begin{equation}
    U_1^{\mathsf{T}} = (199.4\, a_0) \times u_0, \qquad U_1^{\mathsf{B}} = (306.2\, a_0) \times u_0 \qquad (^{173}\text{Yb}). 
\end{equation}
The latter is more favorable as the in-layer interaction strengths  $U_1^{\mathsf{T}}$ and $U_1^{\mathsf{B}}$ are more balanced. 

\vspace{6mm}\toclesslab\subsection{Engineering the Single Particle Terms}{appsubsec-Engineering SingleParticleTerms}

Having discussed the interaction terms, in this section, we describe in detail how we achieve the double Hofstadter hopping and inter-layer tunneling.
Let us recall from Eq.~\eqref{eq-bareprojHamOL}, that
\begin{equation}
    \hat{H}_0 + \hat{H}_{\omega} = - \sum_{\langle n, m\rangle, \alpha, s} t_{\alpha} (\hat{c}_{n \alpha s}^{\dagger}\hat{c}_{m \alpha s} + \text{h.c.}) + \hbar \omega_0 \sum_m (\hat{n}_{m e} - \hat{n}_{m g}).
\end{equation}
In what follows, we will start by first engineering the hopping term $\hat{H}_0$ to realize double Hofstadter hopping, assuming throughout that $t_e = t_g = t$.
We will start by deriving the result carefully in a single layer after which it will be straightforward to determine the results for the double layer system.
Finally, we will discuss the interlayer tunneling term where we will show how to eliminate the $\omega_0$ term.

\subsubsection{Hofstadter Hopping in a Single Layer}

We start by deriving how Hofstadter hopping is engineered for single orbital fermionic atoms.
Imagine that in our experiment, our fermionic atoms live in an optical lattice generated with a standing wave with wavelength $\lambda = 2a$ (where $a$ is the lattice spacing).
Now, imagine that we further impose another standing wave in the $y$-direction with wavelength $\lambda_L = 2 \lambda$.
This will generate a staggered period-2 superlattice along the $y$-direction.
The resulting Hamiltonian will look like:
\begin{equation}
\hat{H}_{\text{lab},0} = -\sum_{n, m}  \left(t_x \hat{c}_{m + 1, n}^{\dagger} \hat{c}_{m, n}  + t_y \hat{c}_{m, n + 1}^{\dagger}\hat{c}_{m, n} + \text{h.c} \right) + \sum_{m, n} \Delta_n \hat{n}_{m, n} ,
\end{equation}
where $\Delta_n = (-1)^n \Delta/2$.
Now, we imagine that we periodically drive with two pairs of lasers (four laser beams in total; see the two colored arrows in Fig.~7(a) of the main text for a visualization.).
We will label the lasers by $i = \{o, e\}$ where $o$ and $e$ refer to odd and even for reasons that will be clear in just a second.
In particular, the electric field due to the one pair of lasers is (e.g., the two blue lasers shown in Fig.~7(a)) 
\begin{equation}
    E_i(x, y, t) = \left[2E_{1i} e^{i \omega_{i1} t + \phi_{1i}} \cos(k_L y + \phii_i) + E_{2i} e^{i (-k_L x + \omega_{2i} t + \phi_{2i}) } \right].
\end{equation}
The optical potential generated is then
\begin{align}
    V(x, y, t) &= |E(x, y, t)|^2 = 4 E_{1i}^2 \cos^2(k_L y + \phii_i) + E_{2i}^2 + 4 E_{1i} E_{2i} \cos\left( k_L y + \phii_i\right) \cos\left( -k_L x + \omega_i t + \phi_i \right),
\end{align}
where $\omega_i =  \omega_{2i} - \omega_{1i}$ and $\phi_i = \phi_{2i} - \phi_{1i}$.
Now, let us translate what this potential would look like on the lattice: 
\begin{equation}
    V^i_{m, n}(t) = V_0^{sw} \cos^2\left(\frac{n\pi}{2} + \phii_i \right) + V_0 \cos\left( \frac{n \pi}{2} + \phii_i \right)  \cos \left( - \frac{m \pi}{2} + \omega_i t + \phi_i \right),
\end{equation}
corresponding to an additional Hamiltonian term:
\begin{equation}
     \delta \hat{H}_{\text{lab}}(t) = \sum_{n, m} \sum_i V_{m, n}^i(t) \hat{n}_{m, n} = \sum_{n, m} \sum_i  \left[V_0^{sw} \cos^2\left( \frac{n \pi}{2} + \phii_i \right) + V_0 \cos\left( \frac{n \pi}{2} + \phii_i \right)  \cos \left( - \frac{m \pi}{2} + \omega_i t + \phi_i \right)  \right] n_{m, n}.
\end{equation}
Now, note that if we pick $\phii_e - \phii_o = \pi/2$, we get that the standing wave contribution cancels out to a constant. 
Dropping the standing wave contribution, further note that:
\begin{align}
V_{m, n + 1}^i - V_{m, n}^i &= V_0 \left[\cos\left( \frac{n + 1}{2} \pi + \phii_i \right) - \cos\left(\frac{n \pi}{2} + \phii_i \right) \right]\cos \left( - \frac{m \pi}{2} + \omega_i t + \phi_i \right) \\
&= V_0 \left[-\sin\left( \frac{n}{2} \pi + \phii_i \right) - \cos\left(\frac{n \pi}{2} + \phii_i \right) \right]\cos \left( - \frac{m \pi}{2} + \omega_i t + \phi_i \right) \\
&= -\sqrt{2} V_0\sin\left( \frac{n}{2} \pi + \phii_i + \frac{\pi}{4} \right) \cos \left( - \frac{m \pi}{2} + \omega_i t + \phi_i \right). 
\end{align}
Note that, when $\phii_i = \pi/4$, then $V^i_{m, n + 1} - V_{m, n}^i$ will vanish when $n$ is odd. 
Conversely, if $\phii = -\pi/4$, then $V^i_{m, n + 1} - V_{m, n}^i$ will vanish when $n$ is even. As such, we will pick
\begin{equation}\label{appoptical-defofphii}
    \phii_e = \pi/4 \quad \phii_o = -\pi/4,
\end{equation}
which meets the criteria necessary to cancel out the standing wave contribution.
This means that $V^i_{m, n}(t)$ will address even and odd bonds separately.
Let us compute the zeroth-order term in the Magnus expansion to get the effective Hamiltonian.
The full, time-dependent Hamiltonian is given by
\begin{align}
    \hat{H}_{\text{lab},0}(t) &= -\sum_{m, n}  \left(t_x \hat{c}_{m + 1, n}^{\dagger} \hat{c}_{m, n}  + t_y \hat{c}_{m, n + 1}^{\dagger}\hat{c}_{m, n} + \text{h.c.} \right) + \sum_{m, n} \Delta_n \hat{n}_{m, n} \\
    & + \sum_{m, n} \sum_i V_0 \cos \left( \frac{n \pi}{2} + \phii_i \right) \cos \left( - \frac{m \pi}{2} + \omega_i t + \phi_i \right) \hat{n}_{m, n}.
\end{align}
Now, we move to a rotating frame via a unitary:
\begin{align}
    \hat{U}(t) &= \exp\left\{ -i \sum_{m, n}\int^t  \left[ \Delta_n + V_0 \sum_j \cos\left( \frac{n \pi}{2} + \phii_{j} \right) \cos\left(- \frac{m \pi}{2} + \omega_j t' + \phi_j \right) \right] \hat{n}_{m, n}\, dt' \right\}  \\
    &=\exp\left\{ - i \sum_{m, n}   \left[\Delta_{n} t +  \sum_j \frac{V_0}{\omega_j} \cos \left( \frac{n \pi}{2} + \phii_j \right) \sin \left(- \frac{m \pi}{2} + \omega_j t + \phi_j \right) \right] \hat{n}_{m, n} \right\}\\
    &= \exp\left\{ -i \sum_{m, n} \chi_{m, n}(t) \hat{n}_{m, n} \right\}. 
\end{align}
This yields an effective Hamiltonian in the rotating frame given by:
\begin{align}
    \hat{H}_0(t) &= \hat{U}^{\dagger}(t)\hat{H}(t) \hat{U}(t) - i \hat{U}^{\dagger}(t) \partial_t \hat{U}(t) \\
    &= -\sum_{n, m} \left(t_x e^{i \eta^x_{m, n}(t)} \hat{c}_{m + 1, n}^{\dagger} \hat{c}_{m, n}  + t_y e^{i \eta^y_{m, n}(t)} \hat{c}_{m, n + 1}^{\dagger}\hat{c}_{m, n} + \text{h.c}\right), 
\end{align}
where the absence of the $\text{lab}$ subscript on the Hamiltonian indicates that we are in the rotating frame,  $\eta^x_{m, n} = \chi_{m + 1, n} - \chi_{m, n}$, and also $\eta^y_{m, n} = \chi_{m, n + 1} - \chi_{m, n}$. 
%
Therefore, we have that:
\begin{align}
    \eta_{m, n}^y &= \chi_{m, n + 1} - \chi_{m, n} \\
    &=(\Delta_{n + 1} - \Delta_n)t + \sum_j \frac{V_0}{\omega_j} \left[ \cos\left(\frac{(n + 1) \pi}{2} + \phii_j \right) - \cos\left(\frac{n \pi}{2} + \phii_j \right)  \right] \sin\left(-\frac{m \pi}{2} + \omega_j t + \phii_j \right) \\
    &= (-1)^{n + 1}\Delta t - \sum_j  \frac{\sqrt{2}V_0}{\omega_j}\sin\left( \frac{n}{2} \pi + \phii_j + \frac{\pi}{4} \right)  \sin \left(- \frac{m \pi}{2} + \omega_j t + \phi_j \right).
\end{align} 
Now, following our discussion surrounding Eq.~\eqref{appoptical-defofphii}, we set $\phii_e = \pi/4$ and $\phii_o = -\pi/4$.
Thus, we can simplify the above to 
\begin{equation}
    \eta_{m, n}^{y} = \begin{cases} -\Delta t + \cfrac{\sqrt{2} V_0}{\omega_e} \sin\left(-\cfrac{ (n + 1)}{2} \pi \right) \sin\left(- \cfrac{m \pi}{2} + \omega_e t + \phi_e \right) & n \text{ even}, \\ \Delta t + \cfrac{\sqrt{2} V_0}{\omega_e} \sin\left(-\cfrac{n}{2} \pi \right) \sin\left(- \cfrac{m \pi}{2} + \omega_o t + \phi_o \right) & n \text{ odd}. \end{cases}
\end{equation}
Now, note that, if $a \in 2\mathbb{Z} + 1$ is odd, then $\sin\left(\frac{a\pi}{2} \right) \sin(x) =   \sin\left(x + \frac{a - 1}{2} \pi \right)$.
%
Therefore, we can re-write the above as
\begin{align}
     \eta_{m, n}^{y} 
&= \begin{cases}
    -\Delta t + \cfrac{\sqrt{2} V_0}{\omega_e} \sin\left(- \cfrac{(m + n)\pi}{2} + \omega_e t + \phi_e \right) & n \text{ even}
    \\
    \Delta t + \cfrac{\sqrt{2} V_0}{\omega_o} \sin\left(- \cfrac{(m + n-1) \pi}{2} + \omega_o t + \phi_o \right) & n \text{ odd},
\end{cases}
\end{align}
where we took $m \to m + 2$ to get this to a more palatable form.
Now, we can compute the effective $y$-hopping by assuming that $|\omega_e| = |\omega_o| = 2\pi/T \gg t_x, t_y$ and performing a Floquet-Magnus expansion in $1/\omega$.
In particular, take $\omega_e = - \Delta$ and $\omega_o = \Delta$ and let
\begin{equation}
    \hat{H}_{y,0} = \hat{H}_{y,0}^{\text{even}}(t) + \hat{H}_{y,0}^{\text{odd}}(t) = -\sum_{m, n \text{ even } } t_y e^{i \eta_{m, n}(t)} \hat{c}_{m, n + 1}^{\dagger} \hat{c}_{m, n}-\sum_{m, n \text{ odd } } t_y e^{i \eta_{m, n}(t)} \hat{c}_{m, n + 1}^{\dagger} \hat{c}_{m, n} + \text{h.c.}
\end{equation}
We handle the even and odd terms separately.
For our purposes, the zeroth order term in the Magnus expansion is sufficient.
For the even bond Hamiltonian, we can compute this term as:
\begin{align}
    \hat{H}_{\text{eff, even}}^{(0)} &= \frac{1}{T} \int_0^{2\pi/|\omega|} \hat{H}_{\text{even}}(t') dt' \\
    &= -\frac{t_y \Delta}{2\pi} \sum_{m, n \text{ even}} \hat{c}_{m, n + 1}^{\dagger} \hat{c}_{m, n} \int_0^{2\pi/\Delta} \exp\left( - i\Delta t' - i\frac{\sqrt{2} V_0}{\Delta} \sin\left( - \frac{(m + n) \pi}{2} - \Delta t' + \phi_e \right) \right) dt' + \text{h.c.}
\end{align}
where we used that $\omega_e = -\Delta$.
To evaluate this integral, we use the Hansen-Bessel formula $\mathcal{J}_{\nu}(x) = \frac{1}{2\pi} \int_0^{2\pi} e^{i (\nu t - x \sin(t))}$ with $\nu = \mathbb{Z}$ and $\mathcal{J}_{\nu}$ being the Bessel function of the first kind.  To do so, we take $\tau = - \frac{(m + n) \pi}{2} - \Delta t' + \phi_e$. 
As such, the limits of integration go from $\frac{- (m + n) \pi}{2} - \phi_e \to -2\pi + \frac{- (m + n) \pi}{2} - \phi_e$.
The resulting integral is:
\begin{align}
    \hat{H}_{\text{eff, even},0}^{(0)} &= -t_y \Delta \sum_{m, n \text{ even}} \hat{c}_{m, n + 1}^{\dagger} \hat{c}_{m, n} \left( - \frac{1}{2\pi \Delta} \int_{-(m + n)\pi/2 - \phi_e}^{-2\pi - (m +n)\pi/2 - \phi_e} d\tau \exp\left( i \tau + i\frac{(m + n) \pi}{2}  - i\phi_e - i \frac{\sqrt{2} V_0}{\Delta} \sin \left( \tau \right)\right) \right) + \text{h.c.}\\
    &= -t_y \Delta \sum_{m, n \text{ even}} \hat{c}_{m, n}^{\dagger} \hat{c}_{m, n} e^{i\left( \frac{(m + n) \pi}{2} - \phi_e\right)} \left( - \frac{1}{2\pi \Delta} \int_{-(m + n)\pi/2 - \phi_e}^{-2\pi - (m +n)\pi/2 - \phi_e} d\tau \exp\left( i \tau - i \frac{\sqrt{2} V_0}{\Delta} \sin \left( \tau \right)\right) \right)   + \text{h.c.}
\end{align}
Now, note that the integrand is periodic with period $2\pi$. Thus, we can replace the bounds with $0$ to $-2\pi$ and reverse the direction of the integral to get
\begin{align}
    \hat{H}_{\text{eff, even},0}^{(0)} &= -t_y \Delta \sum_{m, n \text{ even}} \hat{c}_{m, n + 1}^{\dagger} \hat{c}_{m, n} e^{i\left( \frac{(m + n) \pi}{2} - \phi_e\right)} \left( \frac{1}{2\pi \Delta} \int_{0}^{2\pi} d\tau \exp\left( i \tau - i \frac{\sqrt{2} V_0}{\Delta} \sin \left( \tau \right)\right) \right) + \text{h.c.}\\
    &=-t_y \mathcal{J}_{1}\left( \frac{\sqrt{2} V_0}{\Delta} \right) \sum_{m, n \text{ even}} e^{i  \left(\frac{(m + n)}{2} \pi - \phi_e\right)} \hat{c}_{m, n + 1}^{\dagger} \hat{c}_{m, n} + \text{h.c.}
\end{align}
For the odd terms, we proceed similarly:
\begin{align}
    \hat{H}_{\text{eff, odd},0}^{(0)} &= \frac{1}{T} \int_0^{2\pi/|\omega|} \hat{H}_{\text{odd}}(t') dt' \\
    &=  - \frac{t_y \Delta}{2\pi}  \sum_{m, n \text{ odd}} \hat{c}_{m, n + 1}^{\dagger} \hat{c}_{m, n} \int_{0}^{2\pi/\Delta} \exp\left( i \Delta t' + i \frac{\sqrt{2} V_0}{\Delta} \sin\left( - \frac{(m + n - 1) \pi}{2} + \Delta t' + \phi_o \right) \right) dt'+ \text{h.c.}
\end{align}
Performing a similar change of variables as for the even case, we arrive at
\begin{align}
    \hat{H}_{\text{eff, odd},0}^{(0)} 
    &= -t_y \sum_{m, n}  \hat{c}_{m, n+1}^{\dagger} \hat{c}_{m, n} e^{i \left( \frac{(m + n -1)\pi}{2} - \phi_o\right)} \left( \frac{1}{2\pi} \int_0^{2\pi} d\tau  \exp\left(i \tau + i \frac{\sqrt{2} V_0}{\Delta} \sin(\tau) \right) \right)+ \text{h.c.} \\
    &= -t_y \sum_{m, n}  \hat{c}_{m, n+1}^{\dagger} \hat{c}_{m, n} e^{i \left( \frac{(m + n -1)\pi}{2} - \phi_o\right)} \left( -\frac{1}{2\pi} \int_0^{-2\pi} d\tau'  \exp\left(-i \tau' - i \frac{\sqrt{2} V_0}{\Delta} \sin(\tau') \right) \right)+ \text{h.c.} \\
    &= -t_y \sum_{m, n}  \hat{c}_{m, n+1}^{\dagger} \hat{c}_{m, n} e^{i \left( \frac{(m + n -1)\pi}{2} - \phi_o\right)} \mathcal{J}_{-1}\left(\frac{\sqrt{2} V_0}{\Delta}\right)+ \text{h.c.} \\
    &= -t_y\mathcal{J}_{1}\left(\frac{\sqrt{2} V_0}{\Delta}\right) \sum_{m, n}  \hat{c}_{m, n+1}^{\dagger} \hat{c}_{m, n} e^{i \left(\frac{(m + n +1)\pi}{2} - \phi_o\right)}+ \text{h.c.},
\end{align}
where in the last line, we used the fact that $J_{\nu}(x) = (-1)^{\nu} J_{-\nu}(x)$. 
The coefficient of the $x$-hopping term follows via a similar derivation of the above and is not repeated for sake of brevity.
Thus, in total, the effective Hamiltonian is
\begin{equation} \label{appeq-singleHof}
    \hat{H}_{\text{eff},0}^{(0)} = -t_y\mathcal{J}_1\left(\frac{\sqrt{2} V_0}{\Delta}\right) \sum_{m, n}  e^{i \phi_{m, n}} \hat{c}_{m, n+1}^{\dagger} \hat{c}_{m, n} - t_x \mathcal{J}_0\left(\frac{\sqrt{2} V_0}{\Delta} \right) \sum_{m, n} \hat{c}_{m + 1, n}^{\dagger} \hat{c}_{m, n} + \text{h.c.},
\end{equation}
where
\begin{equation}\label{appeq-singleHof2}
\phi_{m, n} =  \begin{cases} \frac{\pi}{2} (m + n) - \phi_e & n \text{ even}, \\ \frac{\pi}{2} (m + n + 1) - \phi_o & n \text{ odd}. \end{cases}
\end{equation}

\subsubsection{Double Hofstadter Hopping}

As mentioned in the main text, by using an anti-magic wavelength period-2 superlattice, we can make it so that the staggered potential felt by states in the $\ket{g}$ clock state is opposite from the staggered potential felt by states in the $\ket{e}$ clock states.
As such, the full, time-dependent Hamiltonian in the presence of the two clock states can be expressed as:
\begin{align}
    \hat{H}_{\text{lab},0}(t) &= -\sum_{m, n, \ell}  \left(t_x \hat{c}_{m + 1, n, \ell}^{\dagger} \hat{c}_{m, n, \ell}  + t_y \hat{c}_{m, n + 1, \ell}^{\dagger}\hat{c}_{m, n, \ell} + \text{h.c.} \right) + \sum_{m, n, \ell} (-1)^{\ell} \Delta_n \hat{n}_{m, n} \\
    & + \sum_{m, n, \ell} \sum_i V_0 \cos \left( \frac{n \pi}{2} + \phii_i \right) \cos \left( - \frac{m \pi}{2} + \omega_i t + \phi_i \right) \hat{n}_{m, n, \ell},
\end{align}
where $\ell$ labels the $g$ and $e$ clock states, or in the parlance of the double Hofstadter model, the top and bottom layers, $\mathsf{T}, \mathsf{B}$.
Furthermore, $(-1)^{\mathsf{T}/\mathsf{B}} = \pm 1$.
If we go into the rotating frame, analogous to the single layer case, the resulting Hamiltonian is:
\begin{align}
    \hat{H}_{0}(t) &= \hat{U}^{\dagger}(t)\hat{H}(t) \hat{U}(t) - i \hat{U}^{\dagger}(t) \partial_t \hat{U}(t) \\
    &= -\sum_{n, m} \left(t_x e^{i \eta^x_{m, n, \ell}(t)} \hat{c}_{m + 1, n, \ell}^{\dagger} \hat{c}_{m, n, \ell}  + t_y e^{i \eta^y_{m, n, \ell}(t)} \hat{c}_{m, n + 1, \ell}^{\dagger}\hat{c}_{m, n, \ell} + \text{h.c.}\right). 
\end{align}
Here, $\eta_{m, n, \ell}^x$ is independent of layer and is identical to the single layer case. 
Conversely, 
\begin{align}
     \eta_{m, n, \ell}^{y}(t) 
&= \begin{cases}
    -\Delta t + \cfrac{\sqrt{2} V_0}{\omega_e} \sin\left(- \cfrac{(m + n)\pi}{2} + \omega_e t + \phi_e \right) & n \text{ even and } \ell = \mathsf{T}
    \\
    \Delta t + \cfrac{\sqrt{2} V_0}{\omega_o} \sin\left(- \cfrac{(m + n-1) \pi}{2} + \omega_o t + \phi_o \right) & n \text{ odd and }  \ell = \mathsf{T} \\
    \Delta t + \cfrac{\sqrt{2} V_0}{\omega_e} \sin\left(- \cfrac{(m + n)\pi}{2} + \omega_e t + \phi_e \right) & n \text{ even and } \ell = \mathsf{B}
    \\
    -\Delta t + \cfrac{\sqrt{2} V_0}{\omega_o} \sin\left(- \cfrac{(m + n-1) \pi}{2} + \omega_o t + \phi_o \right) & n \text{ odd and }  \ell = \mathsf{B}
\end{cases}
\end{align}
Notice that $\eta_{m, n, \mathsf{T}}^y(t)$ is identical to the single-layer case worked out in detail in the previous subsubsection and the bottom layer has a similar form but with the linear in $t$ term having flipped signs.
Taking $\omega_e = -\Delta$ and $\omega_o = \Delta$ and following the analytical approach used in the previous subsubsection, it is straightforward to go to leading order in the Floquet-Magnus expansion, which yields the following effective Hamiltonian:
\begin{equation} \label{appeq-doubleHof}
    \hat{H}_{\text{eff},0}^{(0)} = -t_y\mathcal{J}_1\left(\frac{\sqrt{2} V_0}{\Delta}\right) \sum_{m, n, \ell}  e^{i \phi_{m, n, \ell}} \hat{c}_{m, n+1, \ell}^{\dagger} \hat{c}_{m, n, \ell} - t_x \mathcal{J}_0\left(\frac{\sqrt{2} V_0}{\Delta} \right) \sum_{m, n, \ell} \hat{c}_{m + 1, n, \ell}^{\dagger} \hat{c}_{m, n, \ell} + \text{h.c.},
\end{equation}
where
\begin{equation}\label{appeq-doubleHof2}
\phi_{m, n, \ell} =  \pi \delta_{\ell, \mathsf{B}} + (-1)^{\ell} \begin{cases} \frac{\pi}{2} (m + n) - \phi_e & n \text{ even,} \\ \frac{\pi}{2} (m + n + 1) - \phi_o & n \text{ odd.} \end{cases}
\end{equation}
In an experiment, it is possible, albeit challenging, to stabilize $\phi_{e/o}$ to be $0$.
The resulting phase pattern is then equivalent up to an electromagnetic gauge transformation to the double Hofstadter hopping presented in the main text.
This can be seen by evaluating gauge invariant primitive fluxes.
The fluxes in the top and bottom layers are
\begin{equation}
    \Phi^{m, n}_{\mathsf{T}} = \phi_{m + 1, n} - \phi_{m, n} = \frac{\pi}{2} = -\Phi^{m, n}_{\mathsf{B}}.
\end{equation}
Moreover, the flux between the layers is
\begin{equation}
    \Phi_{\text{b/w}}^{m, n} = \phi_{m, n, \mathsf{T}} - \phi_{m, n, \mathsf{B}} =  \pi + 2\phi_{m, n} = \pi + \begin{cases} \pi (m + n) & n \text{ even} \\ \pi (m + n +1) & n \text{ odd} \end{cases} = \pi (m + 1),
\end{equation}
where we used the fact that $\pi n$ and $\pi (n +1)$ are some multiple of $2\pi$ for $n$ even and odd respectively.
This flux pattern is identical to the original double Hofstadter model presented in our work.

\subsubsection{Engineering the Interlayer Tunneling}

Having discussed how to engineer $\hat{H}_0$, we now discuss how to engineer the inter-layer tunneling term and suppress the bare splitting between the two clock states.
As mentioned in the main text, inter-layer tunneling can be implemented by using a laser that addresses the clock transition between $\ket{e}$ and $\ket{g}$.
For a laser of frequency $\nu$, the relevant terms in the Hamiltonian will be:
\begin{equation}
    \hat{H}_{\omega} + \hat{H}_g(t) = \hbar \omega_0 \sum_{m} (\hat{n}_{me} - \hat{n}_{mg}) +   2\Omega \widetilde{g}\cos(\nu t)\sum_m (c_{me}^{\dagger} c_{mg} + \text{h.c.}),    
\end{equation}
where $\Omega$ is the Rabi frequency of the laser drive and $\widetilde{g}$ is set by the Wannier overlap of the $e$ and $g$ states:
\begin{equation}
    \tilde{g} = \int w_e^*(\mathbf{r}) w_g(\mathbf{r})\, d^3 \mathbf{r},
\end{equation}
which can be chosen to be real for an appropriate $U(1)$ phase of our fermionic orbitals.
To remove the time dependence, let us go to an appropriate rotating frame via:\footnote{Note that this unitary commutes with the rest of the Hamiltonian as the rest of the Hamiltonian is invariant under a layer $U(1)$ symmetry. Thus, going to this rotating frame does not affect any other term of the Hamiltonian.}
\begin{equation}
    \hat{U}(t) = \exp\left( -i \frac{\nu}{2} \sum_m (\hat{n}_{me} - n_{mg})\right)
\end{equation}
and work at lowest order in the Magnus expansion of the resulting Hamiltonian.
This yields an effective Hamiltonian: 
\begin{align}
    \hat{H}_{\text{eff}, \omega + g} &=  \frac{1}{T}\int_0^{T} dt\, \left(\hat{U}^{\dagger}(t) \left(\hat{H}_{\omega} + \hat{H}_g(t) \right) U(t)  - i U^{\dagger}(t) \partial_t U(t) \right) \\
    &= \hbar (\omega_0 - \nu) \sum_{m} (\hat{n}_{me} - \hat{n}_{mg}) + \Omega \widetilde{g}  \sum_m c_{me}^{\dagger} c_{mg} + \text{h.c.}\\
    &= \hbar (\omega_0 - \nu) \sum_{m} (\hat{n}_{m\mathsf{T}} - \hat{n}_{m\mathsf{B}}) + \Omega \widetilde{g}  \sum_m c_{m\mathsf{T}}^{\dagger} c_{m\mathsf{B}} + \text{h.c.},
\end{align}
where in the last line, we passed to the encoding of Eq.~\eqref{eq-nucspinencoding}.
If the drive is chosen to be on-resonance $(\omega_0 = \nu)$, we have only the inter-layer tunneling terms.
Note that, due to the spatial separation of the Wannier orbitals $w_e$ and $w_g$ used to engineer the interactions (See Sec.~\ref{appsec:engineeringinteractions}), the size of $\widetilde{g}$ is suppressed as well.
For a semi-quantitative estimate of the suppression, we approximate our Wannier orbitals in a harmonic form as in Sec.~\ref{appsec:engineeringinteractions} which yields
\begin{equation}
    \widetilde{g} \sim e^{-z_{\delta}^2/2\xi_{z}^2}. 
\end{equation}
However, given that we can tune the strength of $\Omega$, in principle, the tunneling amplitude can be tuned arbitrarily.

\vspace{8mm}\toclesslab\section{Additional Details on the Relationship Between Double Hofstadter Model and Chiral TBG}{app:Dictionary}

In the main text, we described the similarities between the double Hofstadter model and chiral models of twisted bilayer graphene (TBG).
In this section, we provide additional details regarding these similiarities.
To do so, let us briefly review the form of the interacting Hamiltonian in chiral models of TBG.
For a more complete and pedagogical review, we refer the reader to Ref.~\onlinecite{pedagogical_patrick}.

\vspace{6mm}\toclesslab\subsection{Brief Review of Chiral TBG}{}

Recall that electrons in TBG carry a conserved electronic spin $(s = \Uparrow, \Downarrow)$ and valley $(\tau = K, K')$ and can occupy one of two energetically close but distinct bands.
The narrow bandwidth of the flat band subspace suggests that other linear combinations of these bands, distinct from the single-particle energy eigenbasis, may be more relevant for studying the interacting physics of TBG. 
One natural choice recognized in Refs.~\cite{kang2019strong,bultinck2020ground} is the sublattice polarized basis which
distinguishes bands in the subspace based on their weight on the two sites $\sigma = A, B$ of the honeycomb lattice of graphene.
This sublattice polarization becomes maximal
in the non-physical limit where intervalley hopping can only occur between opposite sublattices (i.e. only between $A$ sites on the top layer to $B$ sites on the bottom, or $B$ on top to $A$ on the bottom). This is called the ``chiral limit of TBG'' \cite{TarnoChiral,BernevigTBGIV, pedagogical_patrick}.

A crucial property of these sublattice polarized bands is that they carry a definite Chern number.
In particular, let $\sigma^{\alpha}, \tau^{\alpha}, s^{\alpha}$ for $\alpha = 0, x, y, z$ correspond to the Pauli matrices of the sublattice, valley, and spin respectively.
Then, the Chern number is labeled by $\mathcal{C} = \sigma^z \tau^z$.
This motivates organizing the bands by their Chern sector $\sigma \tau$, their valley $\tau$, and their electronic spin $s$.
A complete set of commuting Pauli matrices, that define how to measure and flip each such degree of freedom independently, can then be defined as:
\begin{equation}
    \boldsymbol{\eta} = (\sigma^x \tau^x, \sigma^x \tau^y, \tau^z) \qquad \widetilde{\boldsymbol{\eta}} = (s^x, s^y, s^z) \qquad \boldsymbol{\gamma} = (\sigma^x , \sigma^y \tau^z, \sigma^z \tau^z),
\end{equation}
corresponding to a valley pseudospin, the electronic spin, and the Chern label.
In this ``strong coupling basis'' of the flat bands, the band-projected Hamiltonian of chiral-limit TBG takes a particularly simple form:
\begin{align} \label{eq-chiralTBGHam}
    \hat{H}_{\text{TBG}} = \hat{h}_{\text{TBG}} + \hat{\mcV}_{\text{TBG}}, \qquad 
    \hat{h}_{\text{TBG}} = \sum_{\mathbf{k} \in \text{BZ}} \hat{\phii}_{\mathbf{k}}^{\dagger} \left[\gamma^x h^{\text{TBG}}_x(\mathbf{k}) + \gamma^y h^{\text{TBG}}_y(\mathbf{k})\right] \hat{\phii}_{\mathbf{k}},\qquad 
    \hat{V}_{\text{TBG}} = \frac{1}{2A}\sum_{\mathbf{q} \in \mathbb{R}^2} V_{\mathbf{q}} \delta \hat{\rho}_{\mathbf{q}}\delta \hat{\rho}_{-\mathbf{q}},   
\end{align}
where $\phii^{\dagger}$ is a fermionic creation operators organized into a spinor in the Chern, spin, and valley pseudospin, evidently the dispersion is purely off-diagonal in the Chern sector, $A$ is the area of unit cell of TBG, and the interactions are written in terms of relative densities $\delta \hat{\rho}_{\mathbf{q}} = \rho_{\mathbf{q}} - \bar{\rho}_{\mathbf{q}}$.
Here, the relative densities are defined from:
\begin{equation}
    \hat{\rho}_{ \mathbf{q}} = \sum_{\mathbf{k}} \hat{\phii}_{\mathbf{k}}^{\dagger} \Lambda_{\mathbf{q}}(\mathbf{k}) \hat{\phii}_{\mathbf{k} + \mathbf{q}}, \qquad \bar{\rho}_{\mathbf{q}} = \frac{1}{2} \sum_{\mathbf{k}, \mathbf{G}} \delta_{\mathbf{q}, \mathbf{G}} \operatorname{Tr} \left[\Lambda_{\mathbf{G}}(\mathbf{k})\right], \qquad
    \Lambda_{\mathbf{q}}(\mathbf{k}) = F^S_{\mathbf{q}}(\mathbf{k}) e^{i \Phi^S_{\mathbf{q}}(\mathbf{k}) \gamma^z},
\end{equation}
where the form of the density operator arises from projecting the microscopic number density operator into the flat bands.
Note that, manifestly, the interacting portion of the Hamiltonian is diagonal in the Chern number and is independent of spin or valley pseudospin.
As a consequence, this term in the Hamiltonian has a $U(4)_{+} \times U(4)_{-}$ symmetry corresponding to independent rotations of both the spin and valley pseudospin in either Chern sector.
At the magic angle, when the dispersion $\hat{h}_{\text{TBG}}$ identically vanishes, this enlarged symmetry is precisely the symmetry of the chiral TBG Hamiltonian.
However, a departure from the magic angle re-introduces the dispersion which, while independent of both spin or valley pseudospin, mixes the two Chern sectors.
As such, its inclusion breaks the $U(4)_{+} \times U(4)_{-}$ symmetry down to a single $U(4)$.

As stated in the text, a number of experiments in TBG suggest some degree of flavor polarization.
As such, in literature, this $U(4) \times U(4)$ symmetric model is typically reduced to a $U(2) \times U(2)$ symmetric model with half the number of degrees of freedom.
%
Here $U(2)$ describes rotations of the unpolarized degree of freedom, which are generated by either $\boldsymbol{\eta}$ for ``Route 1" or $\tilde{\boldsymbol{\eta}}$ for ``Route 2".
The form of this $U(2) \times U(2)$ symmetric model is identical to Eq.~\eqref{eq-chiralTBGHam} but where the electrons $\hat{\phii}^{\dagger}_{\mathbf{k} \gamma \eta \wilde{\eta}}$ are replaced with $\hat{\phii}^{\dagger}_{\mathbf{k} \gamma \eta}$ or $\hat{\phii}^{\dagger}_{\mathbf{k} \gamma \tilde{\eta}}$.


\vspace{6mm}\toclesslab\subsection{Comparison Between the Double Hofstadter Model and Chiral TBG}{}

In this section, we establish a correspondence between the chiral TBG Hamiltonian and that of the band-projected double Hofstadter model. Recall from Sec.~\ref{app:strong_coup} that the latter can be expressed in the ``strong-coupling'' form
\begin{equation}
    \hat{\mathcal{H}} = \delta \hat{h} + \hat{\mcV} , \qquad 
    \delta \hat{h} = \sum_{\mathbf{k} \in \text{BZ}} \hat{\phii}_{\mathbf{k}}^{\dagger} \left\{ \underbrace{\left[l^x h_x(\mathbf{k}) + l^y h_y(\mathbf{k}) \right]}_{\delta h_g(\mathbf{k})} + \underbrace{l^0 h_0(\mathbf{k})}_{\delta h_0(\mathbf{k})} \right\} \hat{\phii}_{\mathbf{k}},\qquad 
   \hat{\mcV} = \sum_{\mathbf{q} \in \text{EBZ}} V_{\ell \ell'}(\bq) \delta \hat{\rho}_{\mathbf{q}, \ell}\delta \hat{\rho}_{-\mathbf{q}, \ell'},  
\end{equation}
where the interactions are written in terms relative densities $\delta \hat\rho_{\mathbf{q}\ell} = \hat{\rho}_{\mathbf{q}\ell} - \bar{\rho}_{\mathbf{q} \ell}$ defined from:
\begin{equation}
    \hat{\rho}_{\bq\ell} = \sum_{\mathbf{k}} \hat{\phii}_{\mathbf{k}}^{\dagger} \Lambda_{\bq\ell}(\mathbf{k}) \hat{\phii}_{\mathbf{k} + \mathbf{q}},
    \qquad
    \bar{\rho}_{\mathbf{q} \ell} = \frac{1}{\sqrt{qN}} \sum_{\mathbf{k}, \mathbf{G}} \delta_{\mathbf{q}, \mathbf{G}} \text{Tr} \left[\Lambda_{\mathbf{G}\ell}(\mathbf{k})\right].
\end{equation}
When $U_1 = U_2$ and $g = 0$, the Hamiltonian is greatly simplified.
In particular, it can be re-expressed as
\begin{equation}
    \hat{\mathcal{H}} = \sum_{\mathbf{k} \in \text{BZ}} \hat{\phii}_{\mathbf{k}}^{\dagger} l^0 h_0(\mathbf{k}) \hat{\phii}_{\mathbf{k}}  + \sum_{\mathbf{q} \in \text{EBZ}} V(\mathbf{q}) \delta \hat{\rho}_{\mathbf{q}}\delta \hat{\rho}_{-\mathbf{q}},
\end{equation}
where $\delta \hat{\rho}_{\mathbf{q}} = \sum_{\ell} \delta \hat{\rho}_{\mathbf{q}\ell}$ and the corresponding layer-summed form factor is constrained to have the following structure:
\begin{align}
    \Lambda_{\mathbf{q}}(\mathbf{k}) = \sum_{\ell} \Lambda_{\mathbf{q}\ell}(\mathbf{k}) = F_{\mathbf{q}}^{S} (\mathbf{k}) e^{i \Phi_{\mathbf{q}}(\mathbf{k}) l^z}.
\end{align}
%
If we neglect the small bare Hofstadter dispersion $h_0(\mathbf{k})$, the resulting Hamiltonian has \textit{precisely} the same form as the chiral TBG Hamiltonian at the magic angle under the mapping
\begin{align}
    \boldsymbol{l} \longleftrightarrow \boldsymbol{\gamma}, \qquad 
    \mathbf{s} \longleftrightarrow \begin{cases}
        \v{\eta} & \text{ Map 1}\\
        \v{\tilde{\eta}} & \text{ Map 2.}
    \end{cases}
\end{align}
Both models have a $U(2) \times U(2)$ symmetry corresponding to (pseudo)-spin rotations in distinct Chern sectors.
As an added remark, in both models, those Chern sectors are related by a spacetime inversion symmetry $\mathcal{I}$, which in TBG is written in the literature as $C_{2z}\mathcal{T}$.
The effect of introducing $g$ into the double Hofstadter model restores the terms $\delta h_g(\mathbf{k})$ in the dispersion and modifies the form factors to $\Lambda_{\mathbf{q}}(\mathbf{k}) = F_{\mathbf{q}}^{S} (\mathbf{k}) e^{i \Phi_{\mathbf{q}}(\mathbf{k}) l^z} + F_{\mathbf{q}}^{A}(\mathbf{k})  l^xe^{i \Phi_{\mathbf{q}}(\mathbf{k}) l^z} $.
The analog of $\delta h_g$ is already present in chiral TBG (not necessarily at the magic angle) as a dispersion.
%
However, the additional term $F_{\mathbf{q}}^{A}(\mathbf{k})  l^xe^{i \Phi_{\mathbf{q}}(\mathbf{k}) l^z}$ in the form factor has no direct analog in chiral TBG and scales linearly with $g$.
Nonetheless, provided that $g$ is small, the double Hofstadter model at $U_1 = U_2$ can be thought of as being ``perturbatively related'' to chiral TBG, even away from the magic angle.
%




\vfill

\end{document}